\documentclass[a4paper,11pt]{article}
\usepackage{setspace}
\usepackage{jheppub} 

\usepackage{dsfont}
\usepackage{amsmath}
\usepackage{tensor}
\usepackage{mathtools}
\usepackage{enumitem}
\usepackage{physics}
\usepackage{hyperref}
\usepackage{tcolorbox}
\usepackage{bbold}
\usepackage{tikz-cd}
\usepackage{rotating}
\usepackage{arydshln}
\usepackage{xcolor}
\usepackage{ulem}
\usepackage{amsthm}

\newcommand{\be}{\begin{equation}}
\newcommand{\ee}{\end{equation}}
\def\bea{\begin{eqnarray}}\def\eea{\end{eqnarray}}
\newcommand{\CR}{\nonumber \\*}
\newcommand{\sLambda}{I\hspace{-0.79mm}I}

\newcommand{\colvec}[2][.8]{\scalebox{#1}{\renewcommand{\arraystretch}{.8}$\begin{matrix}#2\end{matrix}$}}

\newcommand{\R}{\mathbb{R}}
\newcommand{\C}{\mathbb{C}}
\newcommand{\Z}{\mathbb{Z}}

\newcommand{\mM}{\mathcal{M}}

\newcommand{\SL}{\text{SL}}
\newcommand{\U}{\text{U}}
\newcommand{\SO}{\text{SO}}

\definecolor{darkred}{rgb}{0.65,0.15,0}
\definecolor{darkgreen}{rgb}{0.42, 0.46, 0.14}
\definecolor{darkcerulean}{rgb}{0.03, 0.27, 0.49}

\makeatletter
\newcommand{\Rom}[1]{\uppercase\expandafter{\romannumeral #1\relax}}
\newcommand{\rom}[1]{\lowercase\expandafter{\romannumeral #1\relax}}

\def\longeq#1{
\begin{equation}
    \begin{aligned}
    #1
    \end{aligned}
\end{equation}
}

\def\lceq#1{
\begin{equation}
\begin{gathered}
    #1
\end{gathered}
\end{equation}
}

\def\mx#1{
\left(
\begin{matrix}
#1
\end{matrix}
\right)
}

\def\mathtitle#1#2{\texorpdfstring{$#1$}{#2}}

\DeclareUnicodeCharacter{0301}{\'{a}}
\DeclareUnicodeCharacter{0303}{\'{n}}

\numberwithin{thrm}{section}
\numberwithin{defn}{section}
\numberwithin{lem}{section}

\title{Uniqueness of D=8 minimal  supergravity with two vector multiplets}

\author[a]{Guillaume Bossard,}
\author[b]{Bingxin Lao}
\author[c]{and Ruben Minasian}

\affiliation[a]{Centre de Physique Th\'eorique, CNRS, Institut Polytechnique de Paris,
91128 Palaiseau cedex, France}
\affiliation[b]{Department of Physics, Princeton University, Princeton, NJ 08544, U.S.A.}
\affiliation[c]{Institut de Physique Th\'{e}orique, Universit\'{e} Paris-Saclay, CNRS, CEA, F-9119, Gif-sur-Yvette,
France}

\emailAdd{guillaume.bossard@polytechnique.edu}
\emailAdd{bingxin.lao@princeton.edu}
\emailAdd{ruben.minasian@ipht.fr}

\abstract{There exist only four known string theories with minimal supersymmetry in eight dimensions, whose low energy effective descriptions are given by minimal supergravity coupled to $l=18$, $10$, or $2$ vector multiplets. It has been argued that these numbers are uniquely fixed by consistency conditions of the effective theory. In this work, we investigate the protected couplings of minimal supergravity coupled to two vector multiplets under the assumption that the theory admits a duality symmetry of the form $\Gamma_0(N)\times \Gamma_0(N)\subset {\rm SL}(2,\Z) \times {\rm SL}(2,\Z)$ for some positive integer $N$. We show that consistency with anomaly cancellation, higher-dimensional uplifts, and gauge enhancement loci strongly constrains the allowed couplings and isolates a unique consistent theory in the BPS sector. In particular, we find that the Bianchi identity for the three-form flux cannot have gravitational contributions.}

\keywords{String Duality, Superstrings and Heterotic Strings, Supersymmetric Effective Theories, Supersymmetry and Duality}

\begin{document}
\begin{flushright}
CPHT-RR016.052026
\end{flushright}

\vspace{10pt}
\maketitle
\flushbottom

\section{Introduction and discussion} \label{sec:intro}


\paragraph{} \hskip -3mm
Recent years have been marked by efforts to specify the general contours and constraints on the low energy actions that make them admissible candidates for consistent theories of quantum gravity. Mapping out a full picture of admissible theories in all lower dimensions, even simply determining  for which dimensions/numbers of supercharges the number of consistent theories is finite,  would be a remarkable achievement.  In dimensions equal or less than six such a task seems to be pretty far fetched right now. Dimensions $D=7$ and higher, where only theories with 16 and 32 supercharges exist, seem to be a good place to start. 

It is commonly believed that there is a unique ultra-violet completion of maximal supergravity in $D$ dimensions, i.e. with 32 supercharges, realised as type II string theory on a torus. The two-derivative Lagrangian is uniquely determined by supersymmetry in all dimensions \cite{Cremmer:1978km,Cremmer:1979up,Howe:1983sra}. In eleven dimensions there is no moduli space. Assuming that the M5-brane must exist as a supersymmetric extended object coupled to the six-form, the cancellation of the anomaly inflow 
determines the leading correction in the effective action \cite{Duff:1995wd,Witten:1996hc}, which in turn determines by supersymmetry the effective action up to order $\nabla^6 R^4$. 

In lower dimensions, there is a moduli space of vacua determined by the asymptotic values of the scalar fields. The singularities of the effective action in the moduli space must be resolved by the introduction of massless fields. In maximal supergravity, the only one-half BPS  massive multiplet that can consistently become massless is a spin-2 Kaluza--Klein multiplet. In particular, maximal supergravity in eight dimensions has moduli space $SL(2,\R) / SO(2) \times SL(3,\R) / SO(3)$. The cusp of $SL(2,\R)$ can only be interpreted as a decompactification to eleven dimensions and $SL(3,\Z)$ must appear as the diffeomorphisms of the torus $T^3$. The appropriate cusp of  $SL(3,\R)$ is similarly interpreted as a decompactification to ten dimensions type IIB and $SL(2,\Z)$ must appear as the diffeomorphisms of the torus $T^2$.\footnote{The singularity at the conjugate cusp of $SL(3,\R)$ is only resolved by including both one-half and one-quarter massless states that correspond to strings wrapping the torus in type II string theory. The uniqueness of the spectrum does not follow by supersymmetry in this case.} Supersymmetry and duality symmetry eventually allow to fully determine the coupling up to $\nabla^6 R^4$ \cite{Green:1981yb,Green:1997tv,Green:1997di,Berkovits:1997pj,Pioline:1998mn,Green:1998by,Obers:1999um,Green:1999pv,Kazhdan:2001nx,Basu:2008cf,Green:2005ba,Pioline:2010kb,Green:2011vz,Bossard:2014lra,Bossard:2014aea,Gustafsson:2014iva,Bossard:2015uga,Gourevitch:2019knu,Bossard:2020xod}.

\paragraph{} \hskip -3mm
The two-derivative Lagrangian is also completely determined by supersymmetry for theories with 16 supercharges \cite{Cremmer:1977tt,Chapline:1982ww,Bergshoeff:1985ms}. In ten dimensions, the cancellation of anomalies determines the gauge group to be either $E_8 \times E_8$ or Spin$(32)/\Z_2$ \cite{Green:1984sg,Adams:2010zy}, and the Green--Schwarz mechanism moreover determines the protected couplings of type $R^2$, $B F^4$ and $B R^4$.

In this paper we concentrate on (minimal) supergravity in eight dimensions. It is widely believed that all the string theories with local supersymmetry in eight dimensions have been constructed. 

Classically, the 8D $\mathcal{N} = 1$ supergravity multiplet, made of a graviton, $B$-field, dilaton, two vector fields as well as spin-$\frac{3}{2}$  and spin-$\frac{1}{2}$ Majorana fermions (gravitino and a dilatino), can be coupled to any number of vector multiplets each comprising a vector field (photon), a gaugino (spin-$\frac{1}{2}$ Majorana fermion) and one complex scalar \cite{1985_PLB_Salam}. Denoting the number of vector multiplets by $l$, the $l$ complex scalars in the matter sector  parametrize  the locally symmetric K\"ahler manifold
\lceq{\label{eq:modsp}
\mM =\bigl( \U(1) \times \SO(l)\bigr) \backslash  \SO(2,l) / \Gamma \, ,
}
where $\Gamma$ is a discrete subgroup of $\SO(2,l)$. These $l$ vectors together with two vectors in the gravity multiplet transform in the vector representation of $\SO(2,l)$.

Several physical arguments lead to the conclusion that the theory can only be consistent at the quantum level if $l = 2$, $10$ or $18$. One may for example argue that there must exist a consistent worldvolume field theory for each charged BPS object coupled to each gauge field of the theory \cite{Banks:2010zn}. Considering in particular a 1/2 BPS string coupled to the supergravity $B$ field in $D$ dimensions, one argues that the number of vector multiplets is bounded from above as $26-D$ \cite{Kim:2019ths}.  For D=8 this would imply $l \leq 18$. Compactifying eight-dimensional theories on particular backgrounds and notably using six-dimensional anomaly cancellation (in spacetime and string world-sheet) for theories obtained this way, it has been argued that  the only admissible values of $l$ are $l=2$, $l=10$ and $l=18$ \cite{ Montero2021}. These values of $l$ are also the only ones for which string theoretic constructions exist.

The landscape of string theories is quite well studied, and notably a full list of maximal symmetry enhancements, i.e. loci in moduli space where  the gauge group is fully non-abelian (not simple) has been tabulated  \cite{Font:2020rsk, Font:2021uyw}. 
The symmetry enhancement (as well as the rank of the gauge algebra coupled to string probes) as predicted by the consistency in  supergravity  \cite{Hamada2021} agrees with the string landscape. Constraints on the global structure of the gauge groups have been deduced from the absence of anomalies between large gauge transformations of $B_4$ and 1-form symmetries (using the formulation of the theory with a four-form potential in the gravity multiplet)  \cite{Cvetic:2020kuw, Cvetic:2021sjm}. Global anomalies for Yang-Mills theory on a D7-brane and topological analogues of the Green--Schwarz mechanism  in eight dimensions have been discussed in \cite{Garcia-Etxebarria:2017crf}.

\paragraph{} \hskip -3mm
While the $l=18$ and $l=10$ theories arise for the torus compactifications of heterotic string (without or with $\mathbb{Z}_2$ holonomy turned on), the known constructions of  the theory with $l=2$ are in type II string theories. As an immediate consequence, the correction to the Bianchi identity \be\label{eq:BI}
d H = \frac{\alpha'}{4 (2 \pi)^2} \Bigl[ \kappa \tr R^2 - \tr F^2 \Bigl]
\ee
comes with $\kappa=0$ for $l=2$, rather than $\kappa = 1$ as for $l=18$ and $l=10$. The known theory with two vector multiplets arises in the circle reduction of two different D=9 theories, obtained respectively from a IIA  or a IIB orientifold  \cite{Dabholkar:1996pc,Aharony:2007du}. This is very much in analogy with a unique D=9 maximal supergravity arising in both IIA and IIB circle reductions. We review these theories in section \ref{sec:Klein_bottle} and compute the one-loop threshold corrections that determine the BPS protected higher derivative couplings.

This survey of known D=8 string theories is almost completely paralleled by the quantum consistency considerations for D=8 minimally supersymmetric effective theories. Anomaly inflow arguments suggest that $\kappa$ is integer quantized. One may further restrict to $\kappa<2$ by making two additional assumptions: first, that compactification of the theory on a circle of radius $R$ admits a well-defined T-dual description in eight dimensions in the limit $R\to 0$, and second, that the resulting Kaluza--Klein tower of states is accounted for by the winding modes of the BPS string \cite{Kim:2019ths}. 
Assuming the cobordism conjecture and  using anomaly cancellations, one gets the $l \leq 18$ bound and $l=2$ mod $8$ for $\kappa =1$ (i.e. $l=18,10,2$) and $l=2$ for $\kappa=0$ \cite{ Montero2021, Hamada2021}. 
In other words, while $\kappa =1$ for $l=2$ has no known string theoretic realization it has not been ruled out by general quantum consistency arguments. Theories with two vector multiplets, and in particular the question of the admissible values of $\kappa$ will be the subject of this paper.

It should be noted that the moduli space of  theories  with $l=2$  has two disconnected components with discrete theta angles \cite{ParraDeFreitas2023, Montero:2022vva}. The choice of this theta angle only affects non-perturbative corrections to observables that are not protected by supersymmetry. These will not be within the scope of our analysis.

In this paper we shall analyse the effective theory with $l=2$ vector multiplets from the consistency of the protected higher derivative couplings. We shall first argue that $\Gamma$ in \eqref{eq:modsp} should be an arithmetic subgroup of SL($2,\R)\times $SL$(2,\R)$ such  that $\Gamma \supset \Gamma_0(N)\times \Gamma_0(N)$ for some natural integer $N$. We shall then assume that $N$ is square-free or $N=4,8,9$ in our analysis.

\paragraph{} \hskip -3mm
One strong constraint on the effective action comes from the cancellation of the anomaly for the modular transformations in $\Gamma\subset$ SO($2,l)$. 
Like for any theory with extended supersymmetry, the numerator of the coset \eqref{eq:modsp} acts on bosonic fields of the theory, while the denominator acts on fermions, and can be regarded as a gauge symmetry. In particular the fermion derivatives appearing in the supersymmetry variation and in the action are covariantised with respect to a composite connection in $\mathfrak{u}(1)\oplus \mathfrak{so}(l)$. Notably, the composite $\mathfrak{u}(1)$ connection couples chirally to the fermions and is anomalous \cite{Marcus1985}. This anomaly can be compensated by a field redefinition, and results in a rigid symmetry anomaly for the numerator group SO$(2,l)$  \cite{deWit:1985bn}. This mixed SL($2,\R)$ anomaly  in type IIB can also be interpreted as a diffeomorphism anomaly in F-theory \cite{Gaberdiel_1998,Minasian2017}. Its cancellation permits to quantize the theory while maintaining the duality symmetry \cite{Bossard:2010dq,Bossard:2014lra}. In eight dimensions, the counter-term that allows to cancel the mixed anomaly takes the form \cite{Lao:2023tuj}
\be\label{eq:countert}
 \frac{1}{9 (16\pi)^4}  \int \Bigl( \tfrac{1}{5} \arg \psi(t) \Tr\! R^4 +  \tfrac{1}{4} \arg \chi(t)  \Tr \! R^2 \wedge \Tr \! R^2  \Bigr) \ee
where $\psi(t)$ and $\chi(t)$ are meromorphic modular forms, depending on $l$ complex coordinates $t$, for the congruent subgroup $\Gamma$. There is a logarithmic singularity at zeros and poles of these meromorphic modular forms, which signals that some BPS multiplets must become massless at these special loci in moduli space. Requiring that the singularities in the interior of moduli space can be interpreted as gauge enhancement loci implies that the singularities must take place at the special divisors associated to a reflective lattice $L_{2,l}$ such that the duality group is $O(L_{2,l})$ and $\psi(t)$ and $\chi(t)$ are Borcherds products  \cite{Lao:2023tuj}. This is a restrictive condition, since the number of these lattices is finite and the value of $l$ is bounded by $l \leq 26$. These restrictions are much weaker than the ones described above from the anomaly inflow arguments, but also involve fewer hypotheses. The specific lattices appearing in string theory  are of course all reflective. 

In this paper we analyse the constraints on $ \psi(T,U) $ and $\chi(T,U)$ for $l=2$ such that all the singularities can be interpreted physically either as decompactification limits or as gauge enhancements. In particular, the limit at infinite distance must correspond to decompactification limits and must be consistent with the available landscape of consistent theories in ten dimensions. From these requirements, we shall find that the only consistent solution is compatible with the unique explicit string theory \cite{Dabholkar:1996pc,Aharony:2007du}. This predicts in particular $\kappa=0$. We explain in more details our hypothesis and sketch our reasoning  in the next section.

\paragraph{} \hskip -3mm
Our results corroborate that there are only two consistent theories with one vector multiplet in $D=9$ dimensions, by taking the appropriate decompactification limits, as we explain in section~\ref{sec:Klein_bottle}. They also provide an important preliminary step to extend the analysis to seven dimensions. In seven dimensions, the vector multiplet moduli space is the symmetric space SO$(3,l+1)/( {\rm SO}(3) \times {\rm SO}(l+1))$. Explicit perturbative string theories exist for $l=2,6,8,10,18$, with four distinct candidates for $l=2$, and there are two M-theory compactifications with $l=0$  \cite{deBoer:2001wca}. There are in particular two more examples of heterotic compactifications with $l=2$, whose uplift to eight dimensions is explicitly described in \cite{Fraiman:2021soq}. The cobordism conjecture and anomaly cancellation only predict that $l$ should be even \cite{Montero2021}. Extending our analysis to $D=7$ is left for future work.

%

\paragraph{} \hskip -3mm
The rest of the paper is organized as follows. In section~\ref{sec:Generalities}
we review the eight-dimensional effective couplings that are fixed by
supersymmetry and by the composite $U(1)$ anomaly, and we formulate the
consistency conditions imposed by decompactification limits and finite-distance
gauge-enhancement singularities. Section~\ref{sec:Klein_bottle} analyzes the
type-II asymmetric orbifold, including its protected gravitational and abelian
gauge-field couplings, its nine-dimensional limits, and reviews the interpretation of
the two disconnected components from the M-theory point of view. We then compare
with heterotic compactifications with a rank-sixteen semisimple gauge group,
which provide useful examples of the same Borcherds-lift technology but with
non-abelian sectors. In section~\ref{Borcherds} we derive the asymptotics of the
relevant Borcherds products for $\Gamma_0(N)$ and determine the constant terms
from the principal parts at the cusps. These results are applied in
section~\ref{sec:constraint}, where the square-free cases and the remaining
cases $N=4,8,9$ are ruled out except for the known $N=2$ asymmetric orbifold.
Several technical details on modular forms, Atkin--Lehner transformations and
the singular theta lift are collected in the appendices.

\section{Generalities} \label{sec:Generalities}
In this paper we consider theories with minimal supersymmetry in eight dimensions with a maximally supersymmetric Minkowski vacuum. The low energy effective theory is given by $\mathcal{N}=1$ supergravity coupled to $l$ vector multiplets. We will be taking $l=n_{\rm v}+2$, so that when the $n_{\rm v} = {\rm dim}(G)$ vectors are valued in a semi-simple Lie algebra $\mathfrak{g}$ and the adjoint  scalar fields vanish, the moduli space is the locally symmetric space $\R^+\times \bigl(SO(2) \backslash SL(2,\R)\bigr)^{\times 2}  / \Gamma$ where $\Gamma$ is a discrete subgroup of $SL(2,\R)\times SL(2,\R)$. In string theory, $\Gamma $ is always an arithmetic subgroup, i.e. a subgroup of $SL(2,\Z)\times SL(2,\Z)$ or a finite Atkin--Lehner extension of it. We write $\phi$ the dilaton in $\R^+$ and the upper complex half plane moduli $T = T_1 {+} i T_2$ and $U= U_1 {+} i U_2$. With generic Wilson lines, the semi-simple gauge group is broken to $U(1)^{{\rm rk}(G)}$, and the low energy effective theory is described by (ungauged) supergravity coupled to $l=2{+}{\rm rk}(G)$ abelian vector multiplets. 

\vskip 2mm

The moduli space of string theories with $\mathcal{N}=1$ supersymmetry in eight dimensions with a Minkowski vacuum admits four connected components. At a generic point in moduli space, the BPS states have quantized $U(1)^{4+{\rm rk}(G)}$charges in the dual of an even lattice of signature $(2,2+  {\rm rk}(G))$. The three possible lattices for which there is a string theory are 
\be L_{2,18} = I\hspace{-0.6mm}I_{2,2} \oplus E_8\oplus E_8\; , \quad L_{2,10} = I\hspace{-0.6mm}I_{1,1}\oplus I\hspace{-0.6mm}I_{1,1}[2] \oplus E_8[2] \; , \quad L_{2,2} =  I\hspace{-0.6mm}I_{1,1}\oplus I\hspace{-0.6mm}I_{1,1}[2] \; , \ee
where $I\hspace{-0.6mm}I_{d,d}$ is the Lorenzian selfdual lattice of rank $2d$, $E_8$ the root lattice and $L[N]$ is generally the lattice with norm square rescaled by $N\in \mathds{N}^*$. In our notations, the vectors in $I\hspace{-0.6mm}I_{2,2}$ are parametrised by the four integers $m_1,m_2,n_1,n_2$ and the root lattice vector $q \in  E_8\oplus E_8$ or $ E_8[2]$. For $I\hspace{-0.6mm}I_{1,1}\oplus I\hspace{-0.6mm}I_{1,1}[2] $ we take by convention that $m_2$ is an even integer while $m_1,n_1,n_2$ are integers. The mass $M = \frac{1}{\sqrt{\alpha'}} |Z|$ of the BPS states  is determined by the central charge \footnote{It is holomorphic in $U$, $z= a_1 + U a_2$ and $\tilde{T} = T +\tfrac12 (a_2 , z) $.}
\be Z \hspace{-0.2mm} = \hspace{-0.2mm} \frac{1}{\sqrt{T_2U_2}} \Bigl( m_1 + U m_2 + ( a_1 + U a_2, q) + \bigl( T + \tfrac12 (a_2 , a_1+U a_2)\bigr) n_2 + \bigl( - U T+\tfrac12 (a_1 , a_1 + U a_2) \bigr) n_1\Bigr) \label{Z} \ee
where the two real vectors $a_1,a_2$ parametrise the Wilson lines, while the lattice bilinear form is defined to be mostly positive with 
\be (Q,Q) = - 2m_1 n_1 - 2m_2 n_2 + (q,q) \; . \ee

The first lattice $L_{2,18}$ appears in heterotic string theory on $T^2$, with either $Spin(32)/\Z_2$ or $E_8\times E_8$ gauge group, and on the S-dual type I string theory on $T^2$. The strong coupling limit is described in eleven-dimensional supergravity on a cylinder times a circle $I\times T^2$ at low energies, with an $E_8$ Yang--Mills theory on each boundary \cite{Horava:1995qa,Horava:1996ma}. 

The second $L_{2,10}$ appears in the CHL string obtained by the orbifold by the half period shift along the ninth circle combined with the exchange of the two $E_8$ factors \cite{Chaudhuri:1995fk}. By duality it is realised in type I with a half integer $B$ field on $T^2$ \cite{Bianchi:1991eu,Bianchi:1997rf} and in M-theory on the M\"obius strip times a circle with the $E_8$ Yang--Mills theory on the boundary. 

These two theories can be interpreted as two disconnected components of the moduli space of type I string theory. We will refer to them as heterotic strings, as their decompactification limit to ten dimensions gives rise to heterotic string theories.

On the contrary the two other theories with lattice $L_{2,2}$ give rise to type IIA or IIB string theory in the decompactification limit. They can be realised as the asymmetric orbifold of type II string theory by the half period shift along the ninth circle combined with the spacetime left fermion number $(-1)^{F_L}$ \cite{Dabholkar:1996pc}. The moduli space is disconnected, with two distinct values of the RR axion $C_0$ compatible with the orbifold in type IIB, $C_0=0$ or $C_0=\frac12$ \cite{Montero:2022vva}. It is described by the Dabholkar-Park orientifold, with possibly a half integer $C_0$ axion, and in M theory on, either the Klein bottle times $S^1$ or another quotient of $T^3$ for $C_0= \frac12$. The value of $C_0$ does not modify the perturbative observables and would only be visible in the effective action by including non-BPS Euclidean Dp-brane instantons wrapping $T^2$. We will refer to these theories as type II string theories.

In the closed string interpretation, $U \in SO(2) \backslash SL(2,\R) $ parametrises the complex structure of the torus $T^2$, while $T\in SO(2) \backslash SL(2,\R) $ is the complex K\"ahler structure $T = B_{89} + i \frac{{\rm Vol}(T^2)}{4\pi^2 \alpha'}$. The BPS states in $L^*_{2,2+{\rm rk}(G)}$ are perturbative string states, $m_1,m_2$ are the momentum mode numbers, $n_1,n_2$ the winding numbers and $q$ the root lattice vector in heterotic string theory. Since the left-moving sector of the theory is not supersymmetric, there are no BPS Dp branes. On the contrary, $U = C_{89} + i \frac{{\rm Vol}(T^2)}{4\pi^2 \alpha'} e^{- \phi}$ in the orientifold theories and the perturbative BPS states only span a sublattice of rank $(1,1{+}{\rm rk}(G))$. We shall therefore mostly concentrate on the closed string theory interpretation, in which case the terms in the effective action we shall analyse are one-loop exact. 

The leading correction to the effective action appears as the Riemann square correction in the heterotic theories 
\be W_2  = \kappa \int\hspace{-1.4mm} d^8x\frac{\sqrt{-g}}{(4\pi)^5 \alpha'^2} T_2 e^{-2\phi} R_{\mu\nu\rho\sigma} R^{\mu\nu\rho\sigma} \ee
where $\kappa=1$ for heterotic strings and $\kappa=0$ for the type II asymmetric orbifolds. It is related by supersymmetry to the modification of the $H$-flux Bianchi identity \eqref{eq:BI}.

By supersymmetry, the eight-derivative terms quartic in the Riemann tensor appear in the effective action through 
\begin{multline}  W_4 =\int\hspace{-1.8mm} d^8x\frac{\sqrt{-g}}{576 (4\pi)^5} \Biggl( \partial_\phi (\partial_\phi-2)(\partial_\phi-4)(\partial_\phi-6) K(\phi,T,U) \bigl( t_8 t_8 +\tfrac14 \varepsilon \varepsilon \bigr) R^4 \\
+2\pi  {\rm Re}\Bigl[ (t_8+\tfrac{i}{2}\varepsilon ) \Bigl( -\tfrac{1}{5} \log \psi(T,U) \Tr\! R^4 -  \tfrac{1}{4} \log \chi(T,U)  \Tr \! R^2 \Tr \! R^2 -\tfrac{i}{2} \log \theta(T,U) \varepsilon R^4 \Bigr) \Bigr] \Biggr) \label{Wilson}\end{multline}
where $t_8$ is the standard rank eight invariant tensor and $\psi$, $\chi$ and $\theta$ must be holomorphic functions of the complex moduli $T$ and $U$ for $\kappa=0$, away from singular loci where additional fields become massless. For $\kappa{=}1$,  $\psi$ is holomorphic, but $\chi$ is not and $\theta=0$. The first line is associated to a full-superspace integral and is not protected by supersymmetry. Because of the differential operator acting on $K$, it cannot contribute at one, two, three and four loops, but duality with type I string theory suggests that it should get contributions to all orders in perturbation theory beyond five loops.

 In supergravity, the $SL(2,\R)\times SL(2,\R)$ continuous symmetry is anomalous  in the two-derivative approximation. Using supersymmetry and the $U(1)$ R-symmetry anomaly,\footnote{Where the $U(1)$ R-symmetry is defined according to \eqref{eq:modsp}.} one obtains the anomalous term quartic in the Riemann tensor as
 \bea &&  \Gamma^{\scalebox{0.6}{1-loop}}[ \tfrac{a T  + b}{c T + d},\tfrac{a' U  + b'}{c' U + d'}] -  \Gamma^{\scalebox{0.6}{1-loop}}[T,U] \CR
&=&   - \int\hspace{-1.8mm} d^8x  \frac{\sqrt{-g}}{288 (4\pi)^4}{\rm Im}\Bigl[ \log \bigl[ (c T + d) (c'U + d') \bigr] (t_8+\tfrac{i}{2}\varepsilon ) \label{AnomalySL2} \\
&& \hspace{20mm} \times \Bigl( \tfrac{248 +n_{\rm v} }{10}  \Tr\! R^4 +  \tfrac{n_{\rm v}-40}{8}   \Tr \! R^2 \Tr \! R^2  + 6 \Tr\! F^2 \Tr\! R^2 + 24 \Tr\!  F^4 \Bigr) \Bigr] \; , \nonumber \eea
where the traces of the Yang--Mills field strengths are in the adjoint representation. In order for the full effective action to be invariant under a congruent subgroup $\Gamma \subset SL(2,\Z)\times SL(2,\Z)$, the function $\psi$ must be a meromorphic modular form of weight $\frac{248 + {\rm dim}(G)}{2} $ in both $T$ and $U$,  $\chi$ a (quasi)-meromorphic modular form of weight $ \frac{{\rm dim}(G)-40}{2}$, while $\theta$ must be a meromorphic modular function. 

\vskip 2mm

Singularities at infinite distance correspond to decompactification limits where towers of Kaluza--Klein states become massless. The corresponding BPS states have charge $Q \in L^*_{2,2{+}n_{\rm v}}$ of norm square $(Q,Q)=0$. Taking for example the limit $T_2\rightarrow  \infty$, all the states with no winding and no root charge (i.e. $q=0$) become massless and can be interpreted as the Kaluza--Klein modes of the theory in higher dimensions. Their mass formula 
\be \sqrt{\alpha'} M =  \frac{1}{\sqrt{T_2U_2}} \bigl| m_1 + U m_2 \bigr| \ee
allows to reconstruct the torus metric
\be ds^2 = \frac{T_2}{U_2} \bigl| dy^9 - U dy^8\bigr|^2\; .  \ee 
In string theory one has one gravity multiplet for each mode $m_2 = 0 $ mod $N$, with $N=1$ or $2$, so that the torus volume in string length is
\be {\rm Vol}(T^2) = 4\pi^2 \alpha'_{\rm II}   \frac{T_2}{N}\; . \ee

\vskip 2mm

Singularities in the interior of the moduli space correspond to gauge enhancement loci where some vector multiplets become massless. The corresponding BPS states have charge $Q \in L^*_{2,2{+}n_{\rm v}}$ of norm square $(Q,Q)= \frac{2}{k}$ for $k=1$ or $2$, according to level matching in string theory. The integer $k$  is the level of the corresponding worldsheet Kac--Moody algebra in heterotic and type II string theory. 

\vskip 2mm

In this paper we wish to constrain the possible effective theories for $n_{\rm v}=0$ (i.e. $l=2$). We assume therefore that the low energy effective field theory is $\mathcal{N}=1$ supergravity coupled to two abelian vector multiplets. The main argument we use here is that the second line in \eqref{Wilson} is protected by supersymmetry and severely constrained by the anomaly cancellation described above. By analogy with string theory, we expect therefore that it only gets contributions from BPS states with charges valued inside a lattice $L_{2,2}^*$. This justifies the assumption that this effective coupling should be invariant under the automorphism group of an even lattice $L_{2,2}$. Without loss of generality the lattice $L_{2,2}$ can be written as $L_{2,2} = \sLambda_{1,1}[N_1] \oplus \sLambda_{1,1}[N_2]$, for $N_1$ and $N_2$ two positive integers. If we assume that  $N=N_1N_2$ is square-free, the lattice is isomorphic to $\sLambda_{1,1} \oplus \sLambda_{1,1} [N]$.\footnote{See appendix~\ref{app:mathematical_details} for more details.} In this paper we shall analyse completely the case in which $N$ is square-free and will extend the proof to $N=4,8,9$. The general case can in principle be analyzed along similar lines; however, the treatment becomes more technical, and we will not pursue it here.
We will accordingly assume that $\psi$ and $\chi$ are modular forms for the congruent subgroup $\Gamma_0(N)\times \Gamma_0(N)$ for some $N$, where $\Gamma_0(N)$ is defined as the group of $SL(2,\Z)$ matrices $\gamma = \big(\colvec[0.7]{a&b\\ c&d}\big)$ with $c=0$ mod $N$. Due to the anomaly cancellation, $\psi$ and $\chi$ must be of weight $124$ and $-20$ respectively. 

\vskip 2mm

At infinite distance one must always be able to interpret the singularity of BPS protected couplings as coming from a Kaluza--Klein tower in higher dimensions.\footnote{Non-protected couplings can also have singularities at infinity corresponding to weakly coupled strings.} According to the discussion above, when either $U$ or $T$ approaches a cusp of $\Gamma_0(N)$, we will have a decompactification limit to ten dimensions. The only consistent theories are the two heterotic supergravities, type IIA and IIB supergravity, so we conclude that there must be either an enhanced gauge symmetry or an enhanced supersymmetry in this limit. Considering for example the limit $T_2\rightarrow \infty$, we must discriminate for which charge $m_2 = r$ mod $N$ there is either a gravity multiplet or a gravitino multiplet in the type II case, and either a graviton or vector multiplets in the heterotic case. We shall assume completeness of the spectrum of BPS charges in a weak sense, meaning that charges for which there is a given set of multiplets must define a sublattice. The set of multiplets must then be determined by the divisibility of $m_2$ by $\ell$ dividing $N$. 

Let us first consider the type II case. To recover $\mathcal{N}=2$ supersymmetry in ten dimensions, there must be the same number of gravity and gravitino multiplets, while the class $m_2= 0 $ mod $N$ must include a single graviton multiplet to be consistent with the zero charge.  One defines $n_g(\ell)$ for each divisor $\ell$ of $N$ as the number of gravity multiplets for a mode number divisible by $\ell$, but by no strictly greater divisor of $N$, and $n_\psi(\ell)$ the number of gravitino multiplets with the same condition. We have $n_g(N)=1$ and $n_\psi(N)=0$, while the others must satisfy 
\be \sum_{\ell|N} n_g(\ell) \mu(\ell)  =  \sum_{\ell|N} n_\psi(\ell) \mu(\ell) = K \; , \label{Kdefined} \ee
where $\mu(\ell)$ is the number of $r$ mod $N$ divisible by $\ell$ (and no bigger divisor of $N$). Writing the prime factorisation of $\ell'= \prod_{p|\ell'} p^{\nu_p(\ell')}$, $\mu(\ell)$ is given by the Euler product 
\be \mu(\ell) = \prod_{p|\frac{N}{\ell}} p^{\nu_p(\frac{N}{\ell})-1} (p-1) \; . \ee
The total number of gravity multiplets in average determines the volume of the torus as
\be {\rm Vol}(T^2) = 4\pi^2 \alpha'  K \frac{T_2}{N}\; . \ee
The simplest solution to \eqref{Kdefined} is for $K=1$ and $N$ even, with $n_g(\ell) = \delta_{N,\ell}$ and \mbox{$n_\psi(\ell) = \delta_{\frac{N}{2},\ell}$}. More generally one must have the behaviour at large $T_2$
\bea \psi &\sim& e^{30\pi i K \frac{T}{N} + a_\psi \pi i U } \prod_{n=1}^\infty (1-e^{2\pi i N n U})^{248} \prod_{\substack{\ell|N\\ \ell\ne N}} \prod_{\substack{n\ge 1\\ {\rm gcd}(n,N/\ell)=1}} (1-e^{2\pi i \ell n U})^{248 n_g(\ell) +112 n_\psi(\ell) }  \CR
 \chi &\sim& e^{-6\pi i K \frac{T}{N}+ a_\chi \pi i U} \prod_{n=1}^\infty (1-e^{2\pi i N n U})^{-40} \prod_{\substack{\ell|N\\ \ell\ne N}} \prod_{\substack{n\ge 1\\ {\rm gcd}(n,N/\ell)=1}} (1-e^{2\pi i \ell n U})^{-40 n_g(\ell) -32 n_\psi(\ell) }  \CR 
  \theta &\sim&e^{\mp \frac{\pi i}{4} K \frac{T}{N}\pm  a_\theta \pi i U}  \prod_{\substack{\ell|N\\ \ell\ne N}} \prod_{\substack{n\ge 1\\ {\rm gcd}(n,N/\ell)=1}} (1-e^{2\pi i \ell n U})^{\pm 3 n_\psi(\ell) } \label{KKtower} \eea
where the last sign is $+$ for type IIA and $-$ for type IIB and $a_\psi,a_\chi,a_\theta$ are determined by modular invariance as
\bea a_\psi &=& \frac23 \sum_{\ell|N} \ell \Bigl[ \prod_{p|\frac{N}{\ell}} (1-p) \Bigr] \bigl( 31 n_g(\ell) + 14 n_\psi(\ell) \bigr) \; , \CR
a_\chi &=& -\frac23 \sum_{\ell|N} \ell \Bigl[ \prod_{p|\frac{N}{\ell}} (1-p) \Bigr] \bigl( 5 n_g(\ell) + 4 n_\psi(\ell) \bigr) \; ,  \CR
a_\theta &=& \frac14  \sum_{\ell|N} \ell \Bigl[ \prod_{p|\frac{N}{\ell}} (1-p) \Bigr] n_\psi(\ell) \; .  \eea

In the heterotic case one must obtain $496$ vector multiplets in ten dimensions, forming the $E_8\times { E}_8$ or Spin$(32)/ \Z_2$ gauge group. There is no consistent string theory in eight dimensions in which an entire gauge sector can emerge in a decompactification limit. The highest dimensional examples are in seven dimensions and crucially require a triple of Wilson lines along $T^3$ \cite{deBoer:2001wca}. We nevertheless consider this possibility  from the point of view of the low energy effective action. One defines $n_g(\ell)$ as in the type II case and $n_v(\ell)$ the number of vector multiplets. We have $n_g(N)=1$ and $n_v(N)=0$, while the others must satisfy 
\be \sum_{\ell|N} n_g(\ell) \mu(\ell)  =K\; , \qquad   \sum_{\ell|N} n_v(\ell) \mu(\ell) = 496 K \; . \ee
The total number of gravity multiplets in average determines the volume of the torus in the same way. We must then have the behaviour at large $T_2$
\bea \psi &\sim& e^{60\pi i K \frac{T}{N} + a_\psi \pi i U } \prod_{n=1}^\infty (1-e^{2\pi i N nU})^{248} \prod_{\substack{\ell|N\\ \ell\ne N}} \prod_{\substack{n\ge 1\\ {\rm gcd}(n,N/\ell)=1}} (1-e^{2\pi i \ell n U})^{248 n_g(\ell)  + n_v(\ell)}  \CR
  \theta &\sim&1\eea
and we did not write $\chi$ since it is not holomorphic in this case. 

\vskip 2mm

Singularities at finite distance must correspond instead to enhanced gauge symmetry, where some of the BPS vector multiplets become massless. For a complex codimension one singularity, one vector multiplet remains abelian and the only possible gauge enhancements are to a $U(2)$ or $U(1)\times SO(3)$ gauge group. One must therefore have exactly two massless vector multiplets of opposite charges. Assume that a vector multiplet of  primitive charge $Q_{\scalebox{0.6}{$\circ$}}$ becomes massless at 
\be T = \frac{m_2 U + m_1 }{n_1 U - n_2} \ee
with  
\be (Q_{\scalebox{0.6}{$\circ$}},Q_{\scalebox{0.6}{$\circ$}}) = - 2m_1 n_1 - 2m_2 n_2 = \frac{2}{k} \ee
for some $k>0$. For every vector $Q \in L_{2,2}^*$ for which there is a physical state, one must have 
\be k (Q_{\scalebox{0.6}{$\circ$}},Q) \in \Z \; , \ee
such that it can be interpreted as a weight of $SU(2)$. Assuming there is a state for all $Q\in L_{2,2}^*$, one must have 
$ k Q_{\scalebox{0.6}{$\circ$}}  \in L_{2,2}$, and $k$ must  therefore be an integer dividing $N$. Near the singularity, $\psi$ and $\chi$ must have a single zero proportional to the mass of the vector multiplet fields 
\be \psi(T,U) \sim \chi(T,U) \sim \bigl( m_1 + U m_2 + T  n_2 - U T  n_1\bigr) R(T,U) \ee
up to multiplication by a regular function $R$, to reproduce the behaviour of the singularity associated to a single massless vector multiplet.

\vskip 2mm

The main tool in our argument is to construct the modular forms $\psi(T,U)$ and $\chi(T,U)$ as Borcherds lifts of  modular functions $f(\tau)$ and $g(\tau)$ of $\Gamma_0(N)$ for $N$ square-free. For $N=4,8,9$ we shall describe directly $\psi(T,U)$ and $\chi(T,U)$ as meromorphic modular forms with the required properties. We shall find that requiring the consistency conditions described above, the unique solution is the asymmetric orbifold with $N=2$. The analysis of the modular form $\psi(T,U)$ does not directly rule out the cases $N=5$ and $N=6$ with $\kappa=1$, and we shall argue that there is no possible consistent four-photon amplitude in these cases. We will argue that the cases $N=5$ and $N=6$ are spurious solutions to the consistency conditions for $\psi(T,U)$ that come from the existing orbifolds in $D=7$ dimensions.

\vskip 2mm

\section{The asymmetric orbifold} \label{sec:Klein_bottle}
Before giving constraints on possible effective theories, let us discuss the example of the asymmetric orbifold of type II string theory by the half period shift on the ninth circle combined with the left-moving spacetime fermion number $(-1)^{F_L}$. This theory was first introduced in \cite{Dabholkar:1996pc} as a dual of the Dabholkar-Park type I' string obtained by the orientifold of type IIB by the worldsheet reflection combined with the half-period shift on the ninth circle. It provides a simple example of the dualities conjectured in \cite{Vafa:1995gm}. 

\subsection{Partition function and protected couplings}

Let us recall first the partition function \cite{Gutperle:2000bf}. We shall use the notations \cite{Bianchi:1990yu,Angelantonj:2002ct} for the $Spin(8)$ affine characters 
 \be O_8 = \frac{\vartheta_3(0)^4{+}\vartheta_4(0)^4}{2\eta^4}\, , \  V_8 = \frac{\vartheta_3(0)^4{-}\vartheta_4(0)^4}{2\eta^4}\, , \  S_8 = \frac{\vartheta_2(0)^4{+}\vartheta_1(0)^4}{2\eta^4}\, , \  C_8 = \frac{\vartheta_2(0)^4{-}\vartheta_1(0)^4}{2\eta^4}\, , \ee
and the Narain partition function 
\be \Gamma_{2,2}[^s_r] = \tau_2 \hspace{-2.2mm}\sum_{\substack{m_1,m_2,n_1\in \Z\\ n_2 \in \Z+\frac{s}{2}}} (-1)^{r m_2}e^{ i \pi  \tau |p_L|^2 -i \pi  \bar \tau |p_R|^2}  \ee
with the complex left and right projections 
\be p_R = \frac{1}{\sqrt{2T_2 U_2}} ( m_1 + U m_2 + T n_2 - U T n_1) \; , \quad p_L = \frac{1}{\sqrt{2T_2 U_2}} ( m_1 + U m_2 + \bar T n_2 - U \bar T n_1) \; . \ee
With this notation, $(-1)^{F_L}$ simply acts as a minus  sign on the left Ramond characters $C_8$ and $S_8$, and in type IIA  one obtains the partition function 
\be \mathcal{Z}^{\scalebox{0.6}{AOA}} = \frac12  \overline{\frac{V_8-S_8}{\eta^{8}}} \Biggl( \frac{V_8-C_8}{\eta^8} \Gamma_{2,2}[^0_0] +  \frac{V_8+C_8}{\eta^8} \Gamma_{2,2}[^0_1] +  \frac{O_8-S_8}{\eta^8}\Gamma_{2,2}[^1_0] -  \frac{O_8+S_8}{\eta^8} \Gamma_{2,2}[^1_1] \Biggr)  \; .  \ee
The partition function in type IIB is obtained by exchanging the left moving $S_8$ and $C_8$ characters. To exhibit the lattice of states, it is convenient to introduce the notation 
\be \Gamma_{L_{2,2}} = \tau_2 \hspace{-1mm} \sum_{Q\in L_{2,2}}  \hspace{-1mm}e^{ i \pi  \tau |p_L|^2 -i \pi  \bar \tau |p_R|^2} \; , \quad \Gamma_{L_{2,2}}[f(p_L,p_R)]  = \tau_2  \hspace{-1mm} \sum_{Q\in L_{2,2}}  \hspace{-1mm} f(p_L,p_R) e^{ i \pi  \tau |p_L|^2 -i \pi  \bar \tau |p_R|^2} \; , \ee
for any lattice $L_{2,2}$, with by definition 
\be \Gamma_{\sLambda_{2,2}}  =   \Gamma_{2,2}[^0_0] \; , \quad \Gamma_{\sLambda_{1,1}\oplus \sLambda_{1,1}[2]}  = \frac12 \bigl(  \Gamma_{2,2}[^0_0]+\Gamma_{2,2}[^0_1] \bigr) \; , \quad \Gamma_{\sLambda_{1,1}\oplus \sLambda_{1,1}[\frac12]}  =  \Gamma_{2,2}[^0_0]+\Gamma_{2,2}[^1_0]  \; . \ee
One can then write the partition function as a sum of images of a  $\Gamma_0(2)$ invariant function under the action $\tau|_\gamma =  \frac{a \tau+b}{c \tau + d}$ of the cosets $\gamma \in P\hspace{-0.3mm}S\hspace{-0.3mm}L(2,\Z)/ \Gamma_0(2)$, i.e.
  \be \mathcal{Z}^{\scalebox{0.6}{AOA}} =   \sum_{\gamma \in P\hspace{-0.3mm}S\hspace{-0.3mm}L(2,\Z)/ \Gamma_0(2)}\hspace{-4mm}  \Bigl( \overline{\frac{V_8-S_8}{\eta^{8}}}  \frac{V_8 + C_8 }{\eta^{8}}\Gamma_{ \sLambda_{1,1}\oplus \sLambda_{1,1}[2]}  \Bigr)\Big|_{\gamma} +\overline{\frac{V_8-S_8}{\eta^{8}}}   \frac{S_8 - C_8 }{\eta^{8}}\Gamma_{\sLambda_{2,2}}  \; . \ee 

To compute the one-loop contribution to the effective action, it is convenient to use the character valued partition function, introducing four fugacities $e^{2\pi i \upsilon_a}$ for the Cartan subgroup of the rotation group $Spin(8)\subset Spin(1,9)$. We have then
\bea \frac{V_8 - S_8}{\eta^8} &=& \sum_{i=1}^4 (-1)^{1+\delta_{i,3}} \prod_{a=1}^4 \frac{2\sin \pi \upsilon_a}{\vartheta_1(\upsilon_a)} \vartheta_i(\upsilon_a) \CR
\frac{V_8 +C_8}{\eta^8} &=& \sum_{i=1}^4 (-1)^{i+1} \prod_{a=1}^4 \frac{2\sin \pi \upsilon_a}{\vartheta_1(\upsilon_a)} \vartheta_i(\upsilon_a)\CR\frac{S_8 -C_8}{\eta^8} &=&-32  \prod_{a=1}^4 \sin \pi \upsilon_a\eea
From this formula one obtains the spectrum of BPS states with $(Q,Q)\ge 0$. There is a ten-dimensional graviton multiplet for each isotropic charge $Q \in \sLambda_{1,1}\oplus \sLambda_{1,1}[2]$, meaning a eight-dimensional graviton multiplet plus two vector multiplets. One finds  similarly a gravitino multiplet for each isotropic charge in the dual lattice but not in the lattice, and a vector multiplet for each charge of norm 1 in the dual lattice, i.e. 
\bea &\mbox{a graviton multiplet:}\qquad   &Q \in  \sLambda_{1,1}\oplus \sLambda_{1,1}[2]\; , \quad (Q,Q) = 0\; , \\
&\mbox{a gravitino multiplet:}\qquad   &Q \in \sLambda_{1,1}\oplus \sLambda_{1,1}[\tfrac12]\smallsetminus \sLambda_{1,1}\oplus \sLambda_{1,1}[2]\; , \quad (Q,Q) = 0\; , \CR
&\mbox{a vector multiplet:}\qquad   &Q \in \sLambda_{1,1}\oplus \sLambda_{1,1}[\tfrac12]\smallsetminus \sLambda_{1,1}\oplus \sLambda_{1,1}[2]\; , \quad (Q,Q) = 1\; . \nonumber 
\eea
To obtain the effective action, we use the relation between the character valued partition function and the partition function in the background of a constant curvature proposed in \cite{Schellekens:1986xh}. We shall then crosscheck the consistency of the couplings and find that they are uniquely determined by consistency.  The single trace and double trace $R^4$ couplings, as defined in \eqref{Wilson}, are then given by the (singular) theta lifts 
\begin{multline} k_\psi \log( U_2 T_2) +  \log| \psi(T,U)|^2 \\
 =-\frac{5}{64\pi^8}  \int_{\mathcal{F}} \hspace{-1.2mm} \frac{d^2\tau}{\tau_2^2}   \biggl( \frac{\partial^4}{\partial \upsilon_1^4} {+}\frac{\partial^4}{\partial \upsilon_2^4} {-}6 \frac{\partial^2}{\partial \upsilon_1^2}\frac{\partial^2}{\partial \upsilon_2^2} {+}960 \pi^4 \biggr)   \frac{\partial^4}{\partial \bar \upsilon_1^4} \mathcal{Z}^{\scalebox{0.6}{AOA}} \Bigg|_{\upsilon_a=0}\; ,  \label{EllR4}  \end{multline}
 and 
 \begin{multline}
k_\chi  \log( U_2 T_2) +  \log| \chi(T,U)|^2 \\
 = - \frac{1}{16\pi^2}  \int_{\mathcal{F}} \frac{d^2\tau}{\tau_2^2}   \biggl( 3   \frac{\partial^2}{\partial \upsilon_1^2}\frac{\partial^2}{\partial \upsilon_2^2}   {-}\pi^2   \left( \frac{\partial^2}{\partial \upsilon_1^2}{+}\frac{\partial^2}{\partial \upsilon_2^2}  \right) + \frac{\pi^4}{3} \biggr)   \frac{\partial^4}{\partial \bar \upsilon_1^4} \mathcal{Z}^{\scalebox{0.6}{AOA}}  \Bigg|_{\upsilon_a=0}  \label{EllR22} \; . \end{multline}
The logarithmic term comes from the logarithmic divergence of these integrals as $\tau \rightarrow i \infty$, where the integrals are understood to be regularised and renormalised. We postpone the precise definition of the regularisation to section \ref{Borcherds}. Note that because  the right-moving side is supersymmetric, the amplitude factorises the $t_8 + \tfrac{i}{2} \varepsilon $ structure and we only need to introduce one fugacity $\bar \upsilon_1$ to determine the coupling. The specific combination of the derivatives with respect to $\upsilon_1,\upsilon_2$ can be derived from the worldsheet elliptic genus \cite{Schellekens:1986xh}, using that 
\bea \prod_{a=1}^4 \frac{\pi \upsilon_a}{\sin \pi \upsilon_a} \frac{V_8 - S_8}{\eta^8} &=& \pi^4 \Biggl( 2\sum_{a=1}^4 \upsilon_a^4 - \frac14 \Bigl( 2 \sum_{a=1}^4 \upsilon_a^2\Bigr)^2 - 8 \prod_{a=1}^4 \upsilon_a \Biggr)   \CR
&& \hspace{10mm} \times \exp\Biggl( \sum_{k=2}^\infty\sum_{a=1}^4 \biggl(  \upsilon_a^{2k} -\Bigl(\frac{\sum_{b\ne a}\upsilon_b - \upsilon_a}{2}\Bigr)^{2k} \biggr)  \frac{\zeta(2k)}{k} E_{2k}(\tau) \Biggr)  \; , \CR
\prod_{a=1}^4 \frac{\pi \upsilon_a}{\sin \pi \upsilon_a} \frac{V_8 +C_8}{\eta^8} &=& \pi^4 \Biggl( 2\sum_{a=1}^4 \upsilon_a^4 - \frac14 \Bigl( 2 \sum_{i=1}^4 \upsilon_a^2\Bigr)^2 - 8 \prod_{a=1}^4 \upsilon_a \Biggr)   \CR
&& \hspace{10mm} \times \exp\Biggl( \sum_{k=2}^\infty\sum_{a=1}^4 \biggl(  \upsilon_a^{2k} -\Bigl(\frac{\sum_{b\ne a}\upsilon_b - \upsilon_a}{2}\Bigr)^{2k} \biggr)  \frac{\zeta(2k)}{k} E_{2k}(\tau) \Biggr) \CR
&& + \frac{16\eta(2\tau)^8}{ \eta(\tau)^{16}}\exp\Biggl( \sum_{k=1}^\infty\sum_{a=1}^4  \upsilon_a^{2k}  \frac{\zeta(2k)}{k} \bigl( 2E_{2k}(\tau)- 2^{2k} E_{2k}(2\tau)\bigr) \Biggr) \; .  \ \label{EllipGenus} \eea
The factor of $\bigl| \prod_{a=1}^4 \frac{1}{\sin \pi \upsilon_a}\bigr|^2 $ comes from the integral over the zero modes \cite{Angelantonj:2002id} in the complete partition function, while the factor $\bigl|\prod_{a=1}^4 \pi \upsilon_a \bigr|^2$ can be understood as a regulator to ensure that one retrieves the unit normalisation in the limit $\upsilon_a\rightarrow 0$. 

One checks that the same formula gives the correct $R^4$ coupling for type II and heterotic string theory, although for $\kappa=1$ one needs to modify the second differential operator as \cite{Lerche:1987qk}
\be   \biggl( 3   \frac{\partial^2}{\partial \upsilon_1^2}\frac{\partial^2}{\partial \upsilon_2^2}   -\pi^2   \Bigl(1-\frac{6 }{\pi \tau_2}\Bigr) \left( \frac{\partial^2}{\partial \upsilon_1^2}{+}\frac{\partial^2}{\partial \upsilon_2^2}  \right) + \frac{\pi^4}{3}  \Bigl(1-\frac{3}{\pi \tau_2}\Bigr)^2 \biggr) \ee
to take into account the non-holomorphic part in the integral of the worldsheet Green function $G(z)$
\be \int \frac{d^2z}{\tau_2} \partial G(z) \partial G(z) = - \frac{\pi^2}{3} \hat{E}_2(\tau)= - \frac{\pi^2}{3} {E}_2(\tau)  + \frac{\pi}{\tau_2}   \; . \ee
In the type II asymmetric orbifold we have $\kappa=0$ and \eqref{EllipGenus} is automatically modular invariant.

 Because the differential operator in \eqref{EllR4} and \eqref{EllR22} annihilate $\frac{S_8-C_8}{\eta^8}$, we can simply write the result as an integral over the $\Gamma_0(2)$ fundamental domain 
 \be \mathcal{F}_2 = \bigcup_{\gamma \in SL(2,\Z) / \Gamma_0(2)} \mathcal{F}\big|_\gamma \ee
so that
\bea  - 248 \log( U_2 T_2) - 2 \log| \psi(T,U)|^2 &=& \int_{\mathcal{F}_2} \frac{d^2\tau}{\tau_2^2}  f(\tau)  \Gamma_{\sLambda_{1,1}\oplus \sLambda_{1,1}[2]} \CR
40 \log( U_2 T_2) - 2 \log| \chi(T,U)|^2 &=&  \int_{\mathcal{F}_2} \frac{d^2\tau}{\tau_2^2}  g(\tau)  \Gamma_{\sLambda_{1,1}\oplus \sLambda_{1,1}[2]}  \; , \eea
with \footnote{Here we have written first the direct outcome of the expansion of the partition function, while the last step uses the uniqueness of the holomorphic modular function with no pole at $\tau\rightarrow i \infty$ and a single pole at $\tau \rightarrow 0$.}
 \bea f(\tau) &=&  \frac{5}{32\pi^8}  \frac{\partial^4}{\partial \bar \upsilon_1^4} \overline{ \frac{V_8 - S_8}{\eta^8} }  \biggl( \frac{\partial^4}{\partial \upsilon_1^4} +\frac{\partial^4}{\partial \upsilon_2^4} -6 \frac{\partial^2}{\partial \upsilon_1^2}\frac{\partial^2}{\partial \upsilon_2^2} +960 \pi^4 \biggr)  \frac{V_8 + C_8 }{\eta^{8}}\Bigg|_{\upsilon_a=0}\CR
&=& 360- \bigl( 16 E_4(2\tau)  - 2 E_4(\tau) \bigr)\frac{16\eta(2\tau)^8}{ \eta(\tau)^{16}} \CR
&=& 136 +4096  \frac{\eta(2\tau)^{24}}{\eta(\tau)^{24}}\; , \label{Tr1} \\
 g(\tau) &=&\frac{1}{8\pi^2}  \frac{\partial^4}{\partial \bar \upsilon_1^4} \overline{ \frac{V_8 - S_8}{\eta^8} }  \biggl( 3   \frac{\partial^2}{\partial \upsilon_1^2}\frac{\partial^2}{\partial \upsilon_2^2}   -\pi^2   \left( \frac{\partial^2}{\partial \upsilon_1^2}{+}\frac{\partial^2}{\partial \upsilon_2^2}  \right) + \frac{\pi^4}{3} \biggr)  \frac{V_8 + C_8 }{\eta^{8}}\Bigg|_{\upsilon_a=0} \CR
 &=&- 72 +4  \bigl(2 E_2(2\tau)- E_2(\tau)\bigr)^2 \frac{16\eta(2\tau)^8}{ \eta(\tau)^{16}}\CR
&=& -8 +4096  \frac{\eta(2\tau)^{24}}{\eta(\tau)^{24}}\; . \label{Tr2} \eea

The evaluation of the function $\theta(T,U)$ is slightly more subtle because the corresponding amplitude vanishes. Therefore we must effectively consider the five-point function and we only determine the derivative of the function as
\bea \frac{\partial\, }{\partial T} \log \theta(T,U) &=&\frac{1}{128\pi^8} \frac{\partial\, }{\partial T} \int_\mathcal{F} \frac{d^2\tau}{\tau_2^2}     \frac{\partial^4 \, }{\partial \upsilon_1\partial \upsilon_2 \partial \upsilon_3 \partial \upsilon_4 } \frac{\partial^4}{\partial \bar \upsilon_1^4}
\mathcal{Z}^{\scalebox{0.6}{AOA}} \Bigg|_{\upsilon_a=0}  \CR
 \frac{\partial\, }{\partial U} \log \theta(T,U) &=&-\frac{1}{128\pi^8} \frac{\partial\, }{\partial U}\int_\mathcal{F} \frac{d^2\tau}{\tau_2^2}     \frac{\partial^4 \, }{\partial \upsilon_1\partial \upsilon_2 \partial \upsilon_3 \partial \upsilon_4 } \frac{\partial^4}{\partial \bar \upsilon_1^4}
\mathcal{Z}^{\scalebox{0.6}{AOA}} \Bigg|_{\upsilon_a=0}  \; ,  \eea
where we get a minus sign because the exchange of $U$ and $T$ is a parity transformation and this contribution comes entirely from the odd spin structure. The derivative of $\mathcal{Z}^{\scalebox{0.6}{AOA}}$ simplifies to a constant and we get 
\bea \frac{\partial\, }{\partial T} \log \theta(T,U) &=&\frac32  \frac{\partial\, }{\partial T}\Biggl[   \int_{\mathcal{F}_2} \frac{d^2\tau}{\tau_2^2}   \Gamma_{\sLambda_{1,1}\oplus \sLambda_{1,1}[2]}-  2  \int_{\mathcal{F}} \frac{d^2\tau}{\tau_2^2} \Gamma_{\sLambda_{2,2}}\Biggr]\; ,  \CR
 \frac{\partial\, }{\partial U} \log \theta(T,U) &=&-\frac32  \frac{\partial\, }{\partial U}\Biggl[   \int_{\mathcal{F}_2} \frac{d^2\tau}{\tau_2^2}   \Gamma_{\sLambda_{1,1}\oplus \sLambda_{1,1}[2]}-  2  \int_{\mathcal{F}} \frac{d^2\tau}{\tau_2^2} \Gamma_{\sLambda_{2,2}}\Biggr] \; .  \eea
 As we shall explain in more detail in the next section, one then computes 
\bea \psi(T,U) &=&  \eta(2U)^{136} \eta(U)^{112} \eta(T)^{136} \eta(T/2)^{112}\biggl( \frac{\eta(T/2)^{24}}{\eta(T)^{24}}-   \frac{\eta(U)^{24}}{\eta(2U)^{24}}\biggr) \; ,  \label{PsiAOA} \CR
 \chi(T,U) &=&  \eta(2U)^{-8} \eta(U)^{-32} \eta(T)^{-8} \eta(T/2)^{-32}\biggl( \frac{\eta(T/2)^{24}}{\eta(T)^{24}}-   \frac{\eta(U)^{24}}{\eta(2U)^{24}}\biggr)\; , \CR
 \theta(T,U) &=& \frac{\eta(2U)^3\eta(T)^3}{\eta(U)^3\eta(T/2)^3}\; . \eea
Note that all these functions are invariant (up to an irrelevant overall sign) under the T-duality along $y^9$, $T\rightarrow 2U,U\rightarrow \frac{T}{2}$. So T-duality along $y^9$ is a symmetry of the effective theory in both type IIA and type IIB. The two functions $\psi$ and $\chi$ are invariant under the combined Fricke duality $T\rightarrow - \frac{2}{T}$ and $U\rightarrow - \frac{1}{2U}$, while $\theta$ transforms into its inverse. This is the T-duality along both $y^8$ and $y^9$ that exchanges the type IIA and the type IIB asymmetric orbifolds. 

Let us define for short 
\be  W^{\scalebox{0.6}{BPS}}_4 =\frac{1}{72(8\pi)^4} \int\hspace{-1.8mm} d^8x \sqrt{-g} \, {\rm Re}\, \Bigl[ (t_8+\tfrac{i}{2}\varepsilon ) \mathcal{A}  \Bigr] \ee
with 
\be \mathcal{A} = -\tfrac{1}{5} \log \psi(T,U) \Tr\! R^4 -  \tfrac{1}{4} \log \chi(T,U)  \Tr \! R^2 \Tr \! R^2 -\tfrac{i}{2} \log \theta(T,U) \varepsilon R^4 \; . \ee

Using the same method one also computes the vector multiplet four-photon terms and the two-graviton two-photon terms as
\begin{multline}  W'_4 =\int\hspace{-1.8mm} d^8x\frac{\sqrt{-g}}{6 (4\pi)^4} \Biggl(  {\rm Re}\Bigl[ (t_8+\tfrac{i}{2}\varepsilon ) \Bigl(4 \mathcal{F}(T,U)_{abcd} F^a F^b F^c F^d - \mathcal{G}(T,U)_{ab} F^a F^b \Tr R^2 \Bigr) \Bigr] \Biggr) \label{WilsonFF}\end{multline}
where 
\bea {\rm Re}[ \mathcal{F}]_{abcd} &=& \int_{\mathcal{F}_2}  \hspace{-1.8mm} \frac{d^2\tau}{\tau_2^{\, 2}} \frac{\eta(2\tau)^8}{ \eta(\tau)^{16}}  \Gamma_{\sLambda_{1,1}\oplus \sLambda_{1,1}[2]}\bigl[p_{L a} p_{L b} p_{L c} p_{L d} {-} \tfrac{3}{2\pi \tau_2} \delta_{(ab} p_{Lc} p_{L d)} {+} \tfrac{3}{(4\pi \tau_2)^2}  \delta_{(ab} \delta_{cd)} \bigr] \; , \CR
{\rm Re}[ \mathcal{G}]_{ab} &=&\int_{\mathcal{F}_2} \hspace{-1.8mm}  \frac{d^2\tau}{\tau_2^{\, 2}}  \frac{\eta(2\tau)^8}{ \eta(\tau)^{16}}  \bigl(2E_2(2\tau){-} E_2(\tau)\bigr) \Gamma_{\sLambda_{1,1}\oplus \sLambda_{1,1}[2]}\bigl[p_{L a} p_{L b} -\tfrac1{4\pi \tau_2} \delta_{ab}\bigr]\, , \label{FFFFandFFRR}  \eea
and $p_{La} = ( {\rm Re}[p_L],{\rm Im}[p_L])$. These four and six derivative couplings are one-loop exact thanks to the supersymmetry Ward identities. 

It is illustrative to take a large radius limit to nine dimensions to see the discontinuities of the effective action between different string theories. 

\vskip 2mm

\subsection{Perturbative and large volume limits}

In the type IIA asymmetric orbifold, the shift is along the circle $S_{y^9}$ of radius $\sqrt{\alpha' \frac{T_2}{U_2}}$, while $S_{y^8}$ has radius $\sqrt{\alpha' T_2 U_2 }$. For large $\frac{T_2}{U_2}$ one gets type IIA in nine dimensions, and for small $\frac{T_2}{U_2}$ one gets type IIB, consistently with the limits 
\bea  \mathcal{A}  &\underset{\substack{T\rightarrow i \infty \\ U\rightarrow  0 }}{\sim}&- \frac18 \Bigl[\pi i T ( t_8 + \tfrac{i}{2} \varepsilon ) R^4  + 2\pi i U( t_8 - \tfrac{i}{2} \varepsilon ) R^4   \Bigr]\; ,  \\
\mathcal{A}  &\underset{\substack{T\rightarrow 0 \\ U\rightarrow  i \infty }}{\sim}&- \frac18 \Bigl[\pi i T ( t_8 - \tfrac{i}{2} \varepsilon ) R^4  + 2\pi i U( t_8 +\tfrac{i}{2} \varepsilon ) R^4   \Bigr]\;    , \nonumber
 \eea
where for the cusps at $T\rightarrow 0$ the right-hand-side is written with $T\rightarrow - \frac{2}{T}$ and for $U\rightarrow 0$ with $U\rightarrow - \frac{1}{2 U}$.   Keeping $\frac{T_2}{U_2}$ fixed instead, the large type IIA radius limit $T_2 U_2 \rightarrow \infty $ gives the amplitude of the type IIA asymmetric orbifold theory in nine dimensions, while $T_2 U_2 \rightarrow 0 $ gives the asymmetric orbifold in type IIB 
\begin{eqnarray*}  \mathcal{A} &\underset{\substack{T\rightarrow i \infty \\ U\rightarrow i \infty \\ T_2 > 2 U_2}}{\sim}& -\Bigl( 3\pi i T + \frac{32\pi i U}{5}\Bigr)  \Tr R^4 + \frac14 \Bigl( 3\pi i T + 4 \pi i U\Bigr) \Tr^2 R^2  - \frac{i}{16} ( \pi i T + 2\pi i U)  \varepsilon R^4  \; ,  \CR
\mathcal{A}  &\underset{\substack{T\rightarrow 0 \\ U\rightarrow 0 \\ \frac{T_2}{|T|^2}  > \frac{U_2}{2|U|^2}}}{\sim}&- \Bigl( 3\pi i T + \frac{32\pi i U}{5}\Bigr)  \Tr R^4 + \frac14 \Bigl( 3\pi i T + 4 \pi i U\Bigr) \Tr^2 R^2  + \frac{i}{16} ( \pi i T + 2\pi i U)  \varepsilon R^4 \; \; . \nonumber 
 \end{eqnarray*}
Recall that $T = B_{89} + i \frac{{\rm Vol}(T^2)}{4\pi^2 \alpha'}$ and $U$ is the torus complex structure, so each term in $T_1$ and $U_1$ should be interpreted as a Chern-Simons coupling between the one-forms coming from $B$, the ten-dimensional metric and the $R^4$ Pontryagin  classes
 \bea S^{\scalebox{0.6}{IIA}}_{\scalebox{0.6}{Odd}} &=& - \frac{1}{9 (32\pi)^3} \int \Biggl( B_1 \wedge \bigl( \Tr R^4 - \tfrac14 \Tr R^2 \wedge R^2 \bigr) + A_1 \wedge \bigl( \tfrac{31}{5}\Tr R^4 - \tfrac54 \Tr R^2 \wedge R^2 \bigr)  \Biggr) \CR
S^{\scalebox{0.6}{IIB}}_{\scalebox{0.6}{Odd}} &=& - \frac{1}{9 (32\pi)^3} \int A_1 \wedge \bigl( \tfrac{1}{5}\Tr R^4 + \tfrac14 \Tr R^2 \wedge R^2 \bigr)  . 
 \eea
  One finds that this coupling jumps when one cross the selfdual circle $\frac{T_2}{U_2}= 2$, such that $B_1$ and $A_1$ are exchanged.

 The two other cases $T_2 < 2 U_2$ are obtained with the exchange $T \leftrightarrow 2U$ while the type IIB amplitude is obtained by replacing $T\rightarrow - \frac{1}{U}$ and $U\rightarrow - \frac{1}{T}$.

\vskip 4mm 
These couplings also admit singularities at finite distance. The singularity of $\psi$ and $\chi$ at $T=2U$ corresponds to a level $2$ $U(2)$ gauge enhancement, where the vector multiplets of charges $m_1=n_1=0$ and $m_2=-2n_2=\pm 1$ become massless. The only complete gauge enhancement is at the point  $T=2U = 1+i $ fixed by the $\Gamma_0(2)$ transformation $U\rightarrow \frac{U-1}{2U-1}$, and at which both $\psi$ and its first derivatives vanish. It corresponds to a level $2$ $SO(4) $ gauge enhancement where the additional vector multiplets of charges $m_1=-n_1=-2n_2 = - m_2 = \pm 1 $  become massless. The singularity is manifest in the theta lift, using the limit at the cusp $\tau\rightarrow 0$ and can be interpreted as coming from the field theory amplitude with massive vector multiplets in the loop. One obtains 
\bea  - 248 \log( U_2 T_2) - 2 \log| \psi(T,U)|^2 &\sim &- 2 \hspace{-7mm} \sum_{\substack{Q\in \sLambda_{1,1}\oplus \sLambda_{1,1}[\frac12] \\ (Q,Q) =  {1} , Z(Q)\sim 0 }} \hspace{-7mm}   \log  |Z(Q)|  \; , \CR
40 \log( U_2 T_2) - 2 \log| \chi(T,U)|^2 &\sim & - 2\hspace{-7mm}  \sum_{\substack{Q\in \sLambda_{1,1}\oplus \sLambda_{1,1}[\frac12] \\ (Q,Q) =  {1}, Z(Q)\sim 0 }} \hspace{-7mm}   \log |Z(Q)|   \; , \eea
and
\bea
 {\rm Re}[ \mathcal{F}]_{abcd} &\sim& - \tfrac1{8} \hspace{-7mm} \sum_{\substack{Q\in \sLambda_{1,1}\oplus \sLambda_{1,1}[\frac12] \\ (Q,Q) = {1}, Z(Q)\sim 0 }} \hspace{-7mm}  \log |Z(Q)| \Bigl(p_{L a} p_{L b} p_{L c} p_{L d} {-} \tfrac{3}{2} |Z(Q)|^2\delta_{(ab} p_{Lc} p_{L d)} {-} \tfrac{3}{32}   |Z(Q)|^4 \delta_{(ab} \delta_{cd)}  \Bigr)\; , \CR
 {\rm Re}[ \mathcal{G}]_{ab} &\sim& - \tfrac1{16} \hspace{-6mm} \sum_{\substack{Q\in \sLambda_{1,1}\oplus \sLambda_{1,1}[\frac12] \\ (Q,Q) =  {1}, Z(Q)\sim 0 }} \hspace{-6mm}  \log |Z(Q)|  \Bigl(p_{L a} p_{L b}  {-} \tfrac{1}{4} |Z(Q)|^2\delta_{ab}   \Bigr)\; . \eea
Using the explicit formula \eqref{PsiAOA} and the change of variable $T  = 2 U - U_2 z$ for $z\ll1$ one obtains 
\bea \psi(T,U) &\sim&  \pi i  U_2 z \eta(2U)^{248} \eta(U)^{248}  \bigl( 2 E_2(2U) - E_2(U)\bigr) \; ,  \CR
\chi(T,U) &\sim&  \pi i  U_2 z  \eta(2U)^{-40} \eta(U)^{-40} \bigl( 2 E_2(2U) - E_2(U)\bigr) \; , \eea
at the level 2 $U(2)$ gauge enhancements locus. After removing the singularity, one gets the modified coupling for the $O(2,1)$ model with a single complex modulus $U$ and the additional contribution to the $SL(2,\R)$ anomaly corresponding to the two massless vector multiplets. Note that the regular part $\tilde{\psi}(U)$ of $\psi(T,U) \sim \pi i z U_2 \tilde{\psi}(U)$ has weight $248+2$ and $\tilde{\chi}(U)$ has weight $-40+2$ in agreement with \eqref{AnomalySL2}. One can similarly read the additional contribution to the anomaly from the logarithm term in $z$ in the vector multiplets couplings 
\bea
 {\rm Re}[ \mathcal{F}]_{abcd} &\sim& - \tfrac1{4}  \log  |z|\delta_{a}^2 \delta_{b}^2 \delta_{c}^2 \delta_d^2 \; , \CR
 {\rm Re}[ \mathcal{G}]_{ab} &\sim& - \tfrac1{8}  \log |z| \delta_{a}^2 \delta_{b}^2 \; . \eea
Combining the $\log |z|$ terms together, one gets that the logarithmic divergence in $|Z(Q)|$ gives a coupling proportional to the anomaly polynomial for the two additional massless vector multiplets 
 \be  \frac{1}{72(8\pi)^4}   \hspace{-0.7mm} \int\hspace{-1.6mm} d^8x \sqrt{-g}{\rm Re}\Bigl[ \log| z| (t_8+\tfrac{i}{2}\varepsilon )  \Bigl( \tfrac{1}{5}  \Tr\! R^4 +  \tfrac{1}{4}   \Tr \! R^2 \Tr \! R^2  + 6 \Tr\! F^2 \Tr\! R^2 + 24 \Tr\!  F^4 \Bigr) \Bigr] \; , \label{AnomalyLogDiv} \ee
with $\Tr\! F^{2k} =  (-2)^k (F^{a=2})^{2k} $ for the Cartan component of the massless $SU(2)$ gauge field. 

We conclude therefore that the existence of a gauge enhancement locus implies that the BPS protected couplings must diverge accordingly as above on the locus where the additional vector multiplets become massless. 
 
\vskip 2mm

The singularities at infinite distance can instead be interpreted as decompactification limits. In the large type IIA torus volume limit, i.e. at $T\rightarrow i \infty$, we have 
\bea \psi(T,U) &\sim&  e^{15\pi i T} \eta(2U)^{248} \frac{\eta(U)^{112}}{\eta(2U)^{112}}\; ,  \quad \chi(T,U) \sim e^{-3\pi i T} \eta(2U)^{-40} \frac{\eta(U)^{-32}}{\eta(2U)^{-32}}  \; , \CR
 \theta(T,U) &\sim&e^{\frac{\pi i T}{8}} \frac{\eta(2U)^3}{\eta(U)^3}\; , \eea
up to exponentially suppressed terms in $e^{i \pi T}$. Comparing with \eqref{KKtower}, we find indeed that this is the consistent decompactification limit to type IIA in ten dimensions. The leading term in $T$ gives the ten-dimensional one-loop effective action coming from the eleven-dimensional $R^4$ correction, while the dependence in $U$ can be attributed to the Kaluza--Klein tower of states with one graviton multiplet for $m_2=0$ mod $2$ and one gravitino multiplet for $m_2=1$ mod $2$. By symmetry, the limit $U_2\rightarrow\infty$ gives the same result up to exchanging $T\leftrightarrow 2U$. 

One obtains the limit $\frac{U_2}{|U|^2}\rightarrow \infty$ by the symmetry $U \leftrightarrow - \frac{1}{T}$ that exchanges type IIA and type IIB. This limit can moreover be interpreted as a weak coupling limit in  the Dabholkar--Park orientifold, with the identifications 
\be S' = - \frac{1}{U} = C'_{89} + i e^{- \phi^\prime} V^\prime \; , \quad U' = - \frac{1}{T} \; , \quad e^{\phi'} = e^{-\phi}  \frac{|T U|}{\sqrt{T_2 U_2}} \; , \ee
where $\phi'$ is the type I dilaton and $(2\pi)^2 \alpha'_{\rm I} V'$ the torus volume. One obtains in the large $S_2'$ limit
\bea \psi(T,U) &\sim&  e^{15\pi i S'} \eta(2U')^{248} \frac{\eta(U')^{112}}{\eta(2U')^{112}}\; ,  \quad \chi(T,U) \sim e^{-3\pi i S'} \eta(2U')^{-40} \frac{\eta(U')^{-32}}{\eta(2U')^{-32}}  \; , \CR
 \theta(T,U) &\sim&e^{-\frac{\pi i S'}{8}} \frac{\eta(2U')^3}{\eta(U')^3}\; . \eea
This gives the expected contribution from the Kaluza--Klein tower of type IIB supergravity, with a graviton multiplet for even $m'_2$ and a gravitino multiplet for odd  $m'_2$. The term linear in $S'$ recombines into a $V' e^{-\phi} (t_8 t_8 +\frac14 \varepsilon \varepsilon )R^4$  that looks like a crosscap amplitude. However, we expect the crosscap amplitude to vanish in Dabholkar--Park because the large volume limit must reproduce type IIB supergravity. In order to explain this apparent discrepancy we must recall that the complete four-graviton amplitude is not protected, and in particular the term proportional to $D^4 K (t_8 t_8 +\frac14 \varepsilon \varepsilon )R^4$ in \eqref{Wilson} can also contribute to the same order such that the total contribution of order $V' e^{-\phi}$ vanishes in  Dabholkar--Park. This can only be consistent if $D^4 K (t_8 t_8 +\frac14 \varepsilon \varepsilon )R^4$ gets corrections to all orders in string perturbation theory, similarly as for heterotic -- type I duality \cite{Tseytlin:1995bi}.

The non-analytic logarithmic term in $T_2 U_2$ comes from the one-loop supergravity divergence, taking into account the logarithm of the string coupling constant $\frac{e^{2\phi}}{T_2}$ in eight dimensions in Einstein frame \cite{Green:2010sp}, one checks the  consistency between the asymmetric orbifold and the Dabholkar--Park orientifold. Including the logarithm of the string coupling constant one gets the threshold correction 
\bea  \Gamma^{\scalebox{0.6}{1-loop}}_4 &\underset{{\rm AOA}}{\sim}& \int\hspace{-1.8mm} d^8x\frac{\sqrt{-g}}{576 (4\pi)^4}  \Bigl(  \frac13 \log \frac{e^{2\phi}}{T_2} +  \log U_2 T_2\Bigr) (t_8+\tfrac{i}{2}\varepsilon ) \Bigl( \tfrac{124}{5}   \Tr\! R^4 + 5 \Tr \! R^2 \Tr \! R^2 \Biggr) \; , \CR
&\underset{{\rm DP}}{\sim}& \int\hspace{-1.8mm} d^8x\frac{\sqrt{-g}}{576 (4\pi)^4}  \Bigl(  \frac13 \log \frac{e^{2\phi'}}{V'} +  \log U'_2 V'\Bigr) (t_8+\tfrac{i}{2}\varepsilon ) \Bigl( \tfrac{124}{5}   \Tr\! R^4 + 5 \Tr \! R^2 \Tr \! R^2 \Biggr)\; . \CR
\label{Log}\eea

Because the four and six derivative couplings \eqref{WilsonFF} are 1-loop exact in the asymmetric type II orbifold theory, one can extract the corresponding non-perturbative coupling in the Dabholkar--Park orientifold. One crucial property comes from the fact that 
\bea \tau^4 \Biggl(  \frac{\eta(2\tau)^8}{ \eta(\tau)^{16}} \Biggr) \Bigg|_{\tau\rightarrow - \frac{1}{\tau}} \hspace{-3mm}&=&   \frac{\eta(\frac{\tau}{2})^8}{ 16\eta(\tau)^{16}}\sim \frac{1}{16} e^{- i \pi \tau} - \frac12  + \mathcal{O}( e^{i \pi \tau} )\; , \CR
\tau^2 \Biggl(  \frac{\eta(2\tau)^8\bigl( 2E_2(2\tau) {-} E_2(\tau)\bigr)}{ \eta(\tau)^{16}}   \Biggr) \Bigg|_{\tau\rightarrow - \frac{1}{\tau}} \hspace{-3mm}&=&  - \frac{\eta(\frac{\tau}{2})^8\bigl( 2 E_2(\tau){ -} E_2(\frac{\tau}{2}) \bigr)}{ 32\eta(\tau)^{16}} \CR
&\sim&- \frac{1}{32} e^{- i \pi \tau} - \frac12  + \mathcal{O}( e^{i \pi \tau} )\; ,  \eea
ensuring that in the weak coupling limit 
\bea e^{-2\phi'} {\rm Re}[ \mathcal{F}(T,U)]_{a_1a_2a_3a_4} &\sim &\frac{V'}8  \int_0^\infty  \hspace{-1mm} \frac{d\tau_2}{\tau_2^{\, 6}}  \hspace{-1.8mm}\sum_{\substack{m_1 \in \Z\\ n_2 \in \Z+\frac12}} \hspace{-1.8mm} e^{- \frac{\pi V'}{\tau_2 U'_2} |n_2- U' m_1|^2} \frac{V'{}^2}{U'_2{}^2} \prod_{i=1}^4 (n_2-U' m_1)_{a_i}    \; ,\hspace{10mm} \CR
e^{-\phi'} {\rm Re}[ \mathcal{G}(T,U)]_{a_1a_2} &\sim &\frac{V'}4   \int_0^\infty  \hspace{-1mm} \frac{d\tau_2}{\tau_2^{\, 4}}  \hspace{-1.8mm}\sum_{\substack{m_1 \in \Z\\ n_2 \in \Z+\frac12}}  e^{- \frac{\pi V'}{\tau_2 U'_2} |n_2- U' m_1|^2} \frac{V'}{U'_2} \prod_{i=1}^2 (n_2-U' m_1)_{a_i}   \, , \hspace{10mm}  \eea
can be interpreted as Klein bottle amplitudes in Dabholkar--Park.\footnote{One may see this more directly through the formal direct channel amplitudes 
$$  \int_0^\infty  \hspace{-1mm} \frac{d\tau_2}{\tau_2^{\, 2}}  \hspace{-1.8mm}\sum_{n_1,m_2 \in \Z} \hspace{-1.8mm} e^{- \frac{\pi}{\tau_2 V'U'_2} |n_1+ U' m_2|^2} (-1)^{m_2} \frac{1}{(2V'U'_2)^h} \prod_{i=1}^{2h} (n_1+ U' m_2)_{a_i}   \; . $$} Recall that the shift along the ninth circle in the  Dabholkar--Park orientifold implies that only half-integer winding $n_2$ along $y^9$ contribute to the Klein bottle transverse channel amplitude. Here we used the Weyl rescaling from type II to type I string frame 
\be \alpha'_{\rm II} = e^{ \phi'} \alpha'_{\rm I} \; . \ee
The non-perturbative corrections are by construction exponentially suppressed in $e^{2\pi i S'}$ or $e^{-2\pi i \bar S'}$, consistently with Euclidean D1-brane instantons.

\vskip 2mm

\subsection{M-theory}
At strong coupling in type IIA, one expects the theory to be realised as an orbifold of M-theory on $T^3$, with the M-theory torus metric and three-form
  \be ds_{\scalebox{0.6}{11D}}^2 = e^{\frac{4}{3} \phi} (dy^{10})^2 + e^{-\frac{2}{3} \phi} \frac{T_2}{U_2} | dy^9-U dy^8|^2 + T_2^{- \frac13} ds_{\scalebox{0.6}{8D}}^2 \; , \quad C_{\scalebox{0.6}{11D}} = T_1 dy^8 dy^9 dy^{10} \; . \ee
It can be defined as M-theory on the Klein bottle  \cite{Dabholkar:1996pc}, with the $\Z_2$ orbifold defined by $y^{10}\rightarrow - y^{10}$, $y^9\rightarrow y^9+ \pi \ell$. At low energy, one obtains the mode expansion for the fields $g_{\mu\nu}, \phi,B_{\mu\nu} ,\psi_+ , \lambda_-$, as ten-dimensional fields in type IIA,
\bea&&  \cos (\tfrac{m_3 y^{10}}{\ell}) e^{i \frac{m_1 y^{8}}{\ell} + i \frac{m_2 y^{9}}{\ell} } \; , \qquad m_2=0 \mbox{ mod } 2 \CR
&&  \sin (\tfrac{m_3 y^{10}}{\ell}) e^{i \frac{m_1 y^{8}}{\ell} + i \frac{m_2 y^{9}}{\ell} } \; , \qquad m_2=1 \mbox{ mod } 2  \eea
while for the fields $C_\mu, C_{\mu\nu\rho} , \psi_-, \lambda_+$ one gets the modes 
\bea&&  \cos (\tfrac{m_3 y^{10}}{\ell}) e^{i \frac{m_1 y^{8}}{\ell} + i \frac{m_2 y^{9}}{\ell} } \; , \qquad m_2=1 \mbox{ mod } 2 \CR
&&  \sin (\tfrac{m_3 y^{10}}{\ell}) e^{i \frac{m_1 y^{8}}{\ell} + i \frac{m_2 y^{9}}{\ell} } \; , \qquad m_2=0 \mbox{ mod } 2  \; . \eea
For each charge with $m_3=0$ and $m_2=0$ mod $2$ one gets a short gravity multiplet, for each charge with $m_3=0$ and $m_2=1$ mod $2$ one gets a short gravitino multiplet, while for all charges with $m_3\ge 1$ one gets a long multiplet. The contribution from the BPS multiplets matches the one from the Kaluza--Klein tower in the type IIA asymmetric orbifold. One cannot trust the contribution from the long multiplets because they are not stable states and their masses are renormalised. The naive computation gives rise to a series of instanton corrections associated to Euclidean D0-branes wrapping the torus $T^2$. However, the only stable Euclidean D0-brane is a non-BPS $\Z_2$ brane wrapping the circle along $y^8$ for $T_2> 2U_2$, dual to a non-BPS $\Z_2$ D(-1)-brane in type IIB \cite{Gutperle:2000bf}. For $T_2<2U_2$ it becomes a Euclidean non-BPS D1-brane wrapping the torus and does not have a geometric interpretation in M-theory.

The nine-dimensional string theory obtained from M-theory on the Klein bottle has a discrete $\mathbb{Z}_2$ valued gauge field, and further compactification to eight dimensions (along direction $y^8$) can be carried out with a holonomy of this field $\int C_{1}=\frac12$ turned on. For short, we shall denote it as $C_0$, referring to its T-dual in type IIB as discussed in section \ref{sec:Generalities}.  This reduction gives rise to a new disconnected component of the moduli space with the theta angle.
As proposed in \cite{Montero:2022vva},  it is equivalent to  M-theory on a different non-orientable Bieberbach manifold N$_2^3$ defined through the orbifold  of $T^3$ by $y^{10}\rightarrow - y^{10}-y^{8}$,  $y^{8}\rightarrow  y^{8}$ and $y^{9}\rightarrow y^{9}+ \pi \ell$. The mode expansion of the fields $g_{\mu\nu}, \phi,B_{\mu\nu} ,\psi_+ , \lambda_-$ is then
\bea&&  \cos (\tfrac{m_3 (y^{10} + \frac12 y^{8}) }{\ell}) e^{i \frac{m_1 y^{8}}{\ell} + i \frac{m_2 y^{9}}{\ell} } \; , \qquad m_2=0 \mbox{ mod } 2 \CR
&&  \sin (\tfrac{m_3  (y^{10} + \frac12 y^{8})}{\ell}) e^{i \frac{m_1 y^{8}}{\ell} + i \frac{m_2 y^{9}}{\ell} } \; , \qquad m_2=1 \mbox{ mod } 2  \eea
with $m_1 \in \tfrac{1}{2}\Z$ and $m_1 + \frac{m_3}{2} \in \Z$, and for the fields $C_\mu, C_{\mu\nu\rho} , \psi_-, \lambda_+$
\bea&&  \cos (\tfrac{m_3  (y^{10} + \frac12 y^{8})}{\ell}) e^{i \frac{m_1 y^{8}}{\ell} + i \frac{m_2 y^{9}}{\ell} } \; , \qquad m_2=1 \mbox{ mod } 2 \CR
&&  \sin (\tfrac{m_3  (y^{10} + \frac12 y^{8})}{\ell}) e^{i \frac{m_1 y^{8}}{\ell} + i \frac{m_2 y^{9}}{\ell} } \; , \qquad m_2=0 \mbox{ mod } 2  \; . \eea
The BPS spectrum with $m_3 =0$ is unchanged but the Kaluza--Klein spectrum of long multiplets, for $m_3\ge 1$, admits a fractional mode $m_1  \in \Z + \frac{m_3}{2}$ for odd $m_3$. We find therefore that the BPS protected couplings discussed in this section are unchanged, but the non-perturbative corrections associated to non-BPS Euclidean $\Z_2$ D0-branes along $y^8$ change sign \cite{Montero:2022vva}. Note that, in contrast to K2$\times S^1$, the  Bieberbach manifold N$_2^3$ does not have torsion in the first homology group
  \be H_1({\rm K2}\times S^1,\Z) = \Z^2 \oplus \Z_2 \; , \qquad H_1({\rm N}_2^3,\Z) = \Z^2  \; , \ee
 as well as  in the first  homology group twisted by the orientation bundle \footnote{The other (twisted) homology   groups K2$\times S^1$ and N$_2^3$ are the same.}
    \be \widehat{H}_1({\rm K2}\times S^1,\widehat{\Z}) = \Z \oplus \Z_2 \; , \qquad \widehat{H}_1({\rm N}_2^3,\widehat{\Z})  =  \Z   \; . \ee
A non-BPS D2 brane wrapping $y^{8}$ in the asymmetric orbifold can be interpreted as an M2-brane wrapping the torsion twisted cycle for $C_0=0$  \cite{Gaberdiel:2001ed}. For $C_0=\frac12$, a non-BPS D2 brane wrapping $y^8$ should instead be an M2-brane wrapping the primitive twisted cycle along $y^{10}$, while only M2-branes wrapping the M-theory circle an even number of times are (BPS) fundamental strings in the type IIA asymmetric orbifold.

The naive one loop correction from massive Kaluza--Klein states gives the correction $(t_8 t_8 {+} \frac14 \varepsilon  \varepsilon ) R^4$ with a prefactor:
\bea &&  \frac{2\pi T_2^\epsilon }{\epsilon} ( T_2 e^{-2\phi})^{\frac{\epsilon}{3}} \sum_{m_3=1}^\infty \sum_{m_1,m_2\in \Z} \frac{1}{\bigl( \frac{|m_1 + C_0 m_3  + U m_2 |^{2}}{U_2} + T_2e^{-2\phi} m_3^2 \bigr)^\epsilon }  \\
& \sim&  \zeta(3) T_2e^{-2\phi}  +4\pi e^{-\phi} \sum_{n_1,n_2\in \Z}^\prime e^{2\pi i C_0 n_1}  \frac{\sqrt{T_2U_2}}{|n_2-U n_1|} \sigma_2(n_1,n_2) K_1\Bigl(2\pi e^{-\phi} \sqrt{\tfrac{T_2}{U_2}} |n_2-Un_1|\Bigr)    \; . \nonumber \eea
As described in \eqref{Wilson}, this tensor structure is not protected by supersymmetry, consistently with the property that the non-BPS M2-brane coupled to the theta angle $C_0$ is not BPS and its quantum number is only conserved modulo 2. Note, however, that this computation is not reliable in the small volume limit in which string theory is perturbative. There are a priori corrections to $(t_8 t_8 {+} \frac14 \varepsilon  \varepsilon ) R^4$ to all orders in perturbation theory and  this non-BPS D1-instanton correction is most probably not quantitatively correct. For a similar discussion in the heterotic string context we refer to \cite{Green:2016tfs}.

\section{Heterotic compactifications at enhancement points}
When the rank of the semi-simple gauge group is sixteen in heterotic string (or eight in the CHL string), there are only two abelian vector multiplets and the effective action for the gravity and abelian vector multiplets take the same form \eqref{Wilson}.  In this section we will review some basic properties of heterotic compactifications for comparison with theories without a non-abelian sector. For this purpose, we shall consider the simple example of a discrete Wilson line along the ninth circle. We set $a_1=0$ and $a_2 = \frac{1}{N}\lambda$ for $\lambda \in E_8\oplus E_8 $ a weight of norm square divisible by $2N$, i.e. $N| (\lambda,\lambda)/2$, and redefine the moduli $T$ and $U$ according to
\bea U' &=& \frac{U}{N} \; , \qquad T' = N T \CR
m_2^\prime &=& N m_2 +( \lambda,q) +  \frac{( \lambda,\lambda)}{2 N} n_2\; , \quad q^\prime = q  + \frac{n_2}{N} \lambda \; , \quad n_2^\prime = \frac{n_2}{N} \eea
such that the central charge \eqref{Z} takes the form
\bea Z &=& \frac{1}{\sqrt{T_2U_2}} \Bigl( m_1 + U m_2 + (  U \lambda/N, q) + \bigl( T + \tfrac12 U  (\lambda,\lambda)/N^2 \bigr) n_2  - U T n_1\Bigr)\CR
 &=&\frac{1}{\sqrt{T'_2U'_2}} \Bigl( m_1 + U' m'_2 +T' n'_2  - U' T' n_1\Bigr)  \; , \eea
 while 
 \be \sLambda_{2,2} \oplus E_8\oplus E_8 =\sLambda_{1,1}\oplus  \bigoplus_{r,s\in \Z_N} \bigl( \sLambda_{1,1}^{r,s}[N] \oplus L_{16}^{r,s}\bigr) \ee
 with 
\bea L_{16}^{r,s} &=& \Bigl\{ q' \in E_8 \oplus E_8 + \frac{s}{N} \lambda  \; \big| \; (\lambda,q') = r + \frac{(\lambda,\lambda)}{2N} s\; {\rm mod} N \Bigr\} \; , \CR
 \sLambda_{1,1}^{r,s}[N] &=& \Bigl\{ m'_2 \in N \Z + r\, , \; n'_2 \in \Z + \frac{s}{N} \Bigr\} \; .  \eea
We shall only consider two specific examples that will become relevant later. 

\vskip 2mm

Using the isomorphism 
\be E_8 = \bigoplus_{k\in \Z_5}  (A_4+k{\rm f})\oplus (A_4-2k{\rm f})\ee
where one writes the elements of $ A_4^*/A_4 \cong \Z_5$ as $k{\rm f}$ with ${\rm f}$ the fundamental weight and $k \in \Z_5$, one can define $\lambda = 5 {\rm f}^{\scalebox{0.6}{(1)}} + 5 {\rm f}^{\scalebox{0.6}{(3)}} $ where $ {\rm f}^{\scalebox{0.6}{(1)}}$ and $ {\rm f}^{\scalebox{0.6}{(3)}} $ are in the first $A_4^*$ sublattices in both $E_8$ lattices. Writing the $A_n$ theta series as
\be \vartheta_{A_n,k} = \sum_{q\in A_n } e^{i \pi \tau (q+ k {\rm f},q+ k {\rm f})} \ee
one can write the Narain partition function as
\be\Gamma_{2,18}  = \sum_{k,r,s\in \Z_5} \vartheta_{A_4,k+s}  \vartheta_{A_4,-2k} \vartheta_{A_4,-k+r}  \vartheta_{A_4,2k-2r+2s} \Gamma_{\sLambda_{1,1}\oplus \sLambda_{1,1}^{r,s}[5] } \; . \ee  
The function $\psi$ in the effective action can then be computed as 
\bea  - 344 \log( U_2 T_2) - 2 \log| \psi(T,U)|^2 &=&\int_{\mathcal{F}} \frac{d^2\tau}{\tau_2^2} \frac{E_4(\tau)}{\eta(\tau)^{24} }   \Gamma_{2,18}\CR
&=&  \int_{\mathcal{F}_5} \frac{d^2\tau}{\tau_2^2} \frac{E_4(\tau)}{\eta(\tau)^4 \eta(5\tau)^4}   \Gamma_{\sLambda_{1,1}\oplus \sLambda_{1,1}[5]}  \; . \eea
The pole of $\frac{E_4(\tau)}{\eta(\tau)^4 \eta(5\tau)^4}$ at $0$
\be \frac{E_4(\tau)}{\eta(\tau)^4 \eta(5\tau)^4} \underset{\tau \rightarrow 0}{\sim} 25 e^{  i\frac{2\pi}{5} \frac{1}{\tau} } \ee
can lead to a gauge symmetry enhancement to $SU(5)^2 \times U(10)$. The residue $25 = 5\times 5$ is the dimension of the bi-fundamental of $SU(5)\times SU(5)\subset SU(10)$. It is consistent to have a residue different from $1$ because the gauge group is non-abelian. By construction the term in four abelian gauge field strengths is given by 
\be {\rm Re}[ \mathcal{F}(T,U)]_{abcd} = \tfrac1{16}\int_{\mathcal{F}_5}  \hspace{-1.8mm} \frac{d^2\tau}{\tau_2^{\, 2}} \frac{1}{\Delta_4}  \Gamma_{\sLambda_{1,1}\oplus \sLambda_{1,1}[5]}\bigl[p_{L a} p_{L b} p_{L c} p_{L d} {-} \tfrac{3}{2\pi \tau_2} \delta_{ab} p_{Lc} p_{L d} {+} \tfrac{3}{16 \pi^2 \tau_2^2}  \delta_{ab} \delta_{cd} \bigr] \; , \label{F4N=5}  \ee
with 
\be  \Delta_4(\tau) = \eta(\tau)^4 \eta(5\tau)^4\ee
and the poles of $\frac{1}{\Delta_4}$ can be interpreted as the ones of $\frac{E_4(\tau)}{\Delta_4}$.

\vskip 5mm

For $N=6$ one may use the isomorphism 
\be E_8 = \bigoplus_{k\in \Z_6}  (A_5+k{\rm f})\oplus (A_2+k{\rm f})\oplus (A_1+k{\rm f})\ee
to write the Narain partition function as
\be \Gamma_{2,18}  = \sum_{k,r,s\in \Z_6} \vartheta_{A_5,k}  \vartheta_{A_2,k+s} \vartheta_{A_1,k+s}   \vartheta_{A_5,-k-s+r}  \vartheta_{A_2,-k+r} \vartheta_{A_1,-k+r} \Gamma_{\sLambda_{1,1}\oplus \sLambda_{1,1}^{r,s}[6] } \; . \ee
The function $\psi$ in the effective action can then be computed as 
\be  - 340 \log( U_2 T_2) - 2 \log| \psi(T,U)|^2 = \int_{\mathcal{F}_6} \frac{d^2\tau}{\tau_2^2} \frac{E_4(\tau)}{\eta(\tau)^2 \eta(2\tau)^2\eta(3\tau)^2\eta(6\tau)^2  }   \Gamma_{\sLambda_{1,1}\oplus \sLambda_{1,1}[6]}  \; . \ee
The poles of $\frac{E_4(\tau)}{\eta(\tau)^2 \eta(2\tau)^2\eta(3\tau)^2\eta(6\tau)^2 }$ at the cusps
\bea \frac{E_4(\tau)}{\eta(\tau)^2 \eta(2\tau)^2\eta(3\tau)^2\eta(6\tau)^2  } &\underset{\tau \rightarrow i \infty}{\sim}& e^{- 2\pi i \tau} \; , \quad \frac{E_4(\tau)}{\eta(\tau)^2 \eta(2\tau)^2\eta(3\tau)^2\eta(6\tau)^2  } \underset{\tau \rightarrow 0}{\sim} 36 e^{  \frac{i \pi }{3\tau} }\; ,   \\
 \frac{E_4(\tau)}{\eta(\tau)^2 \eta(2\tau)^2\eta(3\tau)^2\eta(6\tau)^2  } &\underset{\tau \rightarrow \frac13}{\sim}& 4 e^{  i\frac{\pi}{3} \frac{1}{3\tau- 1} }\; , \quad \frac{E_4(\tau)}{\eta(\tau)^2 \eta(2\tau)^2\eta(3\tau)^2\eta(6\tau)^2  } \underset{\tau \rightarrow \frac12}{\sim} 9 e^{  \frac{ i\pi }{6\tau- 2} } \; ,  \nonumber \eea
can respectively lead to gauge symmetry enhancements to $U(2)\times SU(2)^2 \times SU(3)^2 \times  SU(6)^2$, $SU(2)^2 \times SU(3)^2 \times U(12)$,  $U(4)\times SU(3)^2 \times SU(6)^2 $ and $SU(2)^2  \times U(6)\times SU(6)^2$. The  residues $\ell^2$ are respectively the dimensions of the bi-fundamental of $U(\ell)\times U(\ell)\subset U(2\ell)$ for $\ell|6$.

By construction the term in four abelian gauge field strengths is given by 
\be {\rm Re}[ \mathcal{F}(T,U)]_{abcd} =\tfrac1{16} \int_{\mathcal{F}_6}  \hspace{-1.8mm} \frac{d^2\tau}{\tau_2^{\, 2}} \frac{1}{\Delta_4}  \Gamma_{\sLambda_{1,1}\oplus \sLambda_{1,1}[6]}\bigl[p_{L a} p_{L b} p_{L c} p_{L d} {-} \tfrac{3}{2\pi \tau_2} \delta_{ab} p_{Lc} p_{L d} {+} \tfrac{3}{16 \pi^2 \tau_2^2}  \delta_{ab} \delta_{cd} \bigr] \; ,\label{F4N=6} \ee
with 
\be  \Delta_4(\tau) = \eta(\tau)^2 \eta(2\tau)^2\eta(3\tau)^2\eta(6\tau)^2\ee
and the poles of $\frac{1}{\Delta_4}$ can be interpreted in the same way as for $\frac{E_4(\tau)}{\Delta_4}$ above.  

\section{Asymptotics of Borcherds lifts}
\label{Borcherds}
As we have discussed in the previous sections, the known protected couplings in eight dimensions can all be described as one-loop contributions in closed string theories. Using a partial unfolding of the moduli space integral, they all eventually reduce to the theta lift of a meromorphic modular form over the fundamental domain of a congruent subgroup $\Gamma_0 (N)$. The meromorphic modular forms $\psi$, $\chi$ and $\theta$ defining the protected couplings \eqref{Wilson} are then Borcherds lifts  \cite{Borcherds:1995uml,Borcherds:1996uda}. For higher rank groups $O(2,l)$ with $l>2$, all the meromorphic modular forms whose zeros and poles are supported on special divisors are Borcherds lifts \cite{BRUINIER2014315}. The converse theorem extends to $l=2$ if one assumes there is no singularity in the interior of the moduli space \cite{bieker2024}, i.e. if there is no gauge enhancement locus in the theory. The converse theorem is expected to extend to the case including special divisors in the interior. We have explained in section \ref{sec:Generalities} that the meromorphic modular forms defining the $R^4$ Wilsonian coupling must satisfy specific boundary conditions at the cusps and the finite distance singularities. In this section, we shall provide strong evidence  that this implies that it can be obtained from a Borcherds lift when $N$ is square-free.

\subsection{Atkin--Lehner involutions} 
In this section we analyse weight zero theta lifts of a meromorphic modular function with only single poles at the cusps, for the lattice $\sLambda_{1,1} \oplus \sLambda_{1,1}[N]$ with $N$ square-free. 
 We begin by setting up some necessary notations that will be used throughout this section. Assume $f(\tau)$ is a modular invariant meromorphic function of $\Gamma_0 (N)$ with only poles at the cusps, i.e. $f(\gamma \cdot  \tau) = f (\tau)$, for all $\gamma \in \Gamma_0 (N)$ and $\gamma\cdot \tau = (a \tau  +b)/(c \tau + d)$. As mentioned previously, throughout this section (and section~\ref{sec:constraint})  we assume that $N$ is square-free, such that there is a single cusp for each divisor $\ell$ of $N$. 
 
 It will be convenient to introduce $SL(2,\R)$ transformations that map the cusp $\tau \rightarrow i \infty$ to the other cusps at $\tau=0$ and $\tau = \frac{1}{\ell}$ for each divisor $\ell$ of $N$ different from $1$ and $N$. Instead of defining them as $SL(2,\Z)$ matrices in $SL(2,\Z) /\Gamma_0(N)$, it is more convenient to define them as $SL(2,\R)$ matrices normalising $\Gamma_0(N)\subset SL(2,\Z)$. We choose them as 
\be \sigma_1 = \left( \begin{array}{cc} 0 & - \frac{1}{\sqrt{N}}\\ \sqrt{N} & 0 \end{array}\right) \; , \quad \sigma_{\ell} \underset{\ell \ne 1 , N}{=}  \sqrt{\frac{\ell}{N}} \mx{\frac{N}{\ell} & r_\ell \\ N & \frac{N}{\ell} s_\ell} \; , \quad \sigma_N =\left( \begin{array}{cc} 1 & \ 0 \\ 0 & \ 1 \end{array}\right) \; , \label{cosetR} \ee
where $(s_\ell,r_\ell)$ are coprime integers satisfying 
\lceq{
s_\ell \frac{N}{\ell } - r_\ell \ell = 1 \, .
}
The divisors of $N$ satisfying gcd$(\ell,\frac{N}{\ell})=1$ are generally called Hall divisors. Since we assume $N$ is square-free all the divisors of $N$ are Hall divisors. The matrix group generated by $\Gamma_0(N)$ and the $\sigma_\ell$ is the so-called Atkin–Lehner group $\Gamma_{0*}(N)\supset \Gamma_0(N)$ \cite{AtkinLehner}.\footnote{In the original paper \cite{AtkinLehner} the Atkin–Lehner group is defined by the matrices with  integer entries $W_{\ell} =  \sqrt{\ell}\,  \gamma_{\frac{N}{\ell}}$ of determinant equal $\ell$. We find more convenient to write them as matrices in $SL(2,\R)$ such that the group multiplication is the matrix multiplication.} The index of $\Gamma_0(N)$ in the Atkin–Lehner group  is $|\Gamma_{0*}(N) / \Gamma_0(N)|=| \{\ell,  \ell | N\} | $, the number of divisors of $N$. This implies that $r_\ell$ and $s_\ell$ defining $\sigma_\ell$ are determined up to right multiplication by $\Gamma_0(N)$. The coset representatives satisfy the algebra 
\be 
\label{eq:sec5_sigma_multi}
\sigma_\ell \sigma_{\ell'} = \sigma_{\ell*\ell'} \, \gamma(\ell,\ell') \, ,
\quad \text{for some } \gamma(\ell,\ell')\in \Gamma_0(N)\, .
\ee
where the binary operation  $\ell*\ell'$ is given by 
\be \ell * \ell' = \frac{N \gcd(\ell,\ell')^2}{\ell\,\ell'} \; . \label{StarProduct} \ee
The operation is commutative and satisfies $(\ell * \ell')*\ell'=\ell$ for all divisors $\ell$ and $\ell'$. This definition generalises to $N$ not square-free with some additional subtleties, we shall only describe the cases $N=4,8,9$ in appendix \ref{ProofDelta4}.

These matrices are very useful to compute the theta lift of an arbitrary modular function.  This is particularly manifest when one considers $\Gamma_0(N)$ for $N\le 13$ and different from $11$. In these cases there exists a Hauptmodul function $j_N(\tau)$ with a single pole of residue $1$ at $\tau \rightarrow i \infty$ and which is regular at all the other cusps. It is uniquely fixed by setting its constant term at $\tau\rightarrow i \infty$ to vanish. A general modular function with single poles at the cusps can then be defined as
\be f(\tau) = k + \sum_{\ell | N} c_\ell(-1) j_N(\sigma_\ell \cdot \tau) \; .  \label{BasisFromHaupt} \ee
With this definition, for any $\ell|N$ we have
\be f_\ell(\tau) \coloneqq  f(\sigma_\ell \cdot \tau)  = k + \sum_{\ell' | N} c_{\ell'}(-1) j_N\Bigl(\sigma_{\ell'*\ell} \cdot \tau \Bigr) \; .\label{FlbyHaupt} \ee
This definition is justified by the fact that the pole of $f_\ell(\tau)$ at $\tau \rightarrow i \infty$ has residue $c_{\ell}(-1)$. Indeed, the pole comes from $j_N(\tau)$ and  $\ell'*\ell=N$ implies $\ell' = \ell$. We write the $q$ expansion of $f_\ell(\tau)$
\be f_\ell(\tau) = \sum_{n\ge -1} c_\ell(n) e^{2\pi i n \tau} \; , \ee
where by definition $f_N(\tau) \coloneqq f(\tau)$.  In general there is no Hauptmodul function and the minimal modular functions with only simple poles at the cusps must have at least $g{+}1$ poles when the modular surface $X_0(N)$ is of genus $g$. However, the determination of the constant terms from the poles at the cusps can be done as if there was a Hauptmodul function, with additional constraints that we shall not need for our analysis of the asymptotics of the coupling functions.

\subsection{Singular theta lift}
The next step is to evaluate the singular theta lift of a modular function $f$ with at most single poles at the cusps. We define 
\lceq{
\label{eq:sec4_theta_lift}
\Theta [f](T,U) = \int_{\mathcal{F}_N} \frac{d \tau_1 d\tau_2}{\tau_2^2} f (\tau)  \Gamma_{\sLambda_{1,1} \oplus \sLambda_{1,1}[N]} (\tau)\;  ,
}
where the integral kernel is the lattice sum
\longeq{
\Gamma_{\sLambda_{1,1} \oplus \sLambda_{1,1}[N]} &=& \hspace{1.8mm}\tau_2\hspace{-2.8mm} \sum_{\substack{m_1,n_1,n_2 \in \Z\\ m_2 \in  N \Z}} e^{ - \frac{\pi \tau_2}{T_2 U_2} \bigl| ( T,-1) \bigl({}^{n_1}_{m_2}{}^{\ n_2}_{-m_1}\bigr) (^{\; U}_{-1}) \bigr|^2 +2\pi i  \tau \det\bigl({}^{n_1}_{m_2}{}^{\ n_2}_{-m_1}\bigr)} \; . 
}
From the symmetries of the lattice $\sLambda_{1,1} \oplus \sLambda_{1,1}[N]$, it follows that for any divisor $\ell |N$
\lceq{
\Theta\left[f_\ell\right] (T, U) = \Theta \left[f\right] \left(T, \sigma_{\ell} \cdot U\right) = \Theta \left[f\right] \left(\sigma^\intercal_{\ell} \cdot T,  U\right) \, , \label{tauU}
}
and
\lceq{
\Theta\left[f\right] (T, U) =\Theta\left[f\right] ( NU, T/N)  \, .
}

To keep the presentation concise, we move the detailed computations to appendix~\ref{app:mathematical_details} and retain only the final results in the main text. We use the unfolding method assuming $T_2$ is sufficiently large, i.e. 
\be T_2 > N U_2\; , \quad T_2 > \frac{U_2}{N|U|^2} \; , \quad T_2>  \frac{{\rm gcd}(\ell , \frac{N}{\ell})U_2}{\frac{N}{\ell} | \ell  U  + s_\ell|^2} \; , \ee
for all $\ell | N$ different from $1$ and $N$. Assuming $T$ satisfies this condition, we compute in appendix~\ref{app:mathematical_details} that 
\longeq{
\label{DirectTheta}
        \Theta[f](T,U) &= \\
        & \hspace{-1.5cm}- \sum_{\ell |N} c_\ell (0) \log \Bigl[ \frac{\ell^2}{N} U_2 T_2\Bigr]  +\frac{\pi U_2}{3 } \Biggl( \sum_{\ell|N} \ell c_\ell(0) \Biggr) +\frac{\pi T_2}{3 N} \Biggl( \sum_{\ell|N} \frac{N}{ \ell} c_\ell(0)- 24 c_N(-1)\Biggr)  \\
&\hspace{-1.5cm}-2\sum_{\ell | N} \hspace{-2mm} \sum_{\substack{m,n \in  \Z \\ (m,n) \ne (0,0)}} \hspace{-2mm} c_\ell(\ell mn) \log\Bigl[ 1-e^{- 2\pi  \ell | m U_2 + n \frac{T_{\scalebox{0.4}{$2$}}}{N}| + 2\pi i \ell  ( m U_1 + n \frac{T_{\scalebox{0.4}{$1$}}}{N})} \Bigr]  \, . 
}
Using \eqref{eq:sec5_sigma_multi}, one obtains for $d|N$ 
\be f_d(\sigma_\ell\cdot \tau) = f( \sigma_d \sigma_\ell\cdot \tau)  = f\bigl( \sigma_{\ell* d}\cdot \tau\bigr) = f_{\frac{ N{\rm gcd}(d,\ell)^2}{d \ell}}(\tau) \; . \ee
It follows using moreover \eqref{tauU} that
\longeq{
\Theta[ f](T,\sigma_d \cdot U) &= - \sum_{\ell |N} c_\ell (0) \log \Bigl[ \frac{\ell^2}{N} U_2 T_2\Bigr] \\
&\hspace{-12mm}  +\frac{\pi U_2}{3 } \Biggl( \sum_{\ell|N} \ell c_{\frac{N{\rm gcd}( \ell,d)^2}{\ell d}}(0) \Biggr) +\frac{\pi T_2}{3 N} \Biggl( \sum_{\ell|N} \frac{N}{ \ell} c_{\frac{N{\rm gcd}( \ell,d)^2}{\ell d}}(0)- 24 c_d(-1)\Biggr) \\
&\hspace{-12mm}  -2\sum_{\ell | N}  \hspace{-2mm} \sum_{\substack{m,n \in  \Z \\ (m,n) \ne (0,0)}} \hspace{-2mm} c_{\frac{N{\rm gcd}( \ell,d)^2}{\ell d}}(\ell mn) \log\Bigl[ 1-e^{- 2\pi  \ell | m U_2 + n \frac{T_{\scalebox{0.4}{$2$}}}{N}| + 2\pi i \ell  ( m U_1 + n \frac{T_{\scalebox{0.4}{$1$}}}{N})} \Bigr]
}
Because all these functions must have the same leading term linear in $T_2$ at large $T_2$, one concludes that for all divisors  $d| N$ 
\bea  \sum_{\ell|N} \frac{N}{\ell} c_\ell(0) -24 c_N(-1)
= \sum_{\ell|N}\frac{N}{\ell}c_{\frac{ N{\rm gcd}(d,\ell)^2}{d \ell}}(0)    - 24 c_{d}(-1) \; . \label{ConstraintConstant} \eea
Using this equation for $d=1$, one can simplify \eqref{DirectTheta} to
\longeq{
\label{eq:sec5_Theta_lift}
\Theta[ f] &= \\
&\hspace{-0.5cm}- \sum_{\ell |N} c_\ell (0) \log \Bigl[ \ell U_2 \bigl|\eta(\ell U)\bigr|^4 \frac{\ell}{N} T_2 \bigl| \eta(\tfrac{\ell}{N} T)\bigr|^4 \Bigr] 
- 2 c_1(-1)  \log\Bigl[ \Bigl| e^{-2\pi i \frac{ T}{N}} -e^{-2\pi i U} \Bigr|^2 \Bigr]  \\
&\hspace{-0.5cm}  -2\sum_{\ell | N}\sum_{m,n\ge 1} c_\ell(\ell mn) \log\Bigl[ \Bigl| 1-e^{2\pi i \ell ( m U + n \frac{T}{N})}\Bigr|^2 \Bigr]
}
One checks that this formula for the theta lift is valid for all values of the moduli $T$ and $U$, as the logarithm of the corresponding Borcherds product.  For example, if the modular curve $X_0(N)$ is of genus zero, one obtains in this way that the theta lift of the Hauptmodul function $j_d(\tau) \coloneqq j_N(\sigma_d\cdot \tau)$, with $d|N$, is 
\be 
\label{eq:Haupt_Theta_lift}
\Theta[ j_d ] = - \sum_{\ell |N} c_{\frac{N}{\ell}} (0) \log \Bigl[ \frac{\ell^2}{N} U_2 \bigl|\eta(\scalebox{0.8}{$ \frac{d \ell}{{\rm gcd}(d,\ell)^2}$} U)\bigr|^4 T_2 \bigl| \eta(\tfrac{\ell}{N} T)\bigr|^4 \Bigr] - 2 \log \Bigl[ \Bigl| j_N(\tfrac{T}{N}) - j_{\frac{N}{d}}(U) \Bigr|^2 \Bigr] \; . \ee

\subsection{Determination of the constant terms} 
Because the only holomorphic modular function is a constant, any non-constant modular function must be determined by its poles. This implies that all the coefficients $c_\ell(0)$ must be determined in function of the residues $c_\ell(-1)$ up to an additional constant corresponding to the constant function. For $N$ square-free, the equations \eqref{ConstraintConstant} provide one equation for each divisor of $N$ except $N$, and allow one to solve for the $c_\ell(0)$. 

Suppose first that $N\le 13$ and different from $11$ such that there exists a Hauptmodul function $j_N$. Using \eqref{FlbyHaupt} we see that we only need to solve for \eqref{ConstraintConstant} in the case of the Hauptmodul function 
\be j_\ell(\tau) = \sum_{n\ge -1} a_\ell(n) e^{2\pi i n \tau} \; . \ee
Because the Hauptmodul function has only a single pole at $\tau \rightarrow i \infty$, $a_\ell(-1) = \delta_{\ell,N}$ and \eqref{ConstraintConstant} gives for all $\ell|N$
\be \sum_{\ell'|N}\frac{N}{\ell*\ell' }a_{\ell' }(0)    = 24 \delta_{\ell,N}   + K[j] \; ,  \ee
for a constant $K[j]$. We compute the inverse of the matrix $A_{\ell,\ell'} =\frac{N}{\ell * \ell'}$ in \eqref{InverseAB} as
\lceq{
\label{eq:sec6_Bdd}
A^{-1}_{\ell,\ell'} = \frac{1}{\prod_{p|N} (p^2 - 1)} (-1)^{\omega (\ell* \ell') } \frac{N}{\ell * \ell'} \, ,
}
where $\omega(\ell)$ is the number of primes dividing $\ell$, and we consider the set of divisors $\ell$ as labelling the vector components. Using moreover \eqref{TraceB} for $\ell'=N$, one obtains 
\be a_\ell(0) =  \frac{24(-1)^{\omega (\ell ) } \frac{N}{\ell} - \sigma_1(N) K[j]}{\prod_{p|N} (p^2 - 1)}  \; .   \ee
Because $a_N(0)=0$ by convention, $K[j] = \frac{24(-1)^{\omega (N )}}{\sigma_1(N)}$ and  
\be a_\ell(0) =  \frac{24(-1)^{\omega (\ell ) } \frac{N}{\ell} - 24(-1)^{\omega (N )} }{\prod_{p|N} (p^2 - 1)}  \; .   \ee
One obtains directly the general solution using  \eqref{FlbyHaupt} as
\be c_\ell(0) = k' +\frac{24}{\prod_{p|N} (p^2 - 1)}  \sum_{\ell'|N} (-1)^{\omega (\ell*\ell' ) } \frac{ \ell \ell'}{{\rm gcd}(\ell,\ell')^2} c_{\ell'}(-1)  \label{Solc0Nfree} \ee
 where we redefined $k$ to $k'$ to simplify the expression.
 
 Because we only used linear algebra, \eqref{Solc0Nfree} is the general solution to the linear equations \eqref{ConstraintConstant} for arbitrary genus. When the genus $g$ of the modular curve $X_0(N)$ is greater than $1$, there is no Hauptmodul function and there must be at least $g+1$ of the $c_\ell(-1)$ coefficients different from zero. However, we shall not need to solve this additional constraint and \eqref{FlbyHaupt}  will be sufficient to rule out all possible physical couplings with $N\ge 3$.

\subsection{On the converse theorem at genus zero} 
\label{Converse} 
When the modular curve $X_0(N)$ has genus zero, we can construct a general meromorphic function whose divisor is supported on special divisors using the Hauptmodul function and its Atkin--Lehner transforms $j_\ell(\tau)$, for all $\ell| N$. When $N$ is square-free, one checks that the general ansatz takes the form  
\be \psi(T,U) = \prod_{\ell| N} \Bigl( \eta(\tfrac{\ell}{N} T)^{m_\ell} \eta( \ell U)^{n_\ell} \Bigr) \prod_{\ell |N} \Bigl( j_N(\tfrac{T}{N})  - j_{\ell}(U)\Bigr)^{p_{\ell}} \; , \ee
for integers $m_\ell,n_\ell,p_\ell$. 
In particular, any difference of Hauptmodul functions satisfies 
\be  \label{RelationJs} j_\ell(\tfrac{T}{N}) - j_{\ell'}(U) = M(\ell,\ell') \Bigl( j_N(\tfrac{T}{N}) - j_{\ell*\ell'}(U) \Bigr) \ee
with $M(\ell,\ell')$ an appropriate product of Dedekind eta functions, so one can assume only $j_N(\tfrac{T}{N})$ appears. This simplification can be understood from the fact that the Hauptmodul function $j_N$ is an isomorphism between $\overline{\mathcal{F}_N}$ and $\mathbb{CP}^1$ and that the zeros of the meromorphic function are determined by $N T = \sigma_\ell \cdot U$ for some $\ell$. 
The coefficients $p_\ell$ correspond precisely to the coefficients $c_{\frac{N}{\ell}}(-1)$ of the modular function $f(\tau)$ in the Borcherds lift. The coefficients $m_\ell, n_\ell$ are a priori only constrained by the integrality conditions for every $\ell|N$
\be \frac{1}{24} \sum_{\ell'| N} \ell'* \ell  \, m_{\ell'} \in  \Z \; , \qquad \frac{1}{24} \sum_{\ell'| N} \ell'* \ell  \, n_{\ell'} \in  \Z  \; . \ee
Requiring the symmetry $T \leftrightarrow NU$ and the condition that the divergence is the same at every cusp, one requires that all these integers are equal and the coefficients $m_\ell,n_\ell$ are determined in function of the $p_\ell$ and the weight $w$ of the meromorphic modular form. One finds therefore that the general meromorphic form with this property can be obtained from the Borcherd lift formula using \eqref{eq:Haupt_Theta_lift}. Note that this symmetry criterion is necessary for the consistency of the coupling at all cusps.

One can apply the same reasoning when $N$ is not square-free. In this case there are additional generalised Atkin--Lehner generators for each divisor $\ell$ of $N$, which we give explicitly in appendix \ref{ProofDelta4} for $N=4,8,9$. Moreover, when $p^2 | N$, one checks that the fractional element 
\be T_{\frac1p} = \left( \begin{array}{cc} 1 & \ \frac1p \\ 0  & \ 1 \end{array}\right)\; ,  \ee
normalises $\Gamma_0(N)$. There are consequently additional special divisors and one has generally a special divisor at  $NT = \sigma_{\ell}\cdot U + k \frac{N}{p}$ for each $\ell|N$ and each $k$ mod $p$ such that $p^2 |N$. To take these special divisors  into account, one considers the following ansatz \be \label{ansatzNSF}  \psi(T,U) = \prod_{\ell| N} \Bigl( \eta(\tfrac{\ell}{N} T)^{m_\ell} \eta( \ell U)^{n_\ell} \Bigr) \prod_{k=0}^{p-1} \prod_{\ell|N} \Bigl( j_N(\tfrac{T}{N})  -e^{ 2\pi i \frac{k}{p} } j_{\ell}(U)\Bigr)^{q_{\ell,k}} \; . \ee
In this case formula \eqref{RelationJs} generalises schematically to 
\be j_\ell(\tfrac{T}{N}) -e^{ 2\pi i \frac{k}{p} }   j_{\ell'}(U) = M(\ell,\ell',k) \Bigl( j_N(\tfrac{T}{N}) -e^{ 2\pi i \frac{k'(\ell,\ell',k)}{p} }  j_{\ell''(\ell,\ell',k) }(U) \Bigr) \ee
with $M(\ell,\ell',k)$ an appropriate product of Dedekind eta functions. The theta lift associated to the lattice $\sLambda_{1,1} \oplus \sLambda_{1,1}[N]$ would only provide a subset of the possible coefficients, since there is only one coefficient $c_\ell(-1)$ for each divisor $\ell| N$, whereas there are $p>1$ coefficients $ q_{\ell,k}$. The converse theorem may still apply if we combine all the lattices compatible with the $\Gamma_0(N)\times \Gamma_0(N)$ symmetry, but we have not investigated this question. We shall only discuss the examples $N=4,8,9$ of genus zero modular forms using the ansatz \eqref{ansatzNSF}.

\section{Constraints on theories with two abelian vector multiplets}
\label{sec:constraint}
In this section we apply the results of the previous section to check whether a consistent theory can  admit coupling functions with duality group $\Gamma_0(N) \times \Gamma_0(N)$. The three main consistency conditions  are 
\begin{itemize}
    \item Anomaly-cancellation condition: The logarithmic terms in $\log(T_2 U_2)$ must  be consistent with the eight-dimensional composite-anomaly.
        \item Gauge-enhancement limit: The singularities of the couplings in the interior of the moduli space  must be interpreted as a gauge-enhancement, so the corresponding logarithmic divergence must be consistent with the eight-dimensional composite-anomaly with the presence of the additional massless fields. 
    \item Decompactification limit: The limits in which $T_2$ or $U_2$ go to a cusp of the moduli space must correspond to decompactification limits such that the leading contributions to the couplings reproduce the ten-dimensional effective couplings of the heterotic or the type II superstring theories.
\end{itemize}

For this purpose we shall concentrate mostly on the gravitational coupling \eqref{Wilson}. The anomaly-cancellation condition requires that $\psi(T,U)$ is a meromorphic modular form of weight $124$  for $T$ and $U$.  If we assume $\kappa=0$ --  so that the theory must decompactify to type II string theory --  $\chi(T,U)$ must also be a meromorphic modular form of weight $-20$. If $\kappa=1$ --  so that the theory must decompactify to heterotic string theory --  $\chi(T,U)$ admits a non-meromorphicity controlled  by the gauge sector. 

The singularities of the couplings in the interior of the moduli space must be interpreted as the presence of BPS states that become massless at the singular locus. The only available BPS multiplets are the gravitino multiplets and the vector multiplets. Two gravitino multiplets becoming massless would correspond to a supersymmetry enhancement, but there is no gauging deformation of maximal supergravity allowing for a spontaneous supersymmetry breaking to minimal supergravity in eight dimensions \cite{Bergshoeff:2003ri,Catino:2013ppa}. Two vector multiplets becoming massless must correspond to a gauge-enhancement to $SU(2)$ (or $SO(3)$) such that one can interpret the Coulomb branch as giving a mass to the W-boson  supermultiplets. This implies that both $\psi(T,U)$ and $\chi(T,U)$ can only have single zero with a coefficient equal to $+1$ at special divisors. The Yang--Mills anomaly term must also be present so that we get a logarithmic divergence in the protected couplings according to \eqref{AnomalyLogDiv}. 
\subsection{\mathtitle{N}{N} is square-free}
We have argued that for any consistent theory in eight dimensions with lattice $L_{2,2} =\sLambda_{1,1}\oplus \sLambda_{1,1}[N]$ for $N$ square-free, the function $\psi$ in the effective action \eqref{Wilson} can be defined as the  Borcherds lift 
\be - 248 \log( U_2 T_2) - 2 \log| \psi(T,U)|^2 = \int_{\mathcal{F}_N} \frac{d^2\tau}{\tau_2^2}  f(\tau)  \Gamma_{\sLambda_{1,1}\oplus \sLambda_{1,1}[N]}  \ee
where the function $f(\tau)$ has only simple poles at the cusps with residues $c_\ell(-1) =0$ or $1$. We have defined for convenience  
\be f_N(\tau) \coloneq  f(\tau)\;, \quad f_1(\tau) \coloneq f_N\bigl( - \tfrac{1}{N \tau}\bigr) \; , \quad  f_\ell(\tau) \coloneq f_N\biggl( \frac{ \frac{N}{\ell} \tau + r_\ell}{N \tau + \frac{N}{\ell} s_\ell}\biggr) \; .\ee
with \footnote{Here and below, $\omega( \ell)$ denotes the number of distinct prime divisors of $\ell$.}
\be f_\ell(\tau) = \sum_{n\ge -1} c_\ell(n) e^{2\pi i n \tau} \; , \ee
The constant terms are determined in function of the residues and an additional constant $k'$ as \eqref{Solc0Nfree}. 
One can then check that the constant in \eqref{ConstraintConstant} is 
\be  \sum_{\ell|N} \frac{\ell d}{ {\rm gcd}( \ell,d)^2 } c_{\ell }(0)   - 24 c_d(-1) = \sigma_1(N) k'  \; . \label{Prelecc}  \ee
Recall that there may not be a general solution for arbitrary values of $c_\ell(-1)$, there is one  if the modular curve $X_0(N)$ is of genus zero, i.e. if  $N\le 13$ and different from $11$. However, this will not affect our argument, as this gives additional constraints that we do not need to take into account to conclude that there is no physically consistent solution.  

We need to consider separately the possibility that the theory decompactifies to type II string theory or heterotic string theory in ten dimensions. 

\subsubsection*{Type II}

In the type II case we must restore supersymmetry in the large volume limit, so we have the same number of graviton and gravitini multiplets for Kaluza--Klein modes class $\ell= m_2$~mod~$N$ 
\be K =\sum_{\ell|N} n_g(\ell) \prod_{p|\frac{N}{\ell}} (p-1)  =  \sum_{\ell|N} n_\psi(\ell) \prod_{p|\frac{N}{\ell}} (p-1)\; . \ee
By assumption, $n_g(N)=1$ and $n_\psi(N)=0$. For $N$ square-free, $p{-}1$ is even for all $p|\frac{N}{\ell}$ not equal to $2$, so $N$ must be even so that the number of gravitino multiplets can match the number of graviton multiplets. In particular, $n_\psi(\frac{N}{2}) + n_g(\frac{N}{2})$ must be an odd number. 

The first constraint on the modular function $f$ comes from the term proportional to the volume in the effective action. This fixes the constant $K$ as 
\be 360 K= \sum_{\ell|N} \ell c_\ell(0) - 24 c_{1}(-1) = k'\,  \sigma_1(N) \; . \ee
The first equality comes from the identification of the leading term in $\psi(T,U)$ in  $T_2$ \eqref{KKtower}. The second equality is obtained using  \eqref{Prelecc} for $d=1$.

We need also to fix  the number of graviton and gravitino multiplets for each divisor $\ell|N$ of the Kaluza--Klein mode $m_2$. 
Using  
\be \sum_{\ell'|\ell} (-1)^{\omega( \frac{N{\rm gcd}(\ell'',\ell')^2}{\ell' \ell''})} \frac{\ell' \ell''}{{\rm gcd}(\ell',\ell'')^2} = \prod_{p|\ell} (p-1)\times (-1)^{\omega( \frac{N}{\ell  {\rm gcd}(\scalebox{0.6}{$\frac{N}{\ell}$},\ell'')})}{\rm gcd}(\tfrac{N}{\ell},\ell'')\; ,  \ee
one computes that 
\bea \sum_{\ell'|\ell} c_{\ell'}(0) &=& \sigma_0(\ell)  \frac{360 K}{\sigma_1(N)} +\frac{24}{\sigma_1(N) \prod_{p|\frac{N}{\ell}} (p-1)}  \sum_{\ell' |N}(-1)^{\omega( \frac{N}{\ell  {\rm gcd}(\scalebox{0.6}{$\frac{N}{\ell}$},\ell')})}{\rm gcd}(\tfrac{N}{\ell},\ell')  c_{\ell'}(-1)\CR
&=& 248 n_g(\ell) + 112 n_\psi(\ell) \; . \label{ngnpsi} \eea
We shall concentrate on the values $\ell=N$ and $\frac{N}{2}$, for which \eqref{ngnpsi} simplifies to
\bea \sum_{\ell|N} c_{\ell}(0) &=& \sigma_0(N)  \frac{360 K}{\sigma_1(N)} +\frac{24}{\sigma_1(N) }  \sum_{\ell |N}c_{\ell}(-1)= 248 \; , \\ 
\sum_{\ell |\frac{N}{2}} c_{\ell}(0) &=& \frac{\sigma_0(N)}{2}  \frac{360 K}{\sigma_1(N)} +\frac{24}{\sigma_1(N)}  \sum_{\ell |\frac{N}{2} }\bigl( 2  c_{2\ell}(-1)- c_\ell(-1) \bigr)  = 248 n_g(\tfrac{N}{2}) + 112 n_\psi(\tfrac{N}{2})  \, . \nonumber   \eea
Eliminating the unknown $K$, we get the equation
\be \frac{\sigma_1(\frac{N}{2})}{3} \Bigl( 31(2n_g(\tfrac{N}{2}) - 1) + 28 n_\psi(\tfrac{N}{2}) \Bigr) = \sum_{\ell | \frac{N}{2}} \bigl( c_{2\ell}(-1) - c_\ell(-1)\bigr)\; ,  \ee
where the right-hand-side is by assumption bounded by $\sigma_0(N)$ in absolute value. 
The left-hand-side is very large in absolute value unless $n_g(\frac{N}{2})=0$ and $n_\psi(\frac{N}{2})=1$, in which case we have to solve 
\be \sigma_1(\tfrac{N}{2}) = \sum_{\ell | \frac{N}{2}} \bigl( c_\ell(-1)-c_{2\ell}(-1)  \bigr) \; . \label{Sigma1Sigma0} \ee
Using $c_\ell(-1)\in \{ 0, 1\}$, this requires $\sigma_1(\frac{N}{2}) \le \sigma_0(\frac{N}{2})$, which already singles out $N=2$ with $c_1(-1)=1$ and $c_2(-1)=0$.  We conclude that the only solution is $N=2$ with 
\be f(\tau) = 136 +4096  \frac{\eta(2\tau)^{24}}{\eta(\tau)^{24}} \; , \ee
which is the function \eqref{Tr1} defining the coupling in the asymmetric orbifold theory. 
\subsubsection*{Heterotic}
In string theory the highest dimension in which there exists an orbifold with the  rank reduced by $16$ is $D=7$. Although we do not know any mechanism to implement such a rank reduction in $D=8$, we shall  nonetheless consider the possibility that the  non-abelian gauge fields only carry  Kaluza--Klein modes of non-trivial class $\ell = m_2$~mod~N. 

In this case the terms proportional to the volume in the effective action fix 
\be k' = \frac{720 K}{\sigma_1(N)}\; , \ee
and for $K$ graviton multiplets for all Kaluza--Klein modes $1\le \ell \le N$ modulo $N$ we must have $496 K$ vector multiplets 
\be K =\sum_{\ell|N} n_g(\ell) \prod_{p|\frac{N}{\ell}} (p-1)  = \frac{1}{496} \sum_{\ell|N} n_v(\ell) \prod_{p|\frac{N}{\ell}} (p-1)\; . \ee
By assumption $n_g(N)=1$ and $n_v(N)=0$. 

To determine the number of graviton and vector multiplets,  one computes similarly that 
\bea \sum_{\ell'|\ell} c_{\ell'}(0) &=& \sigma_0(\ell)  \frac{720 K}{\sigma_1(N)} +\frac{24}{\sigma_1(N) \prod_{p|\frac{N}{\ell}} (p-1)}  \sum_{\ell' |N}(-1)^{\omega( \frac{N}{\ell  {\rm gcd}(\scalebox{0.6}{$\frac{N}{\ell}$},\ell')})}{\rm gcd}(\tfrac{N}{\ell},\ell')  c_{\ell'}(-1)\CR
&=& 248 n_g(\ell) + n_v(\ell) \; . \label{HeteNgNv} \eea
Using again this equation in the definition of $K$ we obtain the constraint 
\bea 744 K &=& 248 \sum_{\ell|N} n_g(\ell) \prod_{p|\frac{N}{\ell}} (p-1)  +  \sum_{\ell|N} n_v(\ell) \prod_{p|\frac{N}{\ell}} (p-1)
\CR
&=& \sum_{\ell| N} \sum_{\ell'|\ell} c_{\ell'}(0) \prod_{p|\frac{N}{\ell}} (p-1) \CR
&=& 720 K + 24 c_N(-1) \eea
so we conclude that $c_N(-1) = K$. In order to have $c_N(-1)$ at most $1$, we must have $K=1$, implying that $n_g(\ell) = \delta_{\ell,N}$. Using $K=1$ in equation \eqref{HeteNgNv} for $\ell=N$, we obtain 
\be  \sum_{\ell|N} c_{\ell}(0) = \sigma_0(N)  \frac{720 }{\sigma_1(N)} +\frac{24}{\sigma_1(N) }  \sum_{\ell |N}c_{\ell}(-1)= 248 \ee
which gives 
\be  \sum_{\ell |N}c_{\ell}(-1) = \frac{31}{3}\sigma_1(N) - 30 \sigma_0(N) \; . \ee
The left-hand-side is by assumption bounded by $\sigma_0(N)$ in absolute value, while $\frac{31}{3}\sigma_1(N) - 31 \sigma_0(N) >0$ for $N>6$, and  $\frac{31}{3}\sigma_1(N) - 29 \sigma_0(N) <0$ for $N<5$. The only possible solutions are therefore $N=5$ and $N=6$. In both cases this equation requires that $c_\ell(-1)=1$ for all divisors of $N$. Assuming  $c_\ell(-1)=1$ for all $\ell$ we have 
\be \sum_{\ell'|\ell} c_{\ell'}(0) = \sigma_0(\ell)  \frac{744 }{\sigma_1(N)} = n_v(\ell) \; ,  \ee
for all $\ell\ne N$. This gives a consistent solution for $N=5$ and $N=6$ with $c_\ell(0) = \frac{744}{\sigma_1(N)}$ for all $\ell$ and 
\bea \label{N56} N=5: && \quad n_v(1) = 124\; , \quad n_v(5) = 0 \; , \CR
N=6: && \quad n_v(1) = 62 \; , \qquad n_v(2) = n_v(3) = 124\; , \quad n_v(6) = 0\; . \eea

However, this is only the constraint on the  function $\psi$, whereas we must in principle check all the couplings. Because the supersymmetry Ward identities are deformed in the presence of a $R^2$ coupling, we cannot constrain the function $\chi$ without analysing first the $F^4$ coupling. If we assume that the most general solution to the Ward identity can be written as a theta lift 
\be {\rm Re}[ \mathcal{F}(T,U)]_{abcd} \!=\! \int_{\mathcal{F}_N}  \hspace{-1.8mm} \frac{d^2\tau}{\tau_2^{\, 2}} \frac{1}{\Delta_4(\tau)} \Gamma_{\sLambda_{1,1}\oplus \sLambda_{1,1}\![N]}\bigl[p_{L a} p_{L b} p_{L c} p_{L d} {-} \tfrac{3}{2\pi \tau_2} \delta_{ab} p_{Lc} p_{L d} {+} \tfrac{3}{16 \pi^2 \tau_2^2}  \delta_{ab} \delta_{cd} \bigr] \; ,  \ee
for a weight $-4$ modular form $\frac{1}{\Delta_4}$ with poles of order at most $1$ at the cusps, as  in \eqref{FFFFandFFRR}, we find that there would not be any solution. This is a reasonable assumption for $N$ square-free, as was argued in \cite{Bossard:2017wum}. We prove in appendix \ref{ProofDelta4} that the only such modular form  is (up to normalisation) the inverse of the weight four cusp form in both cases 
\be \label{Delta4N56} \Delta_4(\tau) \underset{N=5}{=} \eta(\tau)^4 \eta(5\tau)^4  \; , \quad \Delta_4(\tau) \underset{N=6}{=} \ \eta(\tau)^2 \eta(2\tau)^2 \eta(3\tau)^2 \eta(6\tau)^2 \; . \ee
However, the residue of $\frac{1}{\Delta_4(\tau)}$ at the cusp $\tau\rightarrow \frac{1}{\ell}-1$ for $\ell|N$ not equal to $N$ is $\ell^2$ and not $ 1$, so we conclude that the singularities of such an $F^4$ coupling cannot be compatible with a $U(1)^2$ gauge group without non-abelian factors. This is precisely the function that appears in heterotic string theories with a Wilson line as we saw in \eqref{F4N=5} and \eqref{F4N=6}. It can consistently appear in $D=7$ with $3$ vector multiplets, with the respective lattices $ \sLambda_{3,3}[5]$ and $\sLambda_{3,3}[6]$, but not in $D=8$ with two vector multiplets. We therefore conclude that the partial solutions \eqref{N56} are merely remnants of the $D=7$ consistent string theories, but do not correspond to any consistent theory in eight dimensions.

One may be able to prove rigorously the absence of such a candidate coupling using the pre-potential introduced in \cite{Lerche:1998nx}. This pre-potential is reminiscent of an Eichler integral. It is locally holomorphic and fails to be automorphic by a polynomial cocycle. One expects the set of such `Eichler integrals' of fixed weight and polynomial cocycle to be finite-dimensional, so one may be able to prove that the theta lifts of \eqref{Delta4N56} are indeed the unique candidate couplings compatible with $\Gamma_0(N) \times \Gamma_0(N)$ invariance for $N=5,6$. 

\subsection{\mathtitle{N=4,8,9}{N=4,8,9}}
In this section we consider the cases $N=4,8,9$ separately, following the discussion of section~\ref{Converse}.

One computes using the results of appendix \ref{ProofDelta4} that the general ansatz \eqref{ansatzNSF} for a meromorphic form of $\Gamma_0(4) \times \Gamma_0(4)$ with the same limit at all cusps is 
\begin{multline} \psi = \eta(\tfrac{T}{2})^{12k-4\sum_{\ell|4} ( p_\ell + q_\ell)} \eta(T)^{8 \sum_{\ell|4} ( p_\ell + q_\ell)} \\ \eta(U)^{8(p_1-p_2+q_1-q_2)}  \eta(2U)^{12k+24 (p_2+q_2) -4\sum_{\ell|4} ( p_\ell + q_\ell)}  \eta(4U)^{8(p_4-p_2+q_4-q_2)} \\
 \prod_{\ell | 4} \bigl( j_4(\tfrac{T}{4}) - j_\ell(U) \bigr)^{p_\ell} \prod_{\ell | 4} \bigl( j_4(\tfrac{T}{4}) + j_\ell(U) \bigr)^{q_\ell}\end{multline} 
 with $p_\ell,q_\ell$ equal to $0$ or $1$ and $k \in \mathds{Z}$ such that the leading term $\psi \sim e^{2\pi i k \frac{T}{4}}$ at $T\rightarrow i \infty$ is invariant under $T\rightarrow T+4$. Requiring the weight $w=124$ in order to cancel the gravitational anomaly, one obtains 
\be 12 k + 4 \sum_{\ell|4} (p_\ell+q_\ell) = 248  \qquad  \Rightarrow \qquad k = \frac{62 - \sum_{\ell|4}(p_\ell+q_\ell) }{3}\; . \ee
The only possibilities are then to have either two or five zeroes (non-zero $p_\ell,q_\ell$), that give respectively $k=20$ and $k=19$, whereas the consistency with type II would have required $k$ to be divisible by $15$ (and by $30$ for heterotic). 

For $N=8$ one obtains similarly that 
\begin{multline} \psi = \eta(\tfrac{T}{4})^{4k} \eta(\tfrac{T}{2})^{4k-2\sum_{\ell|8} ( p_\ell + q_\ell)} \eta(T)^{4 \sum_{\ell|8} ( p_\ell + q_\ell)} \eta(U)^{4(p_1-p_2+q_1-q_2)}   \eta(8U)^{4(p_8-p_4+q_8-q_4)}  \\ \eta(2U)^{4k- 2(p_1+q_1- 5p_2-5q_2+2p_4+2q_4)}  \eta(4U)^{4k- 2(2p_2+2q_2- 5p_4-5q_4+p_8+q_8)}   \\
 \prod_{\ell | 8} \bigl( j_8(\tfrac{T}{8}) - j_\ell(U) \bigr)^{p_\ell} \prod_{\ell | 8} \bigl( j_8(\tfrac{T}{8}) + j_\ell(U) \bigr)^{q_\ell}\end{multline} 
with $p_\ell,q_\ell$ equal to $0$ or $1$ and $k \in \mathds{Z}$ such that the leading term $\psi \sim e^{2\pi i k \frac{T}{8}}$ at $T\rightarrow i \infty$ is invariant under $T\rightarrow T + 8$. Requiring the weight $w=124$ one obtains 
\be 8 k + 2 \sum_{\ell|8} (p_\ell+q_\ell) = 248  \qquad  \Rightarrow \qquad k = \frac{124 - \sum_{\ell|8}(p_\ell+q_\ell) }{4}\; . \ee
In this case one finds that $k$ is divisible by $15$ if and only if there are four zeros, in which case $k=30$. There are 70 choices of coefficients $p_\ell,q_\ell$ with four equal to $1$, giving 19 inequivalent solutions, none of which are compatible with either type II or heterotic string theory. 

Using the results of appendix \ref{ProofDelta4} with the general ansatz \eqref{ansatzNSF}, one finds that the general $\Gamma_0(9)\times \Gamma_0(9)$ meromorphic form with the same singularity at each cusp can be expressed as 
\begin{multline} \psi = \eta(\tfrac{T}{3})^{8k-\sum_{a \in \Z_3} \sum_{\ell\in \{ 1,3,3'\!,9\}} p_{\ell,a}}  \eta(T)^{3\sum_{a \in \Z_3} \sum_{\ell\in \{ 1,3,3'\!,9\}} p_{\ell,a}}  \eta(U)^{3\sum_{a \in \Z_3}  ( p_{1,a} - \frac12 p_{3,a} -  \frac12 p_{3'\!,a}) }     \\ \eta(3U)^{8k-\sum_{a \in \Z_3} ( p_{1,a} - 5 p_{3,a} - 5 p_{3'\!,a} + p_{9,a})}   \eta(9U)^{3\sum_{a \in \Z_3}  ( p_{9,a} - \frac12 p_{3,a} -  \frac12 p_{3'\!,a}) }   \\
 \prod_{a \in \Z_3} \prod_{\ell \in \{ 1,3,3'\!,9\} } \bigl( j_9(\tfrac{T}{9}) - e^{\frac{2\pi i a}{3}  } j_\ell(U) \bigr)^{p_{\ell,a}}   \end{multline} 
with $p_{\ell,a}$ equal to $0$ or $1$ and satisfying moreover \footnote{More generally there is a factor of $\bigl( \frac{A_1(U)+ i \sqrt{3} B_1(U)}{A_1(U)- i \sqrt{3} B_1(U)}\bigr)^{\frac{1}{2} \sum_{a\in \Z_3} ( p_{3,a} - p_{3'\!,a})}$ that is not a product of Dedekind eta functions and would be incompatible with the decompactification limit interpretation.} 
\be \sum_{a \in \Z_3} p_{3,a} = \sum_{a \in \Z_3} p_{3'\!,a}\; ,  \ee
while $k \in \mathds{Z}$ such that the leading term $\psi \sim e^{2\pi i k \frac{T}{9}}$ at $T\rightarrow i \infty$ is invariant under $T\rightarrow T + 9$. Here $3$ and $3'$ label the two independent cusps of $\Gamma_0(9)$ for $\ell = 3$. Requiring the weight $w=124$, to cancel the gravitational anomaly, one obtains 
\be 8 k + 2\sum_{a \in \Z_3} \sum_{\ell\in \{ 1,3,3'\!,9\}} p_{\ell,a}   = 248  \qquad  \Rightarrow \qquad k = \frac{124 - \sum_{a \in \Z_3} \sum_{\ell\in \{ 1,3,3'\!,9\}} p_{\ell,a}  }{4}\; . \ee

As for $N=8$, one finds that $k$ is divisible by $15$ if and only if there are four zeros (four $p_{\ell,a}=1$), in which case $k=30$.  There are 159 choices of coefficients $p_{\ell,a}$ compatible with the constraints, giving 7 inequivalent solutions, none of which are compatible with either type II or heterotic string theory.

\section*{Acknowledgments}

Useful conversations with Emilian Dudas, Jens Funke and H\'ector Parra de Freitas are gratefully acknowledged. BL would like to thank the Laboratoire de Physique Th\'eorique et Hautes \'Energies for hospitality at the beginning of this work.



\appendix

\section[Weight w=-4 meromorphic modular forms]{Weight $w=-4$ meromorphic modular forms}
\label{ProofDelta4}
In this appendix we discuss meromorphic modular forms of weight $-4$ with at most a single pole at each cusp. The number of modular forms with at most a single pole at the cusps is equal to the number of cusps minus the number of  cuspidal forms of weight $6$ \cite{Manschot:2007ha}. Using this criterion and the dimension of the space of $\Gamma_0(N)$ cuspidal forms of weight $6$ \cite{Shimura}, one obtains that such a form invariant under $\Gamma_0(N)$ exists only for $N=1,2,3,4,5,6,8,9$. For $N=1$ we have $\frac{E_4(\tau)^2}{\eta(\tau)^{24}}$, for $N=2$, $\frac{16\eta(2\tau)^8}{\eta(\tau)^{16}}$ and $\frac{\eta(\tau)^8}{\eta(2\tau)^{16}}$. In this appendix we give these modular forms for $N=3,4,5,6,8,9$. For each congruent subgroup we define two modular forms $A_w$ and $B_{w'}$ that freely generate the set of modular forms of $\Gamma_0(N)$. We identify the cusps as the zeros of irreducible polynomials in these functions and determine in this way the basis of meromorphic modular forms with poles of order at most $p$. For $N=3,5,6,8,9$ we shall find that there is a single modular form of weight $-4$ that behaves at the cusps as 
\be \frac{1}{\Delta_4} \underset{\tau = i \infty}{\sim} q^{-1} \; , \quad  \frac{1}{\Delta_4} \underset{\tau = \frac{d}{N}\ne \frac1{N}}{\sim} \frac{d^2}{{\rm gcd}(d,\frac{N}{d})^2}  q^{- \frac{{\rm gcd}(d,\frac{N}{d})}{d}} \; . \ee
Only for $N=2$ and $4$ we find modular forms of weight $-4$ and residue $0$ or $1$ at the cusps.

\vskip 2mm

The ring of modular forms of $\Gamma_1(3)$ is freely generated by the modular forms
\be A_1(\tau) = \sqrt{ \frac{3 E_2(3\tau)-E_2(\tau)}{2}}= \vartheta_{A_2}(\tau) \; , \quad B_3(\tau) =\frac{E_4(\tau)-E_4(3\tau)}{240 A_1(\tau)} \; , \ee
and the modular forms of  $\Gamma_0(3)$ are the even polynomials in $A_1$ and $B_3$. We have the action of the Fricke generator  $\sigma_1$
\be f(\tau)|_{\sigma_1} = \frac{1}{(\sqrt{3} \tau)^w} f(- \tfrac{1}{3\tau}) \ee
that gives 
\be A_1|_{\sigma_1} = - A_1\; , \qquad  B_3|_{\sigma_1} =B_3 -\frac{1}{27} A_1^3 \; . \ee
The first cusp form is 
\be \Delta_6(\tau) = \eta(\tau)^6 \eta(3\tau)^6 = B_3\bigl( A_1^3-27 B_3 \bigr) \; . \ee
The unique weight $w=-4$ quasi-modular form with at most a single pole at the cusps is up to normalisation 
\be \frac{3 E_2(3\tau)-E_2(\tau)}{2\Delta_6(\tau)} = \frac{A_1^2}{B_3\bigl( A_1^3-27 B_3 \bigr) }  \; . \ee
The Hauptmodul  function of $\Gamma_0(3)$ is 
\be  j_3(\tau) = \frac{\eta(\tau)^{12}}{\eta(3\tau)^{12}} +12 = \frac{A_1^3}{B_3}-15\; . \ee 

\vskip 2mm

The ring of modular forms of $\Gamma_1(4)$ is freely generated by the modular forms
\be A_1(\tau) =\vartheta_3(2\tau)^2 = \vartheta_{2A_1}(\tau) \; , \quad B_2(\tau) =2E_2(2\tau)-E_2(\tau) \; , \ee
and the modular forms of  $\Gamma_0(4)$ are the  polynomials in $A_1^2$ and $B_2$. For $N=4$ we have the additional Atkin--Lehner involution for $\ell = 2$ dividing $4$ that normalises $\Gamma_0(4)$
\be \sigma_2 = \left(\begin{array}{cc} 1 \, & \, 0\\ 2\, &\, 1\end{array}\right)\; .  \ee
We have the action of $\sigma_1$ and $\sigma_2$
\bea A_1^2|_{\sigma_1} &=&- A^2_1\; , \qquad  B_2|_{\sigma_1} =B_2 -3A_1^2 \; , \CR
A_1^2|_{\sigma_2} &=&A^2_1-B_2\; , \qquad  B_2|_{\sigma_2} =-B_2 \; . \eea
The first cusp form is 
\be \Delta_6(\tau) = \eta(2\tau)^{12}  = \frac1{16} A_1^2 \bigl(A_1^2-B_2\bigr)\bigl( B_2-2 A_1^2 \bigr) \; . \ee
The  weight $w=-4$ quasi-modular forms with at most a single pole at the cusps are linear combinations of 
\be \frac{16}{\bigl(A_1^2-B_2\bigr)\bigl( B_2-2 A_1^2 \bigr) } = \frac{\vartheta_3(2\tau)^4}{\eta(2\tau)^{12}}  \; , \qquad  \;  \frac{16}{A_1^2 \bigl( B_2-A_1^2 \bigr) } = \frac{\vartheta_4(2\tau)^4}{\eta(2\tau)^{12}} \; . \ee
The Hauptmodul  function of $\Gamma_0(4)$ is 
\be  j_4(\tau) = \frac{\eta(\tau)^{8}}{\eta(4\tau)^{8}} +8 = \frac{16A_1^2}{B_2-A_1^2}-8\; . \ee 

\vskip 2mm

The ring of $\Gamma_0(5)$ modular forms is freely generated by the complex modular form $\mathcal{A}_1$ of $\Gamma_1(5)$  that is defined as a root of the two polynomials
\be \mathcal{A}_1 \overline{\mathcal{A}_1}  = \frac{5E_2(5\tau)-E_2(\tau)}{4}\; , \qquad \frac{E_4(5\tau)-E_4(\tau)}{30} =  (1+i) \mathcal{A}_1^4 + (1-i) \overline{\mathcal{A}_1}^4 - 2\mathcal{A}_1^2 \overline{\mathcal{A}_1}^2\; . \ee
One can choose the root (up to multiplication by $i$) such that it expands at the cusp $\tau= i \infty$ as
\be \mathcal{A}_1(\tau) = 1  + (3+i)e^{2\pi i \tau} + (4 - 2 i)e^{4\pi i \tau} + \mathcal{O}(e^{6\pi i \tau} )\; . \ee
The modular forms of $\Gamma_0(5)$ of weight $w$ are the polynomials of order $w$ invariant under $\mathcal{A}_1\rightarrow i  \mathcal{A}_1$. The Fricke generator $\sigma_1$ defined on a weight $w$ modular form as
\be f(\tau)|_{\sigma_1} = \frac{1}{(\sqrt{5} \tau)^w} f(- \tfrac{1}{5\tau}) \ee
acts on $\mathcal{A}_1$ and its complex conjugate as \footnote{Note that $ \mathcal{A}_1$ and $\overline{ \mathcal{A}_1}$ are defined up to multiplication by $i$, so one can consistently act differently on $ \mathcal{A}_1$ and $\overline{ \mathcal{A}_1}$, as long as they are complex conjugate up to a power of $i$.}
\be \mathcal{A}_1|_{\sigma_1} = \frac{\sqrt{11+2i}}{5^{\frac{3}{4}}} \, \overline{\mathcal{A}_1}\; , \qquad \overline{\mathcal{A}_1}|_{\sigma_1} = - \frac{\sqrt{11-2i}}{5^{\frac{3}{4}}}{\mathcal{A}_1}\; . \ee
One has then 
\be \Delta_4 = \eta(\tau)^4 \eta(5\tau)^4 = \frac{11-2i}{16} \bigl( \mathcal{A}_1 + \overline{\mathcal{A}_1}\bigr) \bigl( \mathcal{A}_1 - \overline{\mathcal{A}_1}\bigr)\biggl(\mathcal{A}_1 - \frac{{11+2i}}{5^{\frac{3}{2}}}\overline{\mathcal{A}_1} \biggr) \biggl({\mathcal{A}_1} + \frac{{11+2i}}{5^{\frac{3}{2}}}  \overline{\mathcal{A}_1} \biggr) \; . \ee
The cusp $\tau = i \infty$ is identified with the zero of  $\mathcal{A}_1 = \pm \overline{\mathcal{A}_1} $ and the cusp $\tau = 0$ with the zero of $\mathcal{A}_1 \pm  \frac{{11+2i}}{5^{\frac{3}{2}}}\overline{\mathcal{A}_1}$. 
The modular forms of weight $w=-4$ and poles of order at most $p$ can be written as 
\be f_{p,w} = \frac{\sum_{h=0}^{2p + 2 \lfloor w/4 \rfloor} \alpha_h \,   \mathcal{A}_1^{2h+w/2-2\lfloor w/4 \rfloor} \overline{\mathcal{A}_1}^{4p+w/2+2\lfloor w/4 \rfloor-2h} }{\Delta_4(\tau)^p}  \; .  \ee
It follows that $\frac{1}{\Delta_4}$  is the unique  modular form of weight $w=-4$ and at most single poles at the cusp. The Hauptmodul   function of $\Gamma_0(5)$ can be written as 
\be j_5(\tau) = \frac{(-5+2i) \mathcal{A}_1^2 + (5+2i) \overline{ \mathcal{A}_1}^2}{\mathcal{A}_1^2 -\overline{ \mathcal{A}_1}^2} = \frac{\eta(\tau)^6}{\eta(5\tau)^6}+6 \; . \ee

\vskip 2mm

The modular forms of $\Gamma_1(6)$ are freely generated by the modular form $A_1(\tau)= \vartheta_{A_2}(\tau)$ and $B_1(\tau) = \vartheta_{A_2}(2\tau)$ of $\Gamma_1(6)$. The modular forms of $\Gamma_0(6)$ of weight $w$ are the even polynomials in $A_1$ and $B_1$ of order $w$. The action of the 
Atkin--Lehner group elements is given by
\bea A_1|_{\sigma_1}  &=& \sqrt{2} i B_1 \;, \quad B_1|_{\sigma_1}  =  \frac{ i}{\sqrt{2}} A_1\; , \quad A_1|_{\sigma_3}  = \sqrt{2} B_1 \;, \quad B_1|_{\sigma_3}  =  -\frac{ 1}{\sqrt{2}} A_1\; ,\CR
  A_1|_{\sigma_2}  &=&i  A_1 \;, \quad B_1|_{\sigma_2}  =  - i B_1\; . \eea
One has then 
\be \Delta_4 = \eta(\tau)^2 \eta(2\tau)^2 \eta(3\tau)^2 \eta(6\tau)^2 = -\frac{1}{36}\bigl(A_1+B_1 \bigr) \bigl( A_1-B_1 \bigr)\bigl(A_1+2B_1 \bigr) \bigl( A_1-2B_1 \bigr) \; . \ee
The cusp $\tau= i \infty$ is identified with the zero of $A_1-B_1$, the cusp at $\tau=0$ with the zero of $A_1-2B_1$, the cusp at $\tau = \frac12$ with  the zero of $A_1+B_1$ and the cusp at $\tau = \frac13$ with the zero of  $A_1+2B_1$. 
The modular form of weight $w$ and poles of order at most $p$ can be written as
\be f_{p,w} = \frac{\sum_{h=0}^{4p+w} a_h \, A_1^{h}B_1^{4p+w-h} }{\Delta_4(\tau)^{p}} \; . \ee
There is therefore a unique modular form $\frac{1}{\Delta_4}$ of weight $w=-4$ and at most single poles at the cusp. The Hauptmodul   function of $\Gamma_0(6)$ can be written as 
\be j_6(\tau) = \frac{-A_1 +7 B_1}{A_1-B_1} = \frac{\eta(2\tau)^3 \eta(3\tau)^9}{\eta(\tau)^3\eta(6\tau)^9}-3 \; . \ee

\vskip 2mm

The modular forms of $\Gamma_0(8)$ are the even polynomials in the modular forms $A_1(\tau)= \vartheta_{3}(2\tau)^2$ and $B_1(\tau) = \vartheta_{3}(4\tau)^2$.  For $N=8$ we have the two additional Atkin--Lehner involutions for $\ell = 2$ and $\ell = 4$ dividing $8$  that normalise $\Gamma_0(8)$
\be \sigma_2 = \frac{1}{\sqrt{8}} \left(\begin{array}{cc} 4 \, & \, -1\\ 8\, &\, 0\end{array}\right)\;, \qquad \sigma_4 = \left(\begin{array}{cc} 1 \, & \, 0\\ 4\, &\, 1\end{array}\right)\; .  \ee
The action of the 
generalised Atkin–Lehner group elements is 
\bea A_1|_{\sigma_1}  &=& -\sqrt{2} i B_1 \;, \quad B_1|_{\sigma_1}  = - \frac{ i}{\sqrt{2}} A_1\; , \quad A_1|_{\sigma_2}  = i \sqrt{2} ( A_1-B_1) \;, \CR
  B_1|_{\sigma_2}  &=&  \frac{ i}{\sqrt{2}} (A_1-2B_1)\; , \quad  A_1|_{\sigma_4}  =-  A_1 \;, \quad B_1|_{\sigma_4}  =  B_1-A_1\; . \eea
The first cuspidal form is 
\be \Delta_4 = \eta(2\tau)^4 \eta(4\tau)^4 = - \frac{1}{4} A_1 B_1( A_1-B_1)(A_1-2B_1) \; , \ee
and the unique weight $-4$ quasi-holomorphic modular form with at most single poles at the cusps is $\frac{1}{\Delta_4}$. The Hauptmodul  function is 
\be j_8(\tau) = \frac{\eta(4\tau)^{12}}{\eta(2\tau)^4\eta(8\tau)^8}= \frac{4 B_1}{A_1-B_1}\; . \ee

\vskip 2mm

The modular forms of $\Gamma_0(9)$ are the even polynomials in the modular forms $A_1(\tau)= \vartheta_{A_2}(\tau)$ and $B_1(\tau) =\vartheta_{A_2}(3\tau)$.  For $N=9$ we have the additional Atkin--Lehner generator for $\ell = 3$ dividing $9$ that normalises $\Gamma_0(9)$
\be \sigma_3 = \left(\begin{array}{cc} 1 \, & \, 0\\ 3\, &\, 1\end{array}\right)\; ,  \ee
which squares to its inverse up to right multiplication by $\Gamma_0(9)$. 
 The action of the 
transformations $\tau|_{\sigma_1} = - \frac{1}{9 \tau}$ and $\tau|_{\sigma_3} = \frac{\tau}{3\tau+1}$ is 
\bea A_1|_{\sigma_1}  &=& -\sqrt{3} i B_1 \;, \quad B_1|_{\sigma_1}  = - \frac{ i}{\sqrt{3}} A_1\; , \CR
  A_1|_{\sigma_3}  &=& A_1 \;, \quad B_1|_{\sigma_3}  =  - \frac{1+i \sqrt{3}}{2} B_1+\frac{1+\frac{i}{\sqrt{3}}}{2} A_1\; , \eea
where $\sigma_3$ is unipotent of order $3$. The first cuspidal form is 
\be \Delta_4 = \eta(3\tau)^8 = - \frac{1}{48} (A_1-B_1)(A_1-3 B_1)  (A_1^2 + 3 B_1^2) \; , \ee
which is the product of the four modular forms vanishing at the four cusps related by $\sigma_3$ and $\sigma_1$, with
\bea (A_1-B_1)|_{\sigma_3} &=&  - \frac{1+i \sqrt{3}}{2} \bigl( - B_1 + \frac{i}{\sqrt{3}} A_1\bigr) \; , \CR
 \bigl( - B_1 + \frac{i}{\sqrt{3}} A_1\bigr)|_{\sigma_3} &=& - \frac{1+i \sqrt{3}}{2} \bigl( - B_1 - \frac{i}{\sqrt{3}} A_1\bigr) \; , \CR
 \bigl( - B_1 - \frac{i}{\sqrt{3}} A_1\bigr)|_{\sigma_3} &=& - \frac{1+i \sqrt{3}}{2}(A_1-B_1) \eea
and
\be  (A_1-B_1)|_{\sigma_1} =  \frac{i}{\sqrt{3}} ( A_1-3 B_1) \; . \ee
It follows that the unique weight $-4$ quasi-holomorphic modular form with at most single poles at the cusps is $\frac{1}{\Delta_4}$.  The Hauptmodul  function is 
\be j_9(\tau) = \frac{\eta(\tau)^{3}}{\eta(9\tau)^3}+3= \frac{6 B_1}{A_1-B_1}\; . \ee

\section{Supplementary material on theta lifts}
\label{app:mathematical_details}
To avoid overwhelming the reader with technical details, many lengthy intermediate computations were omitted in the main text, in particular in section~\ref{Borcherds} and section~\ref{sec:constraint}. In this appendix, we provide the necessary mathematical details to assist interested readers in better understanding our analysis. 
\subsection{Lattice isomorphisms}
In this appendix we prove that $L_{2,2} = \sLambda_{1,1}[N_1] \oplus \sLambda_{1,1}[N_2] $ is isomorphic to $\sLambda_{1,1} \oplus \sLambda_{1,1} [N_1 N_2]$ for positive integers $N_1,N_2$, if and only if $\gcd(N_1,N_2) = 1$. One necessary condition is that the discriminant groups of these two lattices are isomorphic, i.e.,
\lceq{
\left(\Z/N_1\Z\right)^2 \times\left(\Z/N_2\Z\right)^2 \cong \left(\Z/ N_1 N_2 \Z\right)^2 \, .
}
By the Chinese remainder theorem, such a ring isomorphism only exists when $\gcd(N_1,N_2) = 1$. Now we need to prove that $\gcd(N_1,N_2) = 1$ is a sufficient condition. Naturally we can construct the basis $(e_1,f_1,e_2,f_2)$ for the lattice $\sLambda_{1,1}[N_1]\oplus \sLambda_{1,1}[N_2]$, satisfying
\lceq{
(e_i,e_j) = (f_i,f_j) = 0, \quad (e_i,f_j) = \delta_{ij} N_i \, ,
}
for all $i,j$. By B\'{e}zout, there exist integers $p,q$ such that $pN_1+qN_2 = 1$. We can define
\lceq{
u\coloneqq e_1 + e_2, \quad v \coloneqq p f_1 + q f_2, 
}
such that $(u,u) = (v,v) = 0$ and $(u,v) = 1$, which is isomorphic to the basis of $\sLambda_{1,1}$. One can also construct 
\lceq{
w \coloneqq q N_2 e_1 - p N_1 e_2, \quad z \coloneqq N_2 f_1 - N_1 f_2,
}
with $(w,u) = (w,v) = (z,u) = (z,v) = (w,w) = (z,z) = 0$ and $(w,z) = N_1 N_2$, therefore the lattice generated by $w$ and $z$ is isomorphic to $\sLambda_{1,1}[N_1 N_2]$ and we have constructed the isomorphism $\sLambda_{1,1}[N_1] \oplus \sLambda_{1,1}[N_2] \cong \sLambda_{1,1} \oplus \sLambda_{1,1}[N_1 N_2]$. One checks in particular that the change of basis is represented by an element of $\SL(4,\Z)$: in the ordered basis $(e_1,f_1,e_2,f_2)$ the columns of $(u,v,w,z)$ form the matrix
\lceq{
\mx{1 & 0 & q N_2 & 0 \\ 0 & p & 0 & N_2 \\1 & 0 & -pN_1 & 0 \\ 0 & q & 0 & -N_1},
}
whose determinant is $(p N_1+q N_2)^2 =1 $. 

\subsection{Atkin--Lehner group}
\label{appB_subB1}
In this appendix we record the elementary properties of the Atkin--Lehner product used in the square-free analysis. 

Since we assume $N$ is square free, all the divisors of $N$ are Hall divisors. Then the coset representatives \eqref{cosetR} are Atkin--Lehner involutions, which normalise $\Gamma_0(N)$ and satisfy $\sigma_\ell^2 \in \Gamma_0(N)$. 
If $\ell$ and $\ell'$ are both Hall divisors of $N$, the coset multiplication rule takes a particularly simple form,
\be 
\label{eq:sec5_sigma_Hall}
\sigma_\ell \sigma_{\ell'} = 
\sigma_{\frac{ N \gcd(\ell,\ell')^2}{\ell\,\ell'}} \, \gamma(\ell,\ell') ,
\qquad \text{for some } \gamma(\ell,\ell')\in \Gamma_0(N)\, .
\ee
It follows that the binary operation defining \eqref{eq:sec5_sigma_multi} is commutative and given by 
\be \ell * \ell' = \frac{N \gcd(\ell,\ell')^2}{\ell\,\ell'} \; . \ee
Because $\sigma_N= \mathds{1}$, one has $N * \ell = \ell * N = \ell$. When $N$ is prime, the only coset representatives are the Fricke involution  $\sigma_1$ and the identity $\sigma_N$. 

\vskip 5mm

We define $N$ such that it is a product of primes $p$ and $\nu_p(n) \in \{0,1\}$ the $p$-valuation of~$n$
\be n = \prod_{p} p^{\nu_p(n)}\; . \ee
One computes that for two divisors $\ell$ and $\ell'$ of $N$ 
\be \ell* \ell' = \frac{N {\rm gcd}(\ell,\ell')^2}{\ell \ell'} = \frac{N}{\ell} \prod_{p|\ell'} p^{2\nu_p(\ell)-1} \; ,  \ee
which  permits to compute that 
\be \sum_{\ell'|N} \frac{N}{\ell*\ell'} = \ell \prod_{p|N} \bigl( 1+ p^{1-2\nu_p(\ell)}\bigr) = \prod_{p|N} (p+1) = \sigma_1(N)\; . \ee
One has also
\be (\ell*\ell'')*\ell' =(\ell*\ell')*\ell''= \frac{N}{\ell*\ell'} \prod_{p|\ell''} p^{(2\nu_p(\ell)-1)(2\nu_p(\ell')-1)}\; , \ee
such that
\bea \sum_{\ell''|N} (-1)^{\omega(\ell*\ell'')} \frac{N}{\ell*\ell''} \frac{N}{\ell''*\ell'}&=&\sum_{\ell''|N} (-1)^{\omega(\ell'')} \frac{N}{\ell''} \frac{N}{(\ell*\ell'')*\ell'} \CR
&=& N \ell * \ell' \prod_{p|N}\Bigl( 1- p^{-1 - (2\nu_p(\ell)-1)(2\nu_p(\ell')-1)}\Bigr) \; , \eea
where $\omega(\ell)$ is the number of primes dividing $\ell$. If $\ell\ne \ell'$ one gets at least one prime such that  $(2\nu_p(\ell)-1)(2\nu_p(\ell')-1)=-1$ and this expression vanishes, whereas if $\ell=\ell'$ one has $(2\nu_p(\ell)-1)(2\nu_p(\ell')-1)=1$ for all primes and so 
\be \sum_{\ell''|N} (-1)^{\omega(\ell*\ell'')} \frac{N}{\ell*\ell''} \frac{N}{\ell''*\ell'}= \delta_{\ell,\ell'} \prod_{p|N}(p^2-1) \; . \label{InverseAB} \ee
One computes similarly that 
\be \sum_{d|\ell'} (-1)^{\omega( \ell*d)} \frac{N}{\ell*d} = (-1)^{\omega(\frac{N}{\ell})} \ell \prod_{p|\ell'} (1-p^{1-2\nu_p(\ell)})  = (-1)^{\omega\bigl(\frac{N}{\ell{\rm gcd}(\ell',\frac{N}{\ell})}\bigr)} {\rm gcd}( \ell, \tfrac{N}{\ell'})   \prod_{p|\ell'} (p-1) \; . \label{TraceB} \ee

\subsection{Computation of the singular theta lift}
In this appendix we compute the theta lift~\eqref{eq:sec5_Theta_lift} for $f$ a modular function of $\Gamma_0(N)$ with $N$ square-free. According to the formula 
\be 
\label{eq:appB_index}
\Bigl| SL(2,\Z) / \Gamma_0(N) \Bigr| = N \prod_{p | N} \Bigl( 1 + \frac{1}{p}\Bigr) = \sum_{\ell |N} \frac{N}{\ell  {\rm gcd}(\ell , \frac{N}{\ell})} \varphi\Bigl( {\rm gcd}(\ell , \frac{N}{\ell }) \Bigr)\ee
the $SL(2,\Z) / \Gamma_0(N)$ coset elements are labelled by the divisor $\ell$ of $N$ as well as a translation in $\Z_{N/\ell}$. We use the Poisson summation formula over $m_1,m_2$ to express the Narain partition function as
\bea  \Gamma_{\sLambda_{1,1}  \oplus \sLambda_{1,1}[N]}(T,U) &=& \hspace{1.8mm}\tau_2\hspace{-2.8mm} \sum_{\substack{m_1,n_1,n_2 \in \Z\\ m_2 \in  N \Z}} e^{ - \frac{\pi \tau_2}{T_2 U_2} \bigl| ( T,-1) \bigl({}^{n_1}_{m_2}{}^{\ n_2}_{-m_1}\bigr) (^{\; U}_{-1}) \bigr|^2 + 2\pi i  \tau \det\bigl({}^{n_1}_{m_2}{}^{\ n_2}_{-m_1}\bigr)} \CR
&=& \hspace{1.8mm} \frac{T_2}{N}\hspace{-2.8mm} \sum_{\substack{m_1,n_1,n_2 \in \Z\\ m_2 \in  \frac1 N \Z}} e^{ - \frac{\pi T_2}{\tau_2 U_2} \bigl| ( 1,\bar \tau ) \bigl({}^{m_2}_{n_2}{}^{m_1}_{n_1}\bigr) (^{-1}_{\; U}) \bigr|^2 - 2\pi i  \bar T  \det \bigl({}^{m_2}_{n_2}{}^{m_1}_{n_1}\bigr) } \, .\eea
One checks that the lattice $\sLambda_{1,1} \oplus \sLambda_{1,1}[N]$ is invariant under the transformations
\lceq{
\mx{n_1 & n_2 \\ m_2 & -m_1} \rightarrow \gamma_{\ell}' \mx{n_1 & n_2 \\ m_2 & -m_1}  \gamma_\ell\; , \qquad \mx{m_2 & m_1 \\ n_2 & n_1 } \rightarrow \gamma_{\ell}' \mx{m_2 & m_1 \\ n_2 & n_1 } \gamma_\ell^\intercal\; , 
}
for a fixed Hall divisor $\ell|| N$, with $\gamma_\ell$ and $\gamma_\ell'$ two  Atkin--Lehner  group elements with the same $\ell$. It follows that the combined transformation
\lceq{
\label{eq:appC_sLambda_prop}
U \rightarrow \frac{\frac{N}{\ell} a U - b}{N c U - \frac{N}{\ell} d }, \quad T \rightarrow \frac{\frac{N}{\ell} a' T - N c'}{b' T - \frac{N}{\ell} d'},
}
is a symmetry of the Narain partition function. 

For completeness we give the equivalent formula with insertion of left-moving momenta 
\bea  && \Gamma_{\sLambda_{1,1}  \oplus \sLambda_{1,1}[N]}(\tau,T,U)[\, p_L^k\, ]  \\
&=& \hspace{1.8mm}\tau_2\hspace{-2.8mm} \sum_{\substack{m_1,n_1,n_2 \in \Z\\ m_2 \in  N \Z}} \hspace{-4mm} e^{ - \frac{\pi \tau_2}{T_2 U_2} \bigl| ( T,-1) \bigl({}^{n_1}_{m_2}{}^{\ n_2}_{-m_1}\bigr) (^{\; U}_{-1}) \bigr|^2 + 2\pi i  \tau \det\bigl({}^{n_1}_{m_2}{}^{\ n_2}_{-m_1}\bigr)}  \frac{1}{(2 T_2 U_2)^{\frac{k}{2}}}  \Bigl( ( \bar T,-1) \bigl({}^{n_1}_{m_2}{}^{\ n_2}_{-m_1}\bigr) (^{\; U}_{-1})  \Bigr)^k \CR
&=& \hspace{1.8mm} \frac{T_2}{N}\hspace{-2.8mm} \sum_{\substack{m_1,n_1,n_2 \in \Z\\ m_2 \in  \frac1 N \Z}}  \hspace{-4mm} e^{ - \frac{\pi T_2}{\tau_2 U_2} \bigl| ( 1,\bar \tau ) \bigl({}^{m_2}_{n_2}{}^{m_1}_{n_1}\bigr) (^{-1}_{\; U}) \bigr|^2 - 2\pi i  \bar T  \det \bigl({}^{m_2}_{n_2}{}^{m_1}_{n_1}\bigr) } \left( \frac{T_2}{2 U_2 \tau_2^2} \right)^{\frac{k}{2}} \Bigl( ( 1,\bar \tau ) \bigl({}^{m_2}_{n_2}{}^{m_1}_{n_1}\bigr) (^{-1}_{\; U})  \Bigr)^k \, ,\nonumber \eea
which allows to prove that 
\bea && \Gamma_{\sLambda_{1,1}  \oplus \sLambda_{1,1}[N]}(\tau,T,U)[\, p_L^k\, ] \\
&=& ( - i \sqrt{N} \tau)^{-k} \left( \frac{-U}{\bar U } \right)^{\frac{k}{2}} \Gamma_{\sLambda_{1,1}  \oplus \sLambda_{1,1}[N]}\Bigl(- \frac{1}{N \tau} ,T, - \frac{1}{N U} \Bigr)[\, p_L^k\, ] \CR 
&=& \Bigl( - i \sqrt{\tfrac{N}{\ell}} ( \ell \tau + s_\ell) \Bigr)^{-k} \! \left( \frac{-\ell U- s_\ell }{\ell \bar U  + s_\ell } \right)^{\frac{k}{2}} \Gamma_{\sLambda_{1,1}  \oplus \sLambda_{1,1}[N]}\Bigl( \frac{\frac{N}{\ell} \tau + r_\ell}{N \tau + \frac{N}{\ell} s_\ell}  ,T,  \frac{\frac{N}{\ell} U + r_\ell}{N U + \frac{N}{\ell} s_\ell} \Bigr)[\, p_L^k\, ] \; , \nonumber \eea
for all Hall divisors $\ell|| N$ different from $1$ and $N$.

To define the singular theta lift, we introduce the $SL(2,\Z)$ continuous modular function defined by continuity from the function $\sigma_2(\tau) = \tau_2$ on the fundamental domain $\tau_1\in [-1/2,1/2]$ and $|\tau|>1$. We introduce the regularised theta lift \cite{Borcherds:1996uda}
\bea \Theta[f_N](T,U)_{L,\epsilon} &=&  \int_{\mathcal{F}_N^L} \frac{d^2\tau}{\tau_2^{\, 2}} \sigma_2(\tau)^{-\epsilon} f_N(\tau) \Gamma_{\sLambda_{1,1} \oplus \sLambda_{1,1}[N]} (\tau,T,U)  \label{RegularisedTheta} \\
&\coloneq& \int_{\mathcal{F} \cap \{ \tau_2< L\}}\frac{d^2\tau}{\tau_2^{\, 2}} \tau_2^{-\epsilon}  \hspace{-4mm} \sum_{\gamma \in SL(2,\Z) / \Gamma_0(N) }\hspace{-4mm} f_N(\gamma\cdot \tau) \Gamma_{\sLambda_{1,1} \oplus \sLambda_{1,1}[N]} (\gamma\cdot \tau, T,U)  \; . \nonumber \eea
For Re[$\epsilon$] sufficiently large and $T_2> N {\rm Im}[ U \sigma_\ell]$ for all $\ell | N$, the limit $L\rightarrow \infty$ converges and the resulting integral admits an analytic continuation to a meromorphic function of $\epsilon \in \C$. We define the singular theta lift as the constant term in the $\epsilon$ Laurent series at $0$, up to an additive constant that we shall fix to get a simple expression. 

We shall use the orbit method \cite{McClain:1986id,MR993311,Dixon:1990pc} with the assumption  $T_2> N {\rm Im}[ U \sigma_\ell]$ for all $\ell | N$, and in particular $T_2> N U_2$. Note that this computation could have been done using the method introduced in \cite{Angelantonj:2012gw,Angelantonj:2013eja}.

For most of the orbits, the contribution will be analytic at $\epsilon=0$ and we will not write the regulator function $\sigma_2(\tau)^{-\epsilon}$ for short. Note that the regulator only affects the identity \eqref{tauU} by an additive constant independent of $U$ and $T$. Importantly, \eqref{tauU} and the combined transformations \eqref{eq:appC_sLambda_prop} are not symmetries if $\ell$ is not a Hall divisor. The symmetry of the theta lift is therefore $\Z_2 \ltimes ( \Gamma_{0*}(N)\times_{\Z_2} \Gamma_{0*}(N))$, where $\Z_2$ exchanges $T  \leftrightarrow NU$. 

\subsection*{Decomposition into orbits}
Notice that, due to the modular properties of the lattice sum and the modular invariance of the integrand, if two integer pairs $(m_1,n_1)$ and $(\hat m_1,\hat n_1)$ are related by a $\Gamma_0(N)$ transformation $\gamma$, namely
\[
(\hat{m}_1,\hat{n}_1)^{\intercal} = \gamma\,(m_1,n_1)^{\intercal},
\]
then their contributions to the sum are identical, being related by a change of the integration variable $\tau \rightarrow \gamma^{-1} \cdot \tau$. 
As a consequence, the summation over $m_1$ and $n_1$ can be organized by decomposing the lattice sum into orbits of $\Gamma_0(N)$ acting on the integer vector $(m_1,n_1)^{\intercal}$. For convenience we define $(m_1,n_1) = \gcd (m_1, n_1) (m_1',n_1')$ and we classify the orbits using the pair $(m_1', n_1')$.

Three classes of orbits can be treated in a particularly simple manner. 
The first is the trivial orbit, for which $m_1' = n_1' = 0$; we denote its contribution by $\Gamma_{(0,0)}$. The second class consists of orbits with $n_1' \equiv 0 \mod N$ while $m_1' \neq 0$. The representative element of this orbit is simply $(1,0)^\intercal$, and we label the corresponding contribution by $\Gamma_{(1,0)}$. To verify this is the case, we need to confirm there exists a $\Gamma_0 (N)$ matrix such that 
\lceq{
\mx{m_1' \\ n_1'} = \mx{a & b \\ c & d} \mx{1 \\ 0} \, , \quad  ad-bc = 1 \quad \text{and} \quad c \equiv 0\mod N\, .
}
One can easily see that we can set $a = m_1'$ and $c = n_1'$. The existence of $b,d$ is guaranteed by B\'{e}zout. The stabilizer of this representative element is the translation group $\Gamma_\infty =  \langle T\rangle$, where $T$ is the modular translation $T\cdot \tau =\tau + 1$. 

The third class of orbits consists of orbits with $n_1' \not\equiv 0 \mod N$ but $\gcd (\hat{n}_1', N) = 1$, and $m_1'\neq 0$. The representative element corresponding to this type of orbit is simply $(0,1)^\intercal$, whose contribution is labeled as $\Gamma_{(0,1)}$. Similarly, we need to prove there exists a $\Gamma_0(N)$ matrix such that 
\lceq{
\mx{m_1' \\ n_1'} = \mx{a & b\\ c & d} \mx{0 \\ 1}, \quad ad- bc = 1 \quad \text{and} \quad c \equiv 0 \mod N \, .
}
One can set $b = m_1'$ and $d = n_1'$. By B\'{e}zout theorem there exist two integers $a_0$ and $c_0$ such that $a_0 d - b c_0  =1$ since $\gcd(m_1',n_1') = 1$. Notice that we can shift the integer $a_0 \rightarrow a_0 - x m_1'$, $c_0 \rightarrow c_0 - x n_1' $ for arbitrary integer $x$ without violating the condition $a_0 d  - b c_0  = 1$. Since $\gcd(n_1',N) = 1$, there exist integers $y,z$ such that $y n_1' + z N = 1$. If we choose $x = c_0 y$, $c = c_0 (1- y n_1') = c_0 z N$, which satisfies the requirement. The stabilizer of this representative is the $S \Gamma_{N,\infty}S^{-1}$, where for positive integer $y$ 
\lceq{
\Gamma_{y,\infty} := \left\{\mx{1 & x \\ 0 & 1}, x \in y \Z\right\} \, .
}
In addition, there exists a final class of orbits, labeled by a divisor $\ell \mid N$ as $\Gamma_{(1,-\ell)}$. 
This case is slightly more subtle, and we therefore examine it in greater detail below. Let us set gcd($n'_1,N)=\ell$. By assumption gcd($m_1',n_1')=1$ so that there exist $b,d$ coprime such that $d  m_1'  + b n_1' = 1$. One can then define $a = m_1' - b \ell$ and $c = - n_1' - d \ell $, such that 
\be \left( \begin{array}{cc} d \ & \ b\\ c \ & \ a \end{array}\right) \left( \begin{array}{c} m_1'\\ n_1' \end{array}\right)  = \left( \begin{array}{c} 1\\- \ell \end{array}\right)\; .  \ee
We need to prove that one can choose $b,d$ such that $c=0$ mod $N$. Note that $d$ and $b$ are chosen up to $d \rightarrow d + x n_1'$ and $b \rightarrow b - x m_1'$, with $x$ arbitrary integer, corresponding to the redefinition 
\be \left( \begin{array}{cc} d \ & \ b\\ c \ & \ a \end{array}\right)\rightarrow \left( \begin{array}{cc} 1-\ell x  \ & -x \\ x \ell^2  \ & \ 1+\ell x  \end{array}\right) \left( \begin{array}{cc} d \ & \ b\\ c \ & \ a \end{array}\right)\; , \ee
that stabilise $(1,-\ell)$. 
By assumption $\ell |n_1'$ so that $\ell|c$. Therefore $c$ is determined up to $c \rightarrow c - n_1' \ell  x = -n_1'-d \ell - n_1' \ell  x$. To satisfy $c \equiv 0 \mod N$, we need to prove there exist $x$ and $y$ such that 
\be d + \frac{n_1'}{\ell}= - x n_1' - y \frac{N}{\ell} \; . \ee
By hypothesis gcd$(n_1',N)=\ell$. Because $N$ is square-free we have $\gcd (n_1',N/\ell)=1$. There exist such $x$ and $y$ by B\'{e}zout. We have shown therefore that the matrix $\gamma \in \Gamma_0(N)$ exists such that $(m_1,n_1)^\intercal = \gcd(m_1,n_1) \gamma  (1,- \ell)^\intercal$ up to $\gamma \in \Gamma_0(N)$ for $\ell|N$. This reads 
\be (m_1,n_1) = {\rm gcd}(m_1,n_1) ( a + b \ell , -c -d \ell ) \ee
for $d  m_1'  + b n_1' = 1$,  $a = m_1' - b \ell$ and $c = - n_1' - d \ell = y N   $. By conjugation 
\be \left( \begin{array}{cc} 1  \ & 0 \\ \ell  \ & \ 1 \end{array}\right)   \left( \begin{array}{cc} 1-\ell  x  \ & -x \\ x \ell^2  \ & \ 1+\ell  x  \end{array}\right)  \left( \begin{array}{cc} 1  \ & 0 \\ -\ell  \ & \ 1 \end{array}\right)= \left( \begin{array}{cc} 1  \ & -x \\ 0  \ & \ 1 \end{array}\right) \ee
so we see that the stabiliser of $(1,-\ell)$ in $\Gamma_0(N)$ is conjugate to the upper triangular subgroup with $x \ell^2= 0$ mod $N$. Since $N$ is square-free, this implies that $x = 0$ mod $\frac{N}{\ell }$ consistently with the size of the cusp.\footnote{For a prime $p|N$ that does not divide ${\rm gcd}(\ell,N/\ell)$ there is no question, since either the prime divides $\ell$ and so does not need to divide $x$ or conversely. If $p|{\rm gcd}(\ell,N/\ell)$ we say that $p^\alpha|\ell$ and $p^\beta |\frac{N}{\ell}$ for some integers $\alpha$ and $\beta$. Then $p^{\alpha+\beta}|N$ and must therefore divide $x \ell^2$. If $\alpha\ge \beta$, $p^{\alpha+\beta}|\ell^2$ there is no constraint and indeed $\frac{N}{\ell {\rm gcd}(\ell,N/\ell)}$ is not divisible by $p$. If instead $\beta>\alpha$ we must have $p^{\beta-\alpha} | x$ and $p^{\beta-\alpha} |\frac{N}{\ell {\rm gcd}(\ell,N/\ell)}$.} 
We have concluded that the sum over $m_1$ and $n_1$ can be split into the orbit contributions: 
\lceq{
\Theta[f_N]  = \Gamma_{(0,0)} + \Gamma_{(1,0)} + \Gamma_{(0,1)} + \sum_{\substack{\ell|N\\ \ell \ne 1,N}} \Gamma_{(1,-\ell)} \, .
}
Now we shall  derive explicitly each contribution. 
\subsubsection*{The orbit with the representative element $(0,0)$} Let us start with the simplest case
\longeq{
\Gamma_{(0,0)} &= \frac{T_2}{N} \int_{\mathcal{F}_N} \frac{d \tau_1 d \tau_2}{\tau_2^2} \sum_{\substack{m_2 \in \Z/N  \\ n_2 \in \Z} } f_N(\tau) \exp \left\{ - \frac{ \pi T_2 }{U_2 \tau_2} |m_2 + \tau n_2|^2 \right\}\\  &=  \frac{T_2}{N} \int_{\mathcal{F}_N} \frac{d \tau_1 d \tau_2}{\tau_2^2} \sum_{\substack{m_2 \in \Z  \\ n_2 \in  N\Z} } f_N(\tau) \exp \left\{ - \frac{ \pi T_2 }{N^2 U_2 \tau_2} |m_2 + \tau n_2|^2 \right\} \, .
}
The integrand is explicitly modular invariant and we can further decompose the sum over  $(m_2,n_2)$  into $\Gamma_0(N)$ orbits. Suppose 
$(m_2, n_2) = g (m_2',n_2')$ where $g = \gcd (m_2,n_2)$. We can identify the contribution of each summation by studying the corresponding orbit. More precisely, we can write
\lceq{
\Gamma_{(0,0)} = \Gamma_{(0,0)}^{(0,0)} + \Gamma_{(0,0)}^{(1,0)} + \Gamma_{(0,0)}^{(0,1)} + \sum_{\substack{\ell | N \\ \ell \neq 1, N}}
\Gamma_{(0,0)}^{(1,-\ell)}.
}
For the trivial orbit $(m_2,n_2) = (0,0)$, the integral is simply
\lceq{
\Gamma_{(0,0)}^{(0,0)} = \frac{T_2}{N} \int_{\mathcal{F}_N} \frac{d \tau_1 d \tau_2}{\tau_2^2} f_N(\tau)  = \frac{T_2}{N}\int_{\mathcal{F}} \frac{d^2\tau}{\tau_{2}^{\, 2}} \sum_{\ell|N} \sum_{i=0}^{\frac{N}{\ell}-1}f_{\ell}\biggl(\frac{\ell}{N}\bigl(  \tau+i-r_\ell \bigr)\biggr) ,
}
where we unfold the integral by using
\be 
\label{eq:appB_f_trans}
f_N\Bigl(- \frac{1}{\tau}\Bigr) = f_1\Bigl(\frac{\tau}{N}\Bigr) \; , \quad  f_N\Bigl(\frac{\tau}{\ell \tau +1}\Bigr) =
f_\ell\Biggl( \frac{\ell}{N}\bigl(  \tau-r_\ell\bigr)  \Biggr)\; .\ee
The $q$-expansion of the modular function 
\bea&&  \sum_{\ell|N} \sum_{i=0}^{\frac{N}{\ell}-1}f_{\ell}\biggl(\frac{\ell}{N}\Bigl(  \tau+i-r_\ell \Bigr)\biggr) = c_N(-1)q^{-1} + \sum_{\ell|N} \frac{N}{ \ell} c_\ell(0)  + \mathcal{O}(q), \, \, q=e^{2\pi i \tau} \, , \eea
which allows one to directly apply the standard formula of~\cite{Lerche:1987qk} for the integration of modular functions,
\lceq{
\int_\mathcal{F} \frac{d \tau_1 d \tau_2}{ \tau_2^2} f(\tau ) = \left. \frac{1}{\pi} G_2 (\tau) f(\tau ) \right|_{q^0 \,  \text{coefficient}}, \quad 
G_2(\tau) = \frac{\pi^2}{3} \left(1 - 24\sum_{n=1}^{+\infty} \frac{nq^{n}}{1-q^{n}}\right) \, .
}
Therefore 
\lceq{
\Gamma_{(0,0)}^{(0,0)}=\frac{\pi T_2}{3 N} \Biggl( \sum_{\ell|N} \frac{N}{\ell} c_\ell(0) - 24 c_N(-1)\Biggr)  \, .
}
The next term is the contribution corresponding to the case when $n_2' \equiv 0 \mod N$ with $m_2' \neq 0$, associated with the representative element $(m_2' , n_2') = (1,0)$. Notice the stabilizer of this kind of orbit is $\Gamma_\infty$, the integration region is unfolded to $\Gamma_\infty \backslash \mathcal{H} = \R^+_{\tau_2} \times [-1/2,1/2)_{\tau_1}$ \longeq{
\Gamma_{0,0}^{(1,0)} = \frac{2 T_2}{N} \sum_{g=1}^\infty \int_{-\frac{1}{2}}^{\frac{1}{2}} d \tau_1 \int_0^{+\infty} \frac{d \tau_2}{\tau_2^2}  f_N(\tau) \exp \left\{ - \frac{ \pi T_2 g^2 }{N^2 U_2 \tau_2} \right\}=\frac{\pi N U_2}{3} c_N (0).
}
When $n_2' \not\equiv 0 \mod N$ but $\gcd(n_2', N) = 1$ with $m_2' \neq 0$, the representative element is $(m_2',n_2') = (0,1)$ with the stabilizer $S \Gamma_{N,\infty}S^{-1}$. Notice that since $n_2 \equiv 0 \mod N$, now $g \in N\Z$. The integration region is unfolded to $S \Gamma_{N,\infty} S^{-1} \backslash  \mathcal{H} $. It is convenient to apply the transformation $\tau \rightarrow -1/\tau$, which gives rise  to the integration region $\R_{\tau_2}^+ \times [-N/2, N/2]$. More explicitly, we have 
\longeq{
\Gamma_{(0,0)}^{(0,1)} = \frac{2T_2}{N} \int_{-N/2}^{N/2} d \tau_1 \int_0^{+\infty} \frac{d \tau_2}{\tau_2^2} f_1 \left(\frac{\tau}{N}\right) \sum_{g = 1}^{+\infty} \exp \left(- \frac{\pi T_2 g^2}{U_2 \tau_2}\right) = \frac{\pi U_2}{3} c_1(0),
}
where we  apply the first equation in~\eqref{eq:appB_f_trans}. For the last type of orbit, we can set $\gcd (n_2', N) = \ell$ while $\ell \neq 1,N$. In this case, $g \in \frac{N}{\ell} \Z$.  We have already determined  the corresponding representative elements  $(m_2',n_2') = (1,-\ell)$ with the stabilizer $\gamma_\ell \Gamma_{\frac{N}{\ell}}\gamma_\ell^{-1}$, where 
\lceq{
\gamma_\ell = \mx{1 & 0 \\ \ell & 1} \, ,
}
and the integration region is unfolded to $\gamma_\ell \Gamma_{\frac{N}{\ell}}\gamma_\ell^{-1} \backslash \mathbb{H}$. Similarly, we change the variable $\tau \rightarrow \frac{\tau}{\ell \tau +1}$ so that the integration domain becomes more tractable,
\longeq{
\Gamma_{(0,0)}^{(1,-\ell)} &= \frac{T_2}{N} \sum_{\substack{g \in \frac{N}{\ell}\Z \\ g \neq 0}} \int_{- \frac{N}{2\ell} }^{ \frac{N}{2\ell} }d\tau_1 \int_0^\infty \frac{d\tau_2}{\tau_2^{2}}  f_{\ell}\biggl(\frac{\ell}{N}\bigl(  \tau-r_\ell \bigr)\biggr) \exp \left(- \frac{\pi T_2 g^2}{U_2 \tau_2 N^2}\right) = \frac{\pi U_2 \ell }{3} c_\ell (0). 
}
In total, the contribution from the orbit  with representative element $(m_1',n_1') = (0,0)$ is 
\longeq{
\label{eq:appB_(0,0)}
\Gamma_{(0,0)} &= \frac{\pi T_2}{3 N} \Biggl( \sum_{\ell|N} \frac{N}{\ell} c_\ell(0) - 24 c_N(-1)\Biggr)  + \frac{\pi U_2}{3} \sum_{\ell|N} \ell c_\ell (0) \, .
}
\subsection*{The orbit with the representative element $(1,0)$}
For later convenience, we define $k=\gcd(m_1,n_1)\ne 0$. We have already determined the stabilizer of the representative element as $\Gamma_{\infty}$ and the integral is simply
\longeq{
\Gamma_{(1,0)} &=  2 \sqrt{T_2 U_2} \sum_{k=1}^{\infty}\sum_{\substack{n_2 \in \Z \\ m_2 \in N \Z}}\int_{-\frac{1}{2}}^{\frac{1}{2}} d \tau_1 \int_{0}^\infty  \frac{d \tau_2}{\tau_2^2} \sqrt{\tau_2} f_N(\tau) \times \\ 
& \hspace{-0.5cm}\exp \left\{- \frac{\pi T_2 U_2 k^2}{\tau_2} + 2 \pi i (U_1 m_2 + T_1 n_2) k 
- 2 \pi i \tau_1 m_2 n_2 - \pi \tau_2 \frac{U_2}{T_2} m_2^2 - \pi \tau_2 \frac{T_2}{U_2} n_2^2\right\} \, .
}
To continue, we split the summation over $m_2,n_2$ into four parts: $m_2 = n_2 = 0$, $m_2 \neq 0$ with $n_2 = 0$, $m_2 = 0$ with $n_2 \neq 0$ and $m_2n_2 \neq 0$. The first term turns out to diverge, so we must keep the regulator introduced in \eqref{RegularisedTheta}
\longeq{
\lim_{L\rightarrow \infty} \Gamma_{(1,0),{L,\epsilon}}^{(0,0)}&=  2 \sqrt{T_2 U_2} \sum_{k=1}^{\infty}\int_{-\frac{1}{2}}^{\frac{1}{2}} d \tau_1 \int_{0}^\infty  \frac{d \tau_2}{\tau_2^{3/2+\epsilon}}  f_N(\tau) \exp \left(- \frac{\pi T_2 U_2 k^2}{\tau_2}\right) \\
&= 2 \sqrt{T_2 U_2} c_N(0) \sum_{k=1}^\infty \int_0^\infty \frac{d\tau_2}{\tau_2^{3/2+\epsilon}}  \exp \left(- \frac{\pi T_2 U_2}{\tau_2} k^2\right) \, . 
}
The analytic continuation in $\epsilon$ gives 
\longeq{
\Gamma_{(1,0)}^{(0,0)}{}_\epsilon &= 2 \sqrt{T_2 U_2} c_N(0) \sum_{k=1}^\infty \int_0^\infty \frac{d\tau_2}{\tau_2^{3/2+ \epsilon} }  \exp \left(- \frac{\pi T_2 U_2}{\tau_2} k^2\right)\\
&= 2 c_N(0) \pi^{-\epsilon - \frac{1}{2}}\Gamma\left(\frac{1}{2} +\epsilon\right) (T_2 U_2)^{-\epsilon} \zeta\left(1 +2 \epsilon\right) \\
&= 2 c_N (0) \left[ \frac{1}{2 \epsilon} - \frac{1}{2} \log (T_2 U_2) + \frac{\gamma}{2} - \frac{1}{2 } \log (4 \pi) + O(\epsilon)\right]\; . 
}
We shall define accordingly the final result as $- c_N(0)\log (T_2 U_2)$ plus a finite constant. The second term gives
\begin{equation*}
\begin{aligned}
&\quad \ \Gamma_{(1,0)}^{(*,0)} \\
&=2 \sqrt{T_2 U_2}\int_{-\frac{1}{2}}^{\frac{1}{2}} \hspace{-2mm} d \tau_1 \int_{0}^\infty  \frac{d \tau_2}{\tau_2^{3/2}} f_N(\tau) \sum_{k = 1}^\infty \sum_{m_2 \in N\Z \smallsetminus\{ 0\} } \hspace{-4mm} \exp \left[- \frac{\pi T_2 U_2 k^2}{\tau_2} + 2 \pi i U_1 m_2 k  - \pi \tau_2 \frac{U_2}{T_2} m_2^2\right] \\
&=  - 2 c_N (0) \sum_{\tilde{m}_2 = 1}^\infty \log \left|1 - e^{2 \pi i N U \tilde{m}_2}\right|^2 \, . 
\end{aligned}
\end{equation*}
One computes in the same way that the third term gives 
\longeq{
\Gamma_{(1,0)}^{(0,*)} = - 2 c_N(0) \sum_{n_2 = 1}^\infty \log \left|1 - e^{2 \pi i T n_2}\right|^2 \, .
}
The last term in $\Gamma_{(1,0)}$ gives
\longeq{
\Gamma_{(1,0)}^{(*,*)} &= 2 \sqrt{T_2 U_2}\int_{-\frac{1}{2}}^{\frac{1}{2}} d \tau_1 \int_{0}^\infty  \frac{d \tau_2}{\tau_2^{3/2}} f_N (\tau) \sum_{k = 1}^\infty\sum_{\substack{m_2 \in N\Z \\ n_2 \in \Z\\ m_2 n_2 \neq 0}}   \\
    &\times \exp  \left[- \frac{\pi T_2 U_2 k^2}{\tau_2} + 2 \pi i (U_1 m_2 + T_1 n_2) k 
- 2 \pi i \tau_1 m_2 n_2 - \pi \tau_2 \frac{U_2}{T_2} m_2^2 - \pi \tau_2 \frac{T_2}{U_2} n_2^2\right] \, .
}
Since $f_N(\tau) = \sum_{n = -1} c_N (n) q^n$, the integration of $\tau_1$ will enforce $n = m_2 n_2$. Notice $m_2 \in N \Z$ and $n_2 \in \Z$, $n \ge - 1$, the pole in $f_N(\tau)$ cannot be level matched and only the terms with  $m_2 n_2 > 0$ contribute. One obtains eventually 
\lceq{
\Gamma_{(1,0)}^{(*,*)} = - 2 \sum_{\substack{ m_2 \in N\Z \\ n_2 \in \Z\\ m_2 n_2 > 0}} c_N(m_2 n_2) \log \left[1 - e^{2 \pi i (U_1 m_2 + T_1 n_2) - 2 \pi |n_2 T_2 + m_2 U_2|}\right] \, .
}
In total, the contribution from the orbit associated with the representative element $(m_1',n_1') = (1,0)$ is
\longeq{
\label{eq:appB_(1,0)}
\Gamma_{(1,0)} &=  - 2 c_N (0) \sum_{\tilde{m}_2 = 1}^\infty \log \left|1 - e^{2 \pi i N U \tilde{m}_2}\right|^2- 2 c_N(0) \sum_{n_2 = 1}^\infty \log \left|1 - e^{2 \pi i T n_2}\right|^2  \\
& \hspace{-0.8cm}- 2 \sum_{\substack{ m_2 \in N\Z \\ n_2 \in \Z\\ m_2 n_2 > 0}} c_N(m_2 n_2) \log \left[1 - e^{2 \pi i (U_1 m_2 + T_1 n_2) - 2 \pi |n_2 T_2 + m_2 U_2|}\right] - c_N (0)\log (T_2 U_2) \, ,
}
where we have renormalised the divergence. 
\subsection*{The orbit with the representative element $(0,1)$}
The computation of the contribution $\Gamma_{(0,1)}$ of the third orbit with representative element $(0,1)$ and stabilizer subgroup $S \Gamma_{N,\infty}S^{-1}$, is analogous to $\Gamma_{(1,0)}$. We simplify the integral through the change  of variable $\tau \rightarrow -1/\tau$ to get 
\longeq{
\Gamma_{(0,1)} &= \frac{2}{N} \sqrt{T_2 U_2} \int_{-N/2}^{N/2} d \tau_1 \int_0^\infty \frac{d \tau_2}{\tau_2^{3/2}} f_1 \left(\frac{\tau}{N}\right) \sum_{k=1}^\infty \sum_{\substack{m_2 \in \Z \\n_2 \in \Z/N}} \\
& \times
\exp  \left[- \frac{\pi T_2 U_2 k^2}{\tau_2} + 2 \pi i (U_1 m_2 + T_1 n_2) k 
- 2 \pi i \tau_1 m_2 n_2 - \pi \tau_2 \frac{U_2}{T_2} m_2^2 - \pi \tau_2 \frac{T_2}{U_2} n_2^2\right].
}
By direct computation one can show
\longeq{
\label{eq:appB_(0,1)}
\Gamma_{(0,1)} &= - c_1 (0) \log (T_2 U_2)- 2 c_1 (-1) \log \left|1 - e^{-2  \pi i \left(U - \frac{T}{N}\right)}\right|^2\\
&-  \sum_{m = 1}^\infty \log \left\{\left|1 - e^{2 \pi i U m}\right|^{4 c_1 (0)} \left|1 - e^{2 \pi i T m /N}\right|^{4 c_1 (0)}\right\}  \\ 
&-2 \sum_{m_2 n_2 > 0, m_2 \in \Z, n_2 \in \Z/N} c_1 (N m_2 n_2) \log \left[1 - e^{2 \pi i (U_1 m_2 + T_1 n_2) - 2 \pi |n_2 T_2 + m_2 U_2|} \right] \\
&= - c_1 (0) \log (T_2 U_2)-  \sum_{m = 1}^\infty \log \left\{\left|1 - e^{2 \pi i U m}\right|^{4 c_1 (0)} \left|1 - e^{2 \pi i T m /N}\right|^{4 c_1 (0)}\right\} \\ 
& \hspace{0.5cm}-2 \sum_{\substack{m_2 \in \Z \\ n_2 \in \Z/N \\m_2n_2\ne 0}} c_1 (N m_2 n_2) \log \left[1 - e^{2 \pi i (U_1 m_2 + T_1 n_2) - 2 \pi |n_2 T_2 + m_2 U_2|} \right] \, . 
}
The main difference in this case is that the argument $N m_2 n_2$ of $c_1(n)$ can be equal to $-1$, when $m_2 = \pm 1$ and $n_2 = \mp \frac{1}{N}$, and one gets a non-zero contribution to the logarithm that diverges when $T\rightarrow NU$. 

\subsection*{The orbit with the representative element $(1,-\ell)$}
The last term comes from the orbits with the representative element $(1,-\ell)$. After using the Poisson summation formula over $m_1$, for each $\ell | N$ ($\ell \neq 1,N$) the lattice sum becomes 
\begin{equation}
\begin{aligned}
\Gamma_\ell(\tau)
={}&
2\sqrt{T_2U_2\tau_2}
\sum_{\substack{n_2\in \Z\\ m_2\in N\Z}}
\sum_{k=1}^{\infty}
\exp\Biggl[
-\frac{\pi T_2U_2}{\tau_2}\,k^2 |1-\ell\tau|^2
\\
&\hspace{22mm}
-\pi\tau_2\frac{U_2}{T_2}
  \bigl(m_2+T_1k\ell\bigr)^2
-\pi\tau_2\frac{T_2}{U_2}
  \bigl(n_2+U_1k\ell\bigr)^2
\\
&\hspace{22mm}
-2\pi i\,\tau_1 m_2n_2
+2\pi i\,k(1-\ell\tau_1)
 \bigl(U_1m_2+T_1n_2+U_1T_1k\ell\bigr)
\Biggr].
\end{aligned}
\end{equation}
One can now use the change of variable associated to $\sigma_\ell$
\be \tau \rightarrow \tau_\ell = \frac{\frac{N}{\ell} \tau + r_\ell}{N \tau + \frac{N}{\ell} s_\ell} \; . \ee
To keep track of the fractional part of $m_2$, we decompose the summation set as
\be \{ m_2\;  | \; m_2 \in \tfrac1N\Z\} = \left\{ m + \frac{r}{\frac{N}{\ell}} + \frac{s}{\ell} \; | \; m\in \Z, r\in \Z_{N/\ell}\; , s\in \Z_{\ell} \right\} \ee
using that $\frac{N}{\ell}$ and $\ell$ are coprime. Because $n_2$ only appears through 
\be n_2 + \ell m_2 = n_2 + \ell m + s + \frac{r\ell^2}{N} \ee
we define $n=n_2+\ell m+s$, $n_r=n+\frac{r\ell^2}{N}$, and
\[
Y_{r,s}(\tau)\coloneqq m+\frac{r\ell}{N}+\frac{s}{\ell}+\Bigl(r_\ell+\frac{N}{\ell}\tau\Bigr)n_r .
\]
Then
\begin{align}
\Gamma_\ell(\tau_\ell)
={}&
\frac{2T_2}{N}
\sum_{s=0}^{\ell-1}
\sum_{r=0}^{\frac{N}{\ell}-1}
\sum_{m,n\in\Z}
\sum_{k=1}^{\infty}
e^{
-\frac{\pi T_2}{U_2\tau_2}\frac{\ell}{N}
\begin{pmatrix}
k & Y_{r,s}(\tau)
\end{pmatrix}
\begin{pmatrix}
|U|^2 & -U_1\\
-U_1 & 1
\end{pmatrix}
\begin{pmatrix}
k\\
Y_{r,s}(\bar\tau)
\end{pmatrix}
+2\pi i T_1 k n_r
}
\nonumber\\[2mm]
={}&
\frac{2\sqrt{T_2U_2\tau_2}}{\sqrt{\ell N}}
\sum_{s=0}^{\ell-1}
\sum_{r=0}^{\frac{N}{\ell}-1}
\sum_{m,n\in\Z}
\sum_{k=1}^{\infty}
\exp\Biggl[
-\pi\tau_2\frac{N}{\ell}
\left(
\frac{U_2}{T_2}m^2
+\frac{T_2}{U_2}n_r^2
\right)
\Biggr]
\nonumber\\
&\hspace{14mm}\times
\exp\Biggl[
-2\pi i\tau_1\frac{Nm}{\ell}n_r
-\frac{\pi}{\tau_2}T_2U_2\frac{\ell}{N}k^2
+2\pi i k\bigl(mU_1+n_rT_1\bigr)
\Biggr]
\nonumber\\
&\hspace{14mm}\times
\exp\Biggl[
-2\pi i m
\left(
\frac{r\ell}{N}
+\frac{s}{\ell}
+r_\ell n_r
\right)
\Biggr]
\nonumber\\[2mm]
={}&
2\sqrt{\frac{\ell T_2U_2\tau_2}{N}}
\sum_{\substack{m\in \ell\Z\\ n'\in \frac{\ell}{N}\Z}}
\sum_{k=1}^{\infty}
\exp\Biggl[
-\pi\tau_2\frac{N}{\ell}
\left(
\sqrt{\frac{U_2}{T_2}}\,m
+\sqrt{\frac{T_2}{U_2}}\,n'
\right)^2
-2\pi i\tau\frac{Nmn'}{\ell}
\Biggr]
\nonumber\\
&\hspace{14mm}\times
\exp\Biggl[
-2\pi i
\left(
\frac{1}{\ell}
+r_\ell
\right)mn'
-\frac{\pi}{\tau_2}T_2U_2\frac{\ell}{N}k^2
+2\pi i k\bigl(mU_1+n'T_1\bigr)
\Biggr].
\end{align}
Here we use Poisson summation in going to the second step. To get the third step one uses that the sum over $s$ projects the sum over $m$ to $m = 0 $ mod $\ell$. We also redefined $n^\prime\equiv  n{+} \tfrac{r \ell^2}{N}$ and, since $\ell$ and $\frac{N}{\ell}$ are coprime, one obtains that $n^\prime $ takes all values in $\frac{\ell}{N}\Z$. Moreover,
\be 1/\ell + r_\ell  = s_\ell \frac{N}{\ell^2} \ee
so the residual phase is trivial for $m\in \ell\Z$ and $n'\in \frac{\ell}{N}\Z$.
We have then the contribution from each orbit labeled by $(1,-\ell)$ 
\longeq{
\label{eq:appB_(1,-l)}
\Gamma_{(1,-\ell)} &= 2\sqrt{\frac{ \ell T_2 U_2  }{N}} \sum_{\substack{m \in \ell \Z \\ n \in \frac{\ell}{N} \Z   }}   \sum_{k=1}^\infty  \int_{- \frac12}^{\frac12 } d\tau_1 \int_0^\infty \frac{d \tau_2}{\tau_2^{3/2}}  f_\ell(\tau)  \times \\
& \exp \Biggl\{ - \pi \tau_2 \frac{N}{\ell}  \Bigl( \sqrt{\frac{U_2}{T_2}} m  +\sqrt{ \frac{T_2}{U_2}} n\Bigr)^2  -2\pi i \tau \frac{N mn }{\ell }   - \frac{\pi}{\tau_2} T_2 U_2 \frac{\ell}{N} k^2 + 2\pi i k ( m U_1 +n T_1 )\Biggr\}  \\
&= -2\sum_{\substack{m \in \ell \Z,  n \in \frac{\ell}{N} \Z  \\ (m,n) \neq (0,0) }} c_\ell(\tfrac{N mn}{\ell}  ) \log\Bigl[ 1-e^{- 2\pi | m U_2 + n T_2| + 2\pi i  ( m U_1 + n T_1)} \Bigr] 
}
Combining the results in~\eqref{eq:appB_(0,0)},~\eqref{eq:appB_(1,0)},~\eqref{eq:appB_(0,1)} and~\eqref{eq:appB_(1,-l)}, the theta lift of arbitrary $f_N$ can be explicitly expressed as
\begin{multline}
\Theta[ f_N] =\frac{\pi T_2}{3 N} \Biggl( \sum_{\ell|N} \frac{N}{ \ell} c_\ell(0)- 24 c_N(-1)\Biggr)
+\frac{\pi U_2}{3 } \Biggl( \sum_{\ell|N} \ell c_\ell(0) \biggr) - \sum_{\ell |N} c_\ell (0) \log \Bigl[ \frac{\ell^2}{N} U_2 T_2\Bigr] \\
-2\sum_{\ell |N}\sum_{\substack{m,n \in  \Z \\ (m,n) \ne (0,0)}} c_\ell(\ell mn) \log\Bigl[ 1-e^{- 2\pi  \ell | m U_2 + n \frac{T_{\scalebox{0.4}{$2$}}}{N}| + 2\pi i \ell  ( m U_1 + n \frac{T_{\scalebox{0.4}{$1$}}}{N})} \Bigr]  \ . 
\end{multline}

\bibliography{ref}
\bibliographystyle{utphys}
\end{document}